\documentclass{aa} 

\usepackage{amssymb} 
\newcommand{\grtsim}{\gtrsim}

\usepackage{graphicx}   
\usepackage{txfonts}    
\usepackage{float}      
\usepackage{natbib}     
\usepackage{subfigure}  
\usepackage{multirow}   
\usepackage{booktabs}   
\usepackage{caption}    
\usepackage{threeparttable}

\usepackage{hyperref}
\hypersetup{
    colorlinks=true,
    linkcolor=blue,
    filecolor=blue,      
    urlcolor=blue,
    citecolor=blue,
}

\usepackage{nameref}
\usepackage{varioref}
\usepackage{cleveref}   

\bibpunct{(}{)}{;}{a}{}{,} 

\begin{document}

   \title{Radio--FIR correlation: A probe into cosmic ray propagation in the nearby galaxy IC~342}
\titlerunning{Radio--FIR Correlation in IC~342}

   \author{M.R. Nasirzadeh
          \inst{1}
          \and
          F. S. Tabatabaei\inst{1,2,3}\fnmsep\thanks{Email:ftaba@ipm.ir} \and R. Beck \inst{2}
          \and V. Heesen\inst{4}  \and P. Howaida\inst{1} \and  M. Reina-Campos \inst {5} \and R. Paladino\inst{6} \and  R.-J.~Dettmar\inst{7} \and K.~T.~Chy\.zy\inst{8}
 } 
          
\institute{School of Astronomy, Institute for Research in Fundamental Sciences (IPM), PO Box 19395-5531, Tehran, Iran
\and Max-Planck Institut f\"ur Radioastronomie, Auf dem Hügel 69, 53121 Bonn, Germany
\and Max-Planck Institut f\"ur Astronomie, K\"onigstuhl 17, 69117 Heidelberg, Germany
\and Hamburger Sternwarte, Universität Hamburg, Gojenbergsweg 112, 21029 Hamburg, Germany  \and Canadian Institute for Theoretical Astrophysics (CITA), University of Toronto, 60 St George St, Toronto, M5S 3H8, Canada \and INAF - Istituto di Radioastronomia, via P. Gobetti 101, 40129 Bologna, Italy \and Ruhr University Bochum, Faculty of Physics and Astronomy, Astronomical Institute (AIRUB), 44780 Bochum, Germany \and Astronomical Observatory of the Jagiellonian University, ul. Orla 171, 30-244
Krak\'ow, Poland\\}

   \date{-}

 \abstract{Resolved studies of the correlation between the radio and far-infrared (FIR) emission from galaxies at different frequencies can unveil the interplay between star formation and relativistic interstellar medium (ISM). Thanks to the LOFAR LoTSS observations combined with the VLA, Herschel, and WISE data, we study the role of the cosmic rays and magnetic fields in the radio--FIR correlation on scales of $\grtsim$200~pc in the nearby galaxy IC342. {The thermal emission traced by the 22 $\mu$m emission, constitutes about 6\%, 13\%, and 30\%
 of the observed radio emission at  0.14, 1.4, 4.8 GHz,  respectively, in star forming regions and less in other parts}. The nonthermal spectral index becomes flatter at frequencies lower than 1.4\,GHz ($\alpha_n$= $-0.51\pm 0.09$, $S_{\nu}\propto \nu^{\alpha_n}$) than between 1.4 and 4.8\,GHz ($\alpha_n$= $-1.06\pm 0.19$) on average and this flattening {occurs not only in star-forming regions but also in diffuse ISM}. The radio--FIR correlation holds at all radio frequencies; however, it is tighter at higher radio frequencies. {A multi-scale analysis shows that this correlation cannot be maintained on small scales due to diffusion of cosmic ray electrons (CREs). The correlation breaks on a larger scale ($\simeq$320~pc) at 0.14~GHz than at 1.4~GHz ($\simeq$200~pc) indicating that those CREs traced at lower frequencies have diffused a longer path in the ISM. We find that the energy index of CREs becomes flatter in star forming regions in agreement with previous studies. Cooling of CREs  due to the magnetic field is evident globally only after compensating for the effect of star formation activity which both accelerate CREs  and amplify magnetic fields. Compared with other nearby galaxies, it is shown that the smallest scale of the radio--FIR correlation is proportional to the CREs propagation length on which the ordered magnetic field has an important effect.}}


\maketitle

\section{Introduction}
\label{sec:intro}

The correlation between the radio and far-infrared (FIR) continuum luminosities from star-forming galaxies is one of the tightest in astronomy, spanning over more than four orders of magnitude in luminosity \citep{intro1&Helou,intro2&dejong,intro3&Gavazzi}. This correlation, which is globally almost linear, is linked to massive star formation and is therefore widely used as a tracer of the star formation rate \citep[SFR, e.g.][]{intro4&Condon} and a tool to distinguish active Galactic nuclei (AGN) from starbursts at high redshifts \citep[e.g.][]{intro5&Sargent}. However, understanding the radio-FIR correlation needs inclusion of several factors and possible sources of the radio and FIR continuum emission. Particularly, this becomes important considering that the FIR emission can emerge from not only a warmer dust component heated by young massive stars but also from a colder dust component heated by non-ionizing UV/optical photons of old, solar mass stars \citep{intro6&Xu,intro7&Suvage}, that is, diffuse interstellar radiation field \citep[ISRF,][]{intro8&Draine}. Investigating the correlation at different FIR frequencies can show us the effect of the dust temperature or the ISRF. 
   
Similarly, the radio continuum emission emerges from two main radiation mechanisms; the thermal free-free emission and the nonthermal synchrotron emission. Although the thermal emission is mainly produced in HII regions and is related to young massive stars \cite{intro9&Osterbrock},  the link between the nonthermal emission and massive stars
 can be complicated by the fact that cosmic ray electrons (CRes) propagate away from their birthplace (supernova remnants, SNRs) along the magnetic fields that control the synchrotron emission in the interstellar medium (ISM). Hence, the synchrotron--FIR correlation can actually manifest the star formation--ISM interplay \citep[see Fig.~3 in][]{intro10&taba} which can be best studied on resolved scales inside galaxies. 
  
 Resolved radio-FIR correlation holds inside galaxies on kpc scales and smaller \citep[e.g.][]{intro11&Hughes,intro12&taba}. The correlation can change depending on galactic region, e.g., arms vs inter-arms, inner vs. outer disks  \citep[e.g.][]{intro13&Dumas}, and star forming regions vs diffuse structures \citep{intro14&taba}.  A study in NGC~6946 already calls for a fine balance between cosmic rays, magnetic fields, and gas pressures as the main physical reason for the observed correlation in between the spiral arms \citep{intro15&taba}. Multiscale analysis of the radio and FIR maps, such as 2D wavelet decomposition, can help inferring diffusion and propagation lengths of CREs as indicated by a break in the radio--FIR correlation towards small spatial scales in few nearby galaxies \citep{intro12&taba}. Similar studies in galaxies with different ISM conditions and star-formation properties must be performed to draw robust conclusions.  
 
Investigating the radio--FIR correlation at low frequencies adds another stringent insight about the CRE propagation. {Unresolved studies with the Low-Frequency Array (LOFAR) find that the low-frequency radio--FIR correlation deviates from linearity \citep[e.g.,][]{intro16&Read,intro17&Wang} and shows a strong dependence on the stellar mass of galaxies globally \citep{Gurkan,Smith,McCheyne}. Resolved LOFAR observations find similar results in a sample of nearby galaxies \citep{Heesen_22}. \cite{Gurkan}  noted that a broken power-law model can better fit the correlation than a single power-law, potentially indicating an additional mechanism for generation of the CREs. On the other hand, as}  CREs propagate over longer distances at lower frequencies, it is expected to see a break in the radio--FIR correlation on scales larger than those found at higher frequencies \citep{intro12&taba}. At low frequencies, electrons radiate away their energy more slowly than at higher frequencies, resulting in a relationship between the age of the electron population and the radio spectral index \citep[e.g.,][]{intro18&Scheuer,intro19&Blundell,intro21&Hans,intro22&Schober}. It should be noted that, even at higher frequencies, the synchrotron-FIR correlation, obtained after correcting the {Radio Continuum} (RC) emission for thermal contamination, deviates from linearity \citep[e.g.,][]{intro23&Niklas,intro24&taba}. Hence, the non-linearity of the low-frequency radio--FIR correlation may be just due to lower thermal contamination at those frequencies. Hence, separating the thermal and nonthermal emission is needed to infer the cause of the non-linearity and further assess the role of the CREs aging/propagation in linearity of the correlation.

IC 342 is the third largest spiral galaxy (SABcd) in the sky at a distance of 3.3~Mpc. It is an intermediate, almost face-on spiral galaxy from the Maffei group  \cite{intro26&Karachentsev}. Thanks to its low inclination ($i=31^{\circ}$), it is ideal for mapping a variety of astrophysical properties (Table~\ref{tab:Tabel1}).  With a dynamical mass of  $2 \times 10^{8}$ \textbf{M$_{\odot}$}, the global star formation rate of IC~342 is about 2.8\,\textbf{M$_{\odot}$}~per year, most of which occurs in its central nuclear star cluster \citep{intro29&Turner}. IC~342 hosts two mini spiral arms connected to an inner molecular ring.  The molecular ring and arms contain five prominent Giant Molecular Clouds (GMC), each of which has a mass of approximately $10^{6}$ \textbf{M$_{\odot}$} \citep[]{intro30&Solomon}. No clear evidence for AGN activity is found in this galaxy \citep[e.g.,][]{intro31&Tsai}.%

IC~342 has been studied at several radio frequencies but mostly at $\nu>$1\,GHz  \citep{intro27&Graeve,intro28&Krause,Data15&rainer}.  
In this paper, we present a high-resolution RC map of this galaxy at a frequency much lower than those studied before ($\nu=$0.14\,GHz). A multi-frequency ($4.8\leq \nu \leq 0.14$\,GHz) and multi-scale ($a>110$\,pc) analysis of the radio--FIR correlation is carried out after correcting the observed radio emission for thermal contamination. We investigated the linearity and the smallest scale of the correlation as a function of dust temperature, SFR, and magnetic field, and further explored the relationships between the energy index of CREs vs. the SFR and magnetic field. This study is only now possible thanks to  
the \textrm{Low-frequency array Two-meter Sky Survey} \citep[LoTSS, ][]{Data8&ShimwellDR1} that provides a large field of view and high sensitivity on both small and large angular scales.

We describe the observations and data sets used in this paper in  Sect.~\ref{sec:data} and present the result of the LoTSS observations in Sect.~\ref{sec:lowfreq}. After separating the the free-free and synchrotron emissions components (Sect.~\ref{sec:thermal}), 
 the total and synchrotron spectral index are mapped (Sect.~\ref{sec:alpha}) and the magnetic field strength is obtained (Sect.~\ref{sec:magnet}). The radio-FIR correlation is calculated using three different approaches, that is, the classical pixel-by-pixel correlation, the ratio of the FIR to the radio emission, and the multiscale correlation (Sect.~\ref{sec:radio-fir}). We then discuss the results in Sect.~\ref{sec:discu} and summarize in Sect.~\ref{sec:summ}. 
   
\begin{table}
\begin{center} 
\caption{Positional data adopted for IC342}\label{tab:Tabel1}
\begin{tabular}{l l} 
\hline\hline
Position of nucleus$^{1}$ & RA= 03$^{h}$ 46$^{m}$ 48.503$^s$ \\ 
         & DEC= +68$^{\circ}$ 05$\arcmin$ 46.92$\arcsec$ \\
Distance$^{2}$ &  3.3\,Mpc ($1\arcsec \simeq$16\,pc) \\ 
Morphology$^{3}$ & SAB(rs)cd \\ 
Inclination angle$^{4}$ & 31$^{\circ}$ \\ 
Position angle of major axis$^{4}$ & 39.4$^{\circ}$ \\
\hline
\end{tabular}
\tablebib{$^1$ \cite{Data9&Krause}, $^2$ \cite{DataRef1&Karachentsev}, $^3$ \cite{DataRef2&Vaucouleurs}, $^4$ \cite{DataRef3&Lucian}}

\end{center}
\end{table}


\section{Data}
\label{sec:data}

\begin{figure*}\hspace{-5.5em} 
\centering

\subfigure{\includegraphics[width=0.33\textwidth]{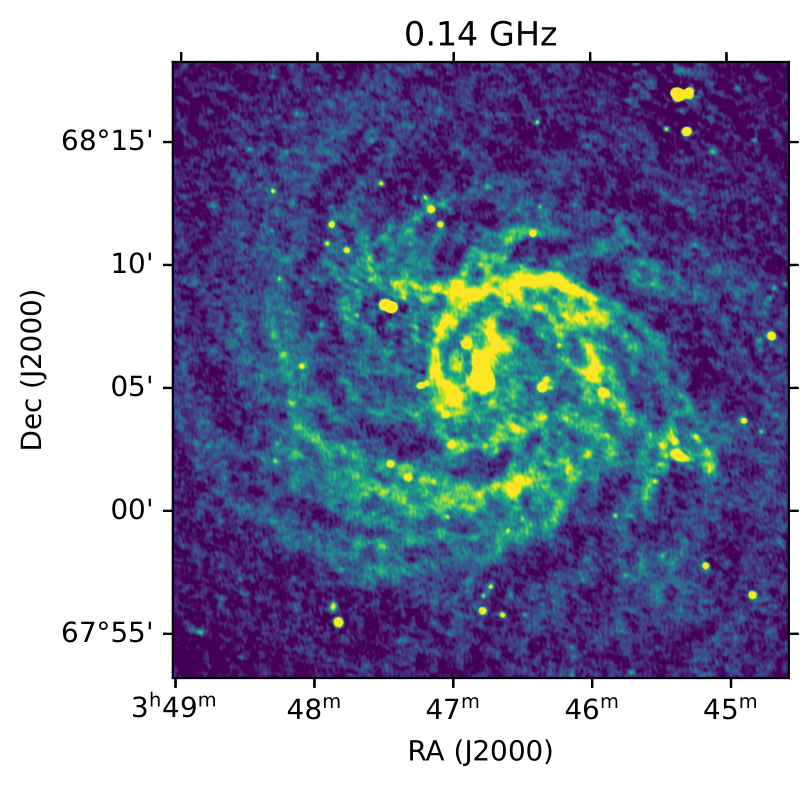}}\hfill
\subfigure{\includegraphics[width=0.33\textwidth]{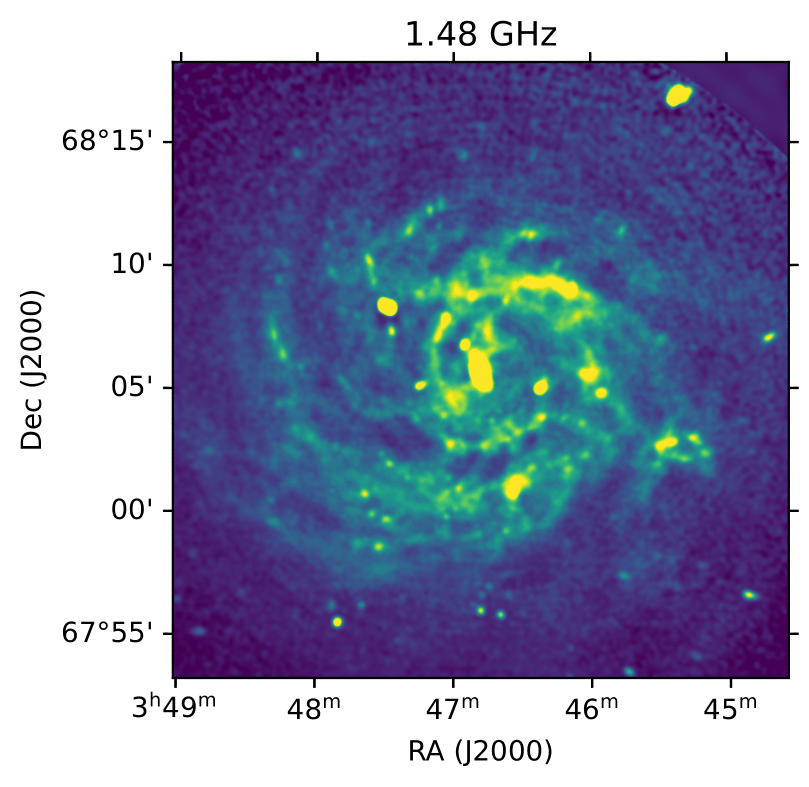}}\hfill
\subfigure{\includegraphics[width=0.33\textwidth]{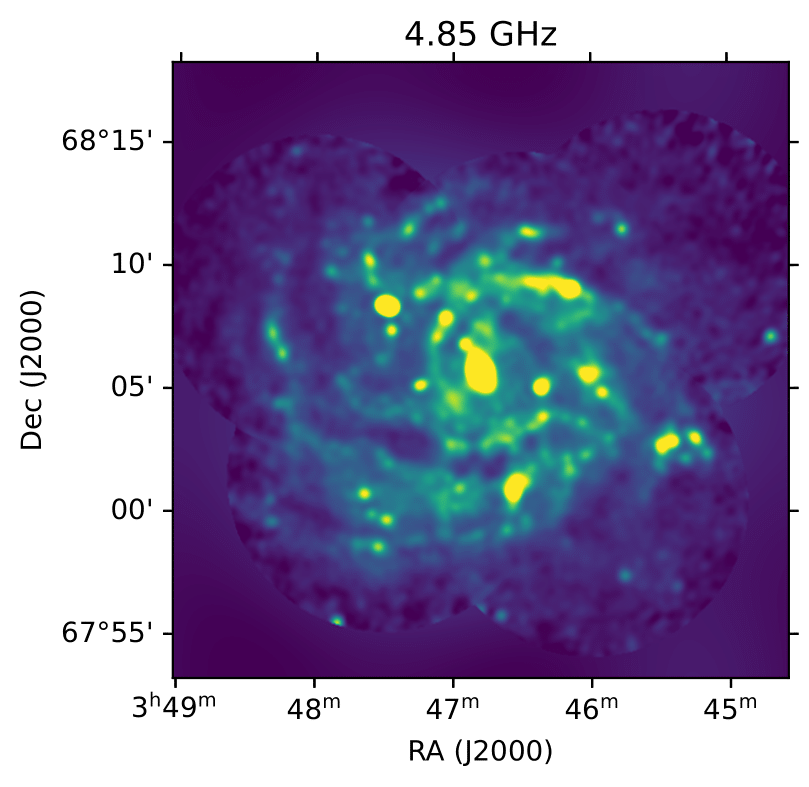}}\\

\subfigure{\includegraphics[width=0.33\textwidth]{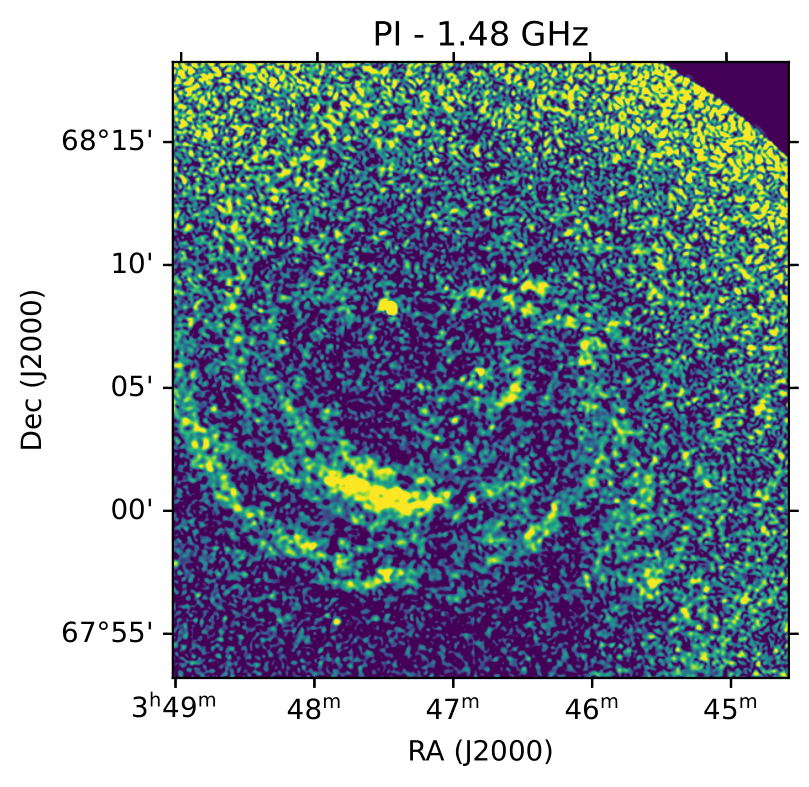}}\hfill
\subfigure{\includegraphics[width=0.33\textwidth]{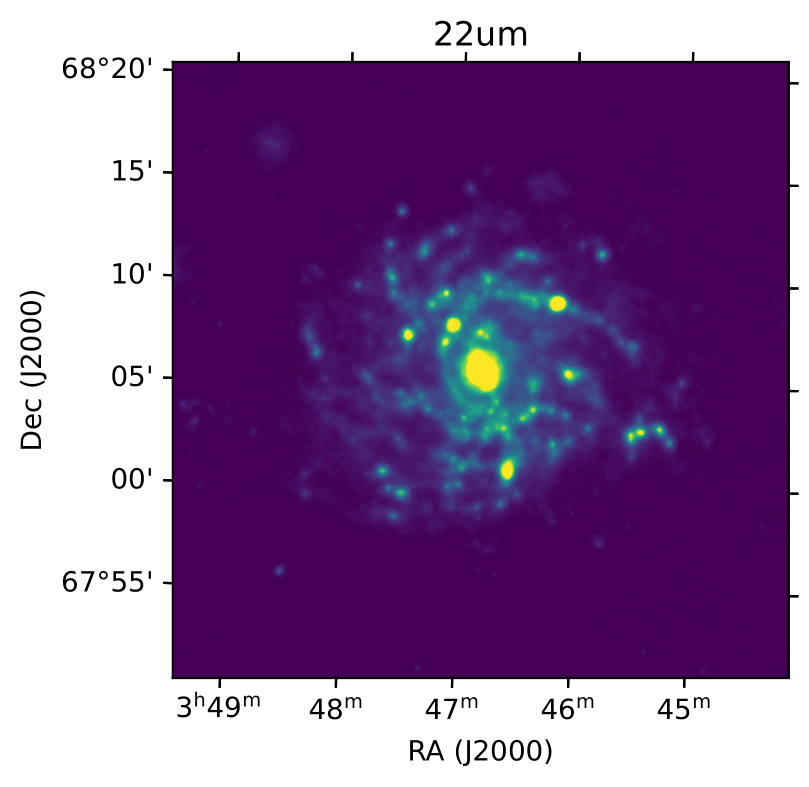}}\hfill
\subfigure{\includegraphics[width=0.33\textwidth]{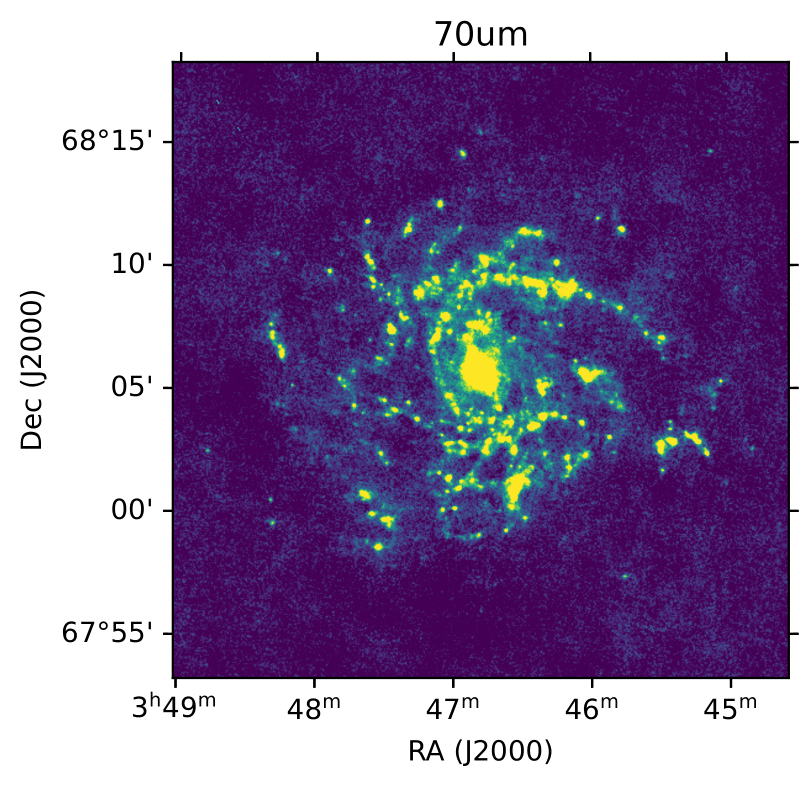}}\\

\subfigure{\includegraphics[width=0.33\textwidth]{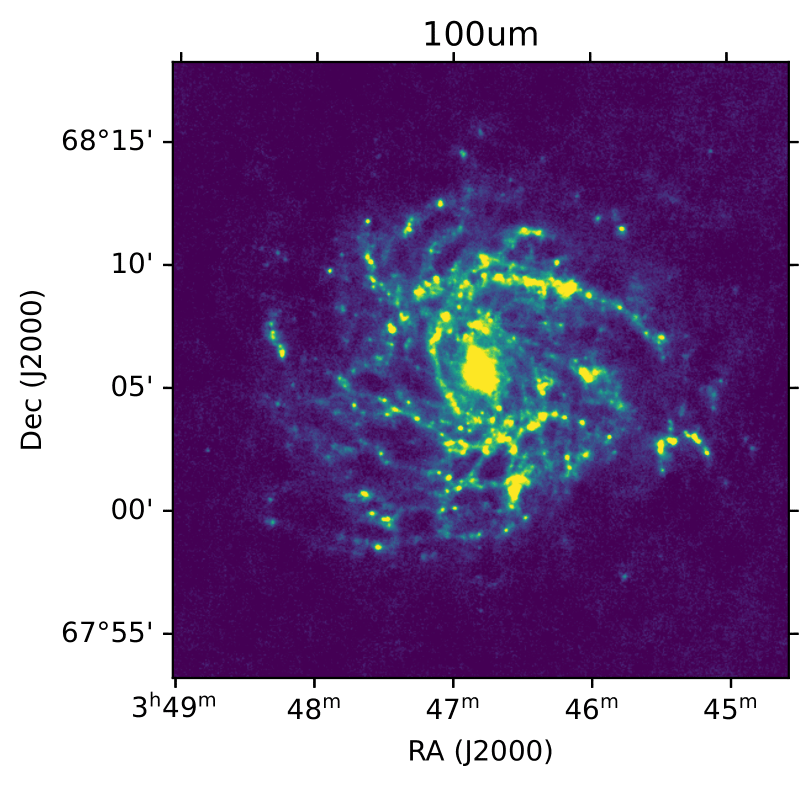}}\hfill
\subfigure{\includegraphics[width=0.33\textwidth]{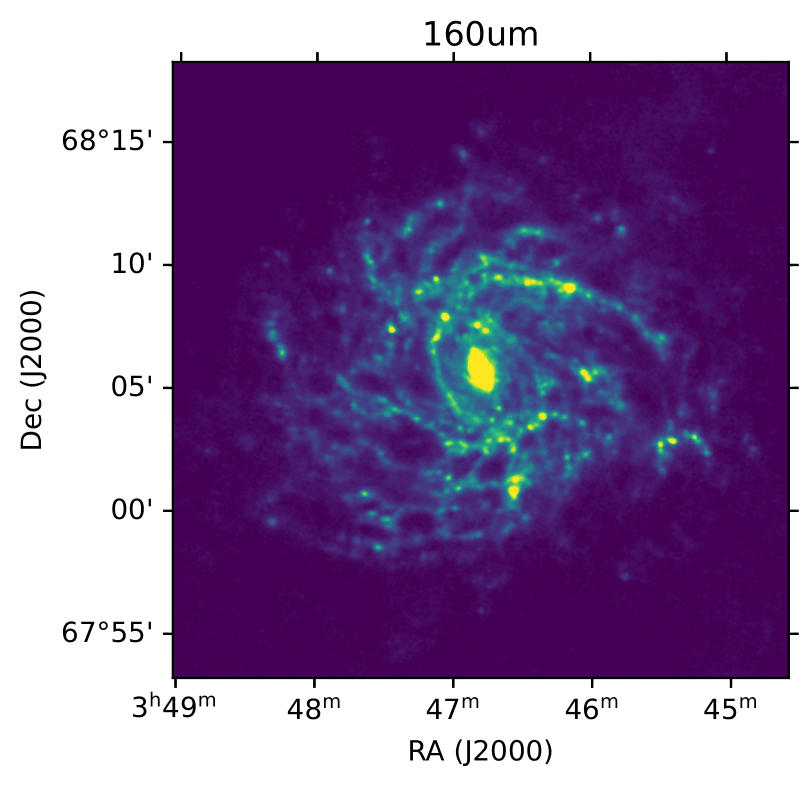}}\hfill
\subfigure{\includegraphics[width=0.33\textwidth]{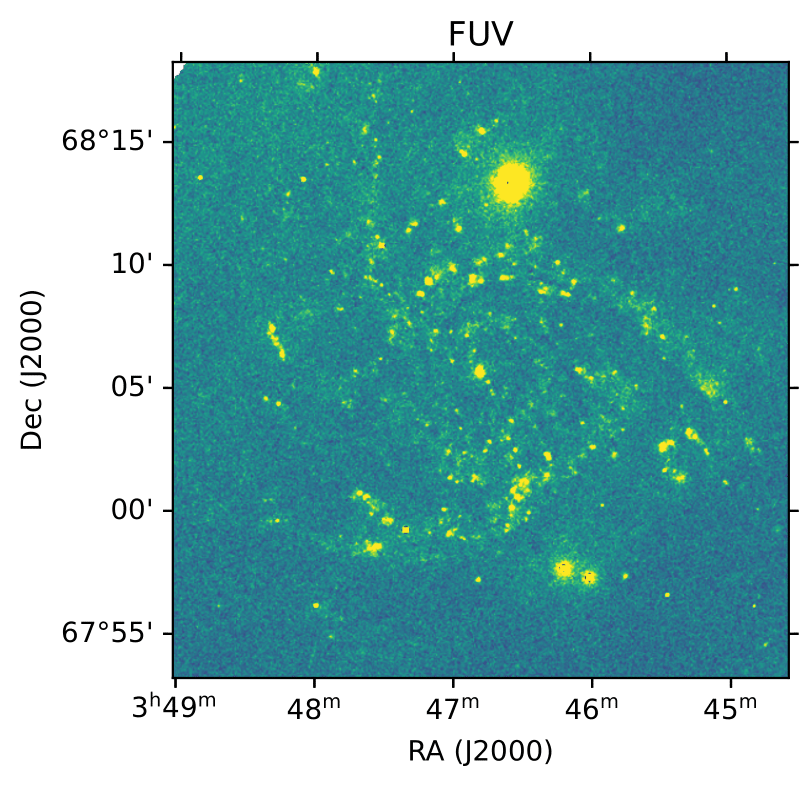}}

\caption{IC 342 images from radio to FUV used in this work (see Table~\ref{tab:Table2}).}
\label{fig:all}
\end{figure*}

Table~\ref{tab:Table2} summarizes the data (Fig.~\ref{fig:all}) used in this study. We explain them in more details as follows.

\subsection{LoTSS Observations}
 \label{sec:data144}
 The low frequency data are taken from the LoTSS Data Release 2 \citep[DR2][]{Data1&Shimwell} conducted at 144 MHz (frequency range 120-168 MHz). The LoTSS DR2 provides two fields centered on latitudes 0h and 13h covering 5700 square degrees. It is processed using an enhanced pipeline compared to the DR1. The LOFAR's High Band Antenna (HBA) system is used to take the LoTSS data. Based on the HBA dual inner configuration, the data were collected with a dwell time of 8 hours and a frequency coverage of 120–168 MHz. With 3168 pointings, the entire northern sky is covered. The publicly available LOFAR direction independent calibration procedure is described in detail by \cite{Data2&vanweeren} and \cite{Data3&Williams}. For averaging and calibration, it uses the LOFAR Default Preprocessing Pipeline \citep[DPPP][]{Data4&vanDiepen} and BlackBoard Self calibration \citep[BBS][]{Data5&Pandey}. It has been committed to improving the treatment of direction-dependent effects (DDEs) during calibration for LOFAR due to their importance. Using the task ddfacet \citep{Data6&tesse}, the LoTSS applies DDE corrections to its pipeline with Kill MS \citep{Data7&SmirandTess}. The LoTSS-DR2 \citep{Data1&Shimwell} provides much improved imaging capabilities for diffuse extended emission compared to the LoTSS-DR1 \citep{Data8&ShimwellDR1}.

\subsection{Auxiliary data}
 \label{sec:dataax}
In addition to the LoTSS DR2 data, we used the radio continuum maps at higher frequencies. IC~342 was observed with the Very Large Array (VLA) in full polarization at 1.4 GHz in D-array \citep{Data9&Krause}  centred on the nucleus (at RA, DEC (J2000) = 03$^{h}$ 46$^{m}$ 48$^s$, +68$^{\circ}$ 5$\arcmin$ 47$\arcsec$).  These data were combined with those observed in C-array in the uv plane and corrected for missing short spacing using single-dish observations taken with the 100-m Effelsberg telescope by \cite{Data15&rainer}. Clean maps were generated using uniform weighting at $15\arcsec$ angular resolution.  The maps were corrected for primary beam attenuation. {The 4.85 GHz  ($\lambda$6.2 cm) data uses four different pointings observed with the VLA D array, located to the south-east, south-west, north-east, and far north-west of the center \citep{beck_15,Krause_93}.
The uv data from these} pointings were naturally weighted into maps with an angular resolution of $19^{\arcsec}$to $23^{\arcsec}$, which were then smoothed to a common resolution of $25^{\arcsec}$.

IC~342 was mapped in FIR with the PACS Camera onboard the Herschel Space observatory at $70\,\mu$, $100\,\mu$m and $160\,\mu$m. Details of the observations and data reduction are presented in \citep{Data10&Davies}. The final PACS maps have pixel sizes of 2\arcsec , 3\arcsec, and 4\arcsec, respectively. 
To trace the thermal free-free emission, we also use the mid-IR data of IC~342 at 22\,$\mu$m observed with the  Wide-field Infrared Survey Explorer (WISE) \citep{Data11&Warner}.
The pixel units of all maps were converted from MJy sr$^{-1}$ to Jy pix$^{-1}$.\\


To trace star formation, we use the far-ultraviolet (FUV) map of IC342 observed with the GALaxy Evolution eXplorer \citep[GALEX, ][]{Data12&Morrissey}. This work adopts the GR6/7 data release presented by \cite{Data13Bianchi}.
%
In contrast to the standard GALEX pixel size of 1.5\arcsec, the final GALEX cutouts have pixels of 3.2\arcsec in FUV. The native GALEX tile pixel units of counts s$^{-1}$ pix$^{-1}$ were converted to Jy pix$^{-1}$ using the conversion factors of $1.076 \times 10^{-4}$Jy counts$^{-1} s$ in the FUV ; this corresponds to the standard GALEX AB magnitude zero points of 18.82 mag \citep[][]{Data12&Morrissey}.
Detailed explanations of Multiwavelength photometry and imagery, including IR and FUV data, are provided by \cite{data&herschel} and \cite{Data14&Clark}.

\begin{table}[!hbt]
\caption{Data of IC~342 used in this work.}
\label{tab:Table2}  
\begin{tabular}{c c c c}
\hline\hline
$\nu$ or $\lambda$& Angular & Telescope & RMS\\
& resolution & & noise\\
\hline
$144$ MHz & 10.5\arcsec & LOFAR$^1$& $180$ ($\mu$Jy/beam)\\ 
$1.4$ GHz & 15\arcsec & VLA+Effelsberg$^2$ & $7$ ($\mu$Jy/beam)\\ 
$4.8$ GHz & 25\arcsec & VLA+Effelsberg$^2$ & $24$ ($\mu$Jy/beam)\\ 
22\,$\mu$m & 12\arcsec & WISE$^{3}$ & $5.1$ ($\mu$Jy/pix) \\ 
70\,$\mu$m & 9\arcsec & Herschel$^{4}$ & $240$ ($\mu$Jy/pix)\\
100\,$\mu$m & 10\arcsec & Herschel$^{4}$ & $420$ ($\mu$Jy/pix) \\
160\,$\mu$m & 13\arcsec & Herschel$^{4}$ & $250$ ($\mu$Jy/pix )\\
153nm  & 4.3\arcsec & GALEX$^{5}$ & $0.9$ ($\mu$Jy/pix)\\
(FUV) & & &\\
\hline
\end{tabular}
\tablebib{$^1$ This work, $^2$ \cite{beck_15}, $^3$ \cite{Data11&Warner}, $^4$ \cite{data&herschel}, $^{5}$ \cite{Data14&Clark}}

\end{table}

\section{Low frequency radio continuum emission}
\label{sec:lowfreq}
	
\begin{figure}[]
    \centering
    \includegraphics[width=\columnwidth]{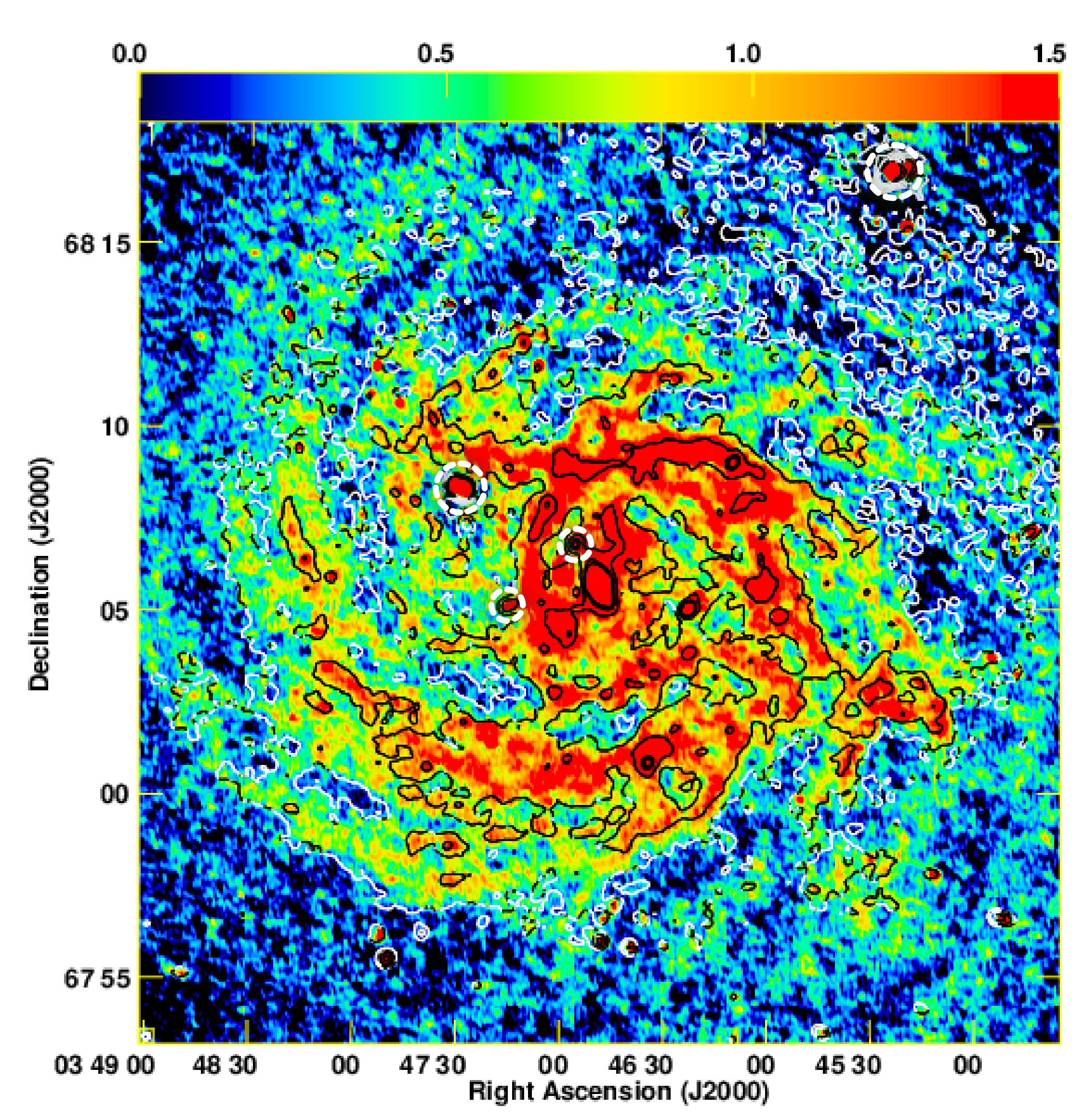}
\caption{The LOFAR radio continuum map of IC~342 observed at $0.144$~MHz overlaid with contours of the VLA $1.4$ GHz emission. Contour levels are 3, 6, 12, 24, 32, $10^{-4}$~Jy/beam. Dashed circles indicate the positions of the background radio sources which are subtracted for this study. The beam size of 10.5" is indicated in the bottom-left corner of the image.} 
    \label{fig:144mhz}
\end{figure}

The resulting LoTSS map reduced has an rms noise level of $\sim$ 180\,($\mu$Jy/beam) at 10.5\arcsec~angular resolution. IC~342 exhibits a sharp spiral structure at 0.14\,GHz, (Fig. \ref{fig:144mhz}). The strongest emission is detected at the center of the galaxy with a mean flux density  of 0.36 Jy at 0.14\,GHz. Moreover,  the  disk appears to be brighter in the inner $\sim$ 7 kpc than in the outer, similar to that at other  wavelengths. The low surface-brightness spiral arms at $>7$\,kpc are not as sharply visible in the 1.4\,GHz map because of its  poorer angular resolution. They are also not visible in the 4.8\,GHz map because of its limited observation coverage. We also note that the distribution of the low-surface brightness emission at 0.14\,GHz differs from that in the FIR maps (at 70, 100, 160\,$\mu$m).    

About 56\% of the low-frequency radio sources are identified as star forming regions by cross-matching them with the 22\,$\mu$m emitting sources. The rest can be background radio source candidates, particularly those in the inner R $<$\,7 kpc disk. Identifying the origin of the radio sources beyond R $\sim$\,7 kpc, is not as straightforward because the 22\,$\mu$m emission is mainly limited to the inner disk. Studying the radio--FIR correlation, 
{we subtracted known background radio sources 
such as the bright, unresolved source located northeast of the nucleus ($\alpha$ = $3^{h}$ $47^{m}$ $29^{s}$, $\delta$ = $68^{\circ}$ $08^{\arcmin}$ $24^{\arcsec}$) was omitted. \cite{Turner} identified this source as having a distinct double-lobed appearance typical of a radio galaxy, with a 21 cm continuum flux of 46 mJy. Furthermore, \cite{Hummel} confirmed this identification, noting a steep spectral index of $\alpha$ = -0.87, consistent with that of a background radio galaxy. Similarly, the unresolved source positioned near the edge of the primary beam to the northwest ($\alpha$ = $3^{h}$ $45^{m}$ $23^{s}$, $\delta$ = $68^{\circ}$ $17^{\arcmin}$ $00^{\arcsec}$), with a 21 cm continuum flux of 26 mJy was also identified as a background galaxy by \cite{Hummel}. The likelihood of this source being within IC~342 is minimal, as its total-power output exceeds that of Cas~A by a factor of six, which is uncharacteristic for sources within the galaxy \citep{Lucian}. Two other radio sources at $\alpha$ = $3^{h}$~$47^{m}$~$14^{s}$, $\delta$ = $68^{\circ}$~$05^{\arcmin}$~$10^{\arcsec}$ and
$\alpha$ = $3^{h}$~$46^{m}$~$54^{s}$, $\delta$ = $68^{\circ}$~$06^{\arcmin}$~$5^{\arcsec}$ were also excluded. They are likely external sources, as they are detected only at 0.14\,GHz and lack counterparts in the IR and FUV maps.
} 

\section{Thermal and nonthermal radio maps}
\label{sec:thermal}

As discussed in Sect.~\ref{sec:intro}, to investigate the origin of the resolved radio--FIR correlation and to assess the role of the magnetic fields and  CREs , the thermal and nonthermal components of the radio continuum emission should be separated. This will also help mapping the pure nonthermal spectral index that is proportional to the energy index of  CREs  to address their cooling across the galaxy.
Although, globally, the thermal emission represents no more than 10\% at 1.4 GHz and about 5\% at 0.33 GHz \citep{thermal3&Basu,intro24&taba},  it can be larger than 20-30\% in star forming regions \citep{intro14&taba} even at low-frequencies \citep{Sep2&Hamid}. Hence, we map the thermal and nonthermal emission at all radio frequencies of study. 

\begin{figure*}
    \centering
     \subfigure{\includegraphics[width=8cm]{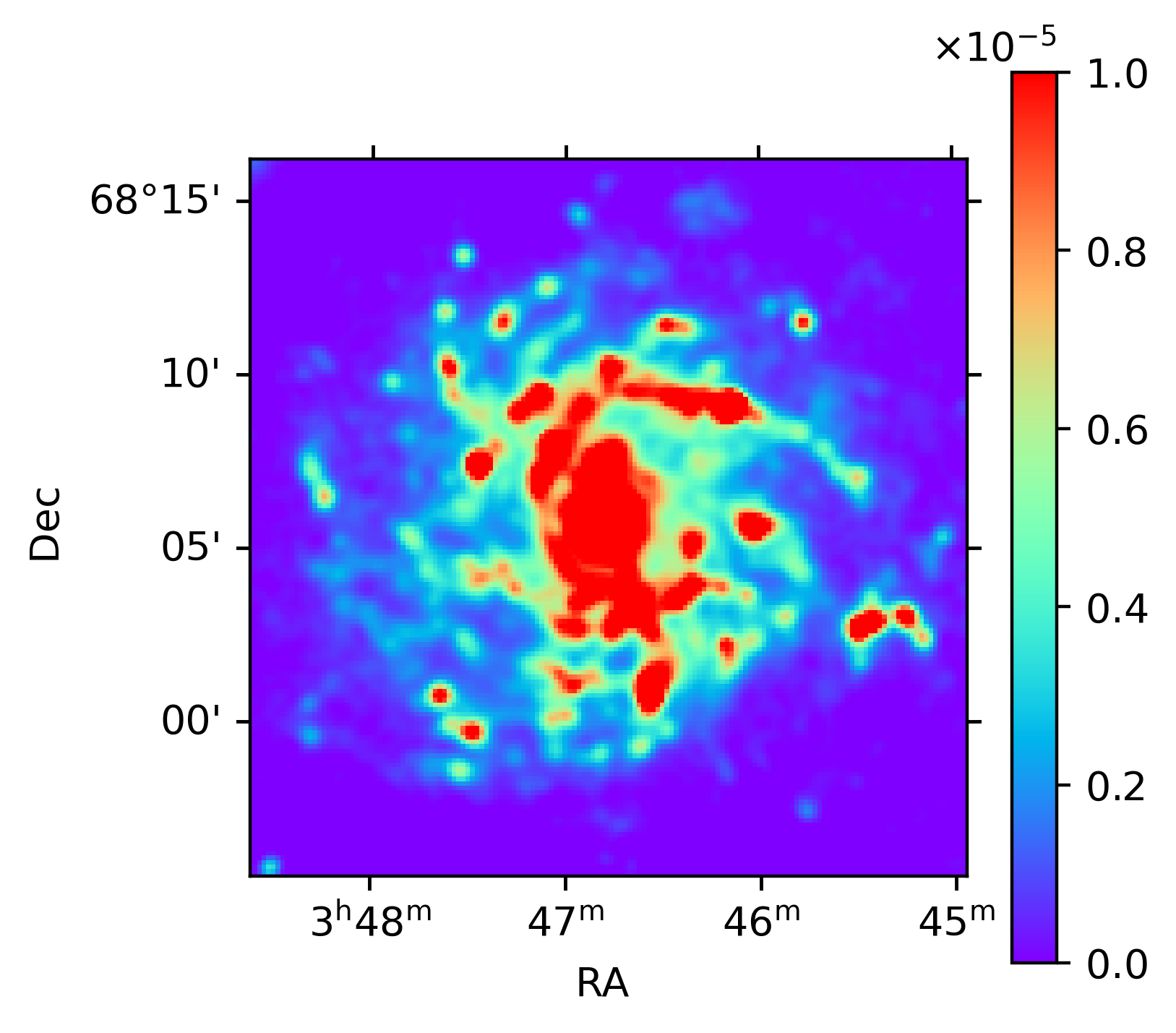}}
     \subfigure{\includegraphics[width=8cm]{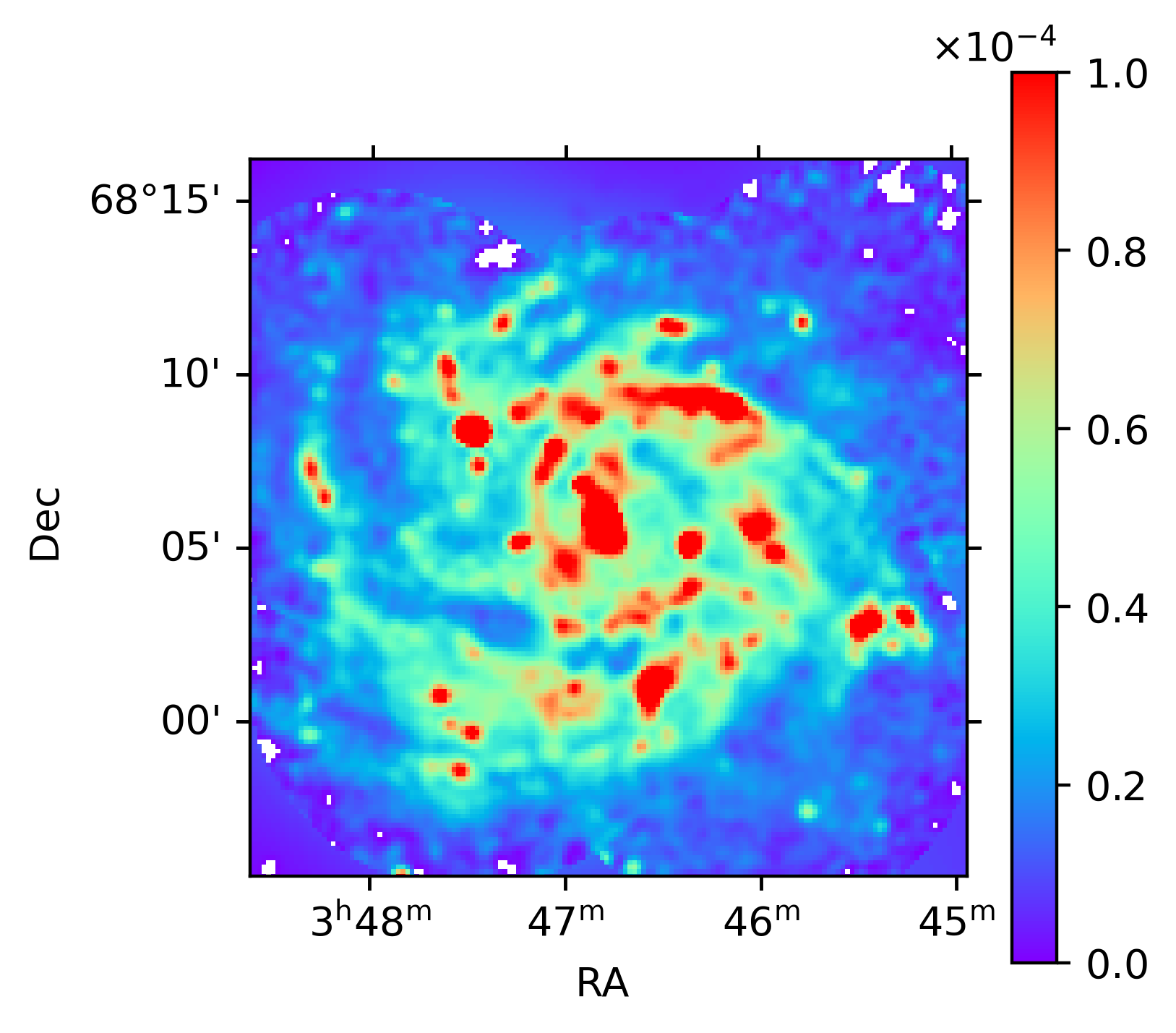}} 
    
      \subfigure{\includegraphics[width=8cm]{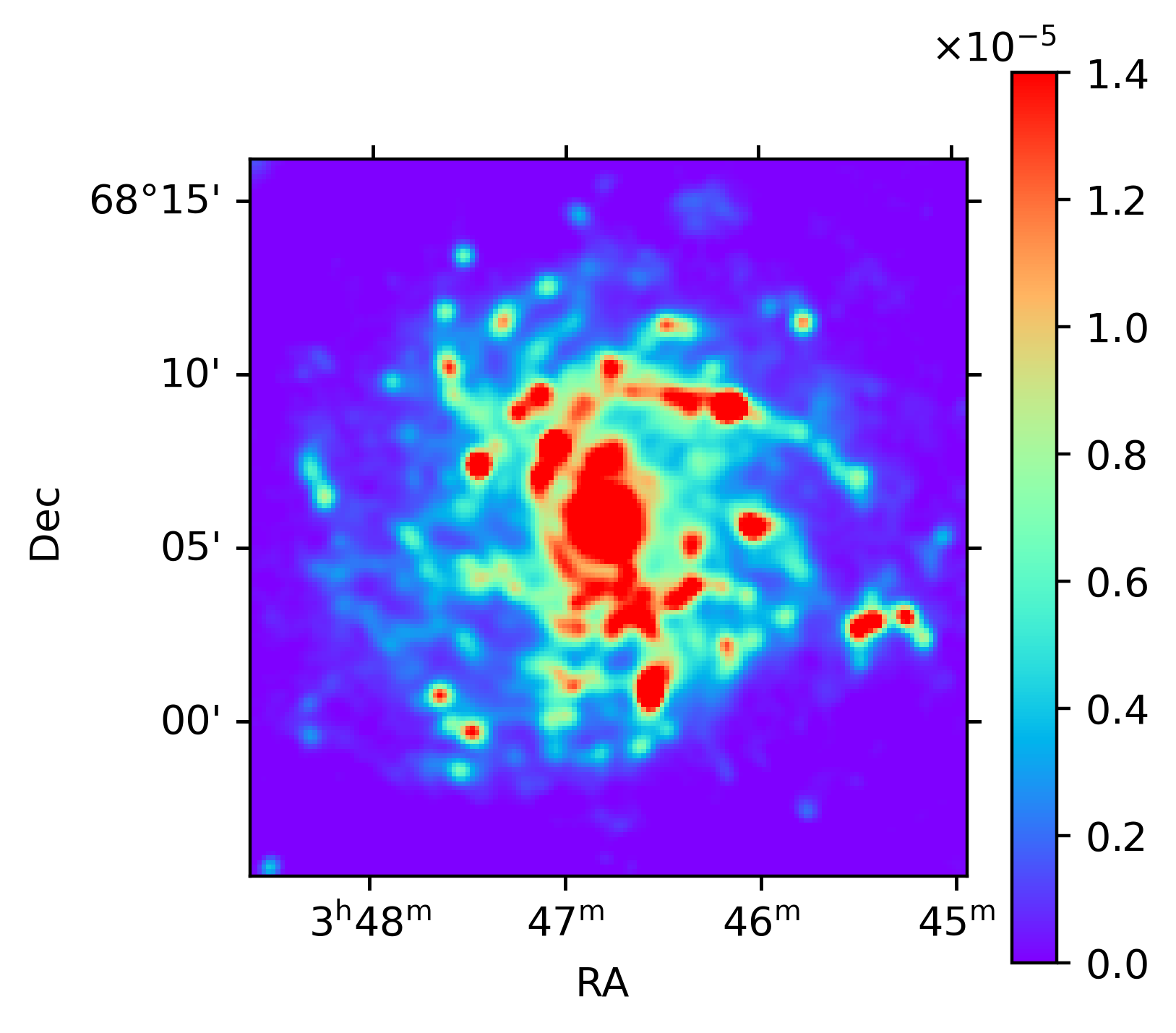}}
      \subfigure{\includegraphics[width=8cm]{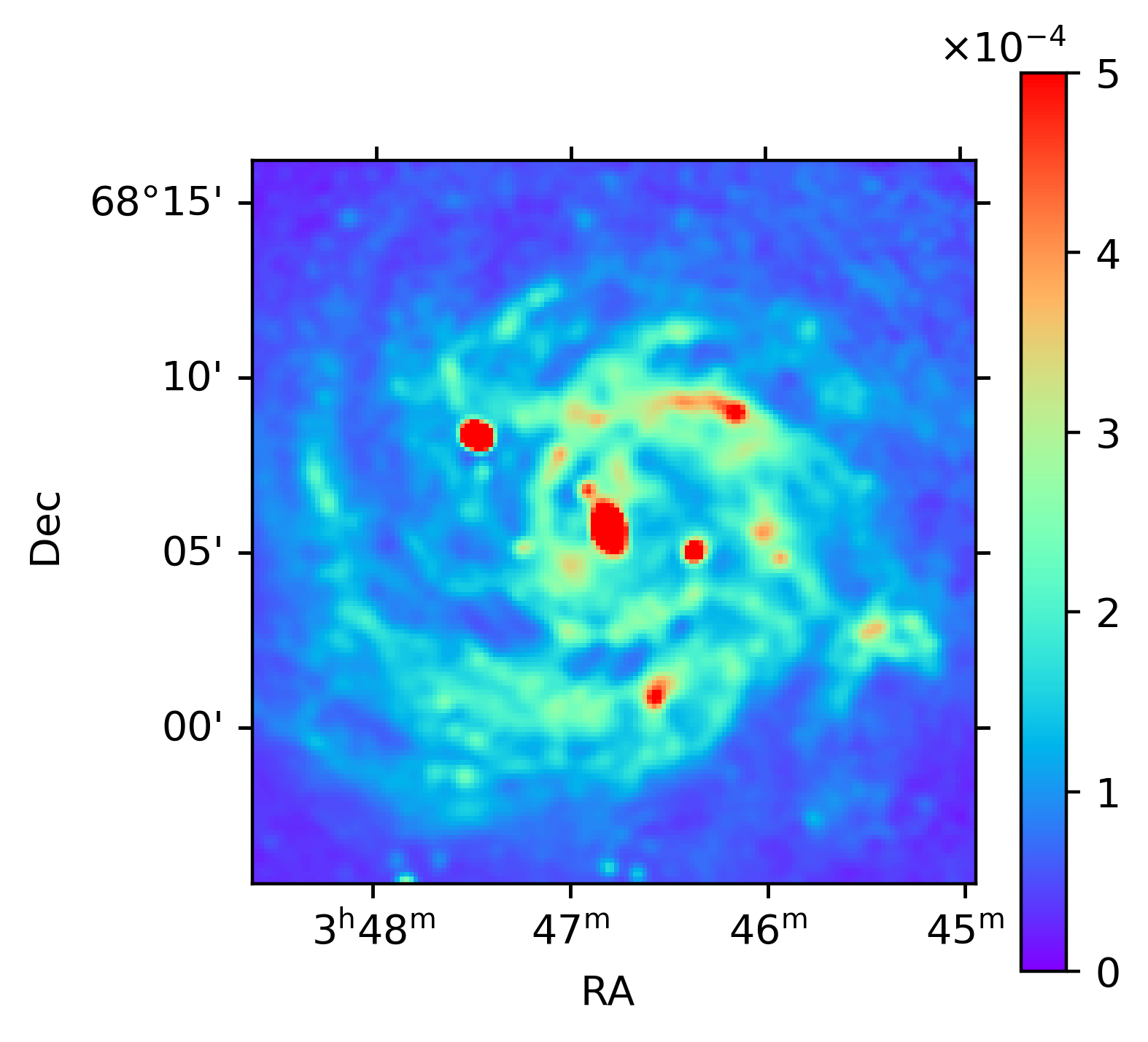}}
    
   		 \subfigure{\includegraphics[width=8cm]{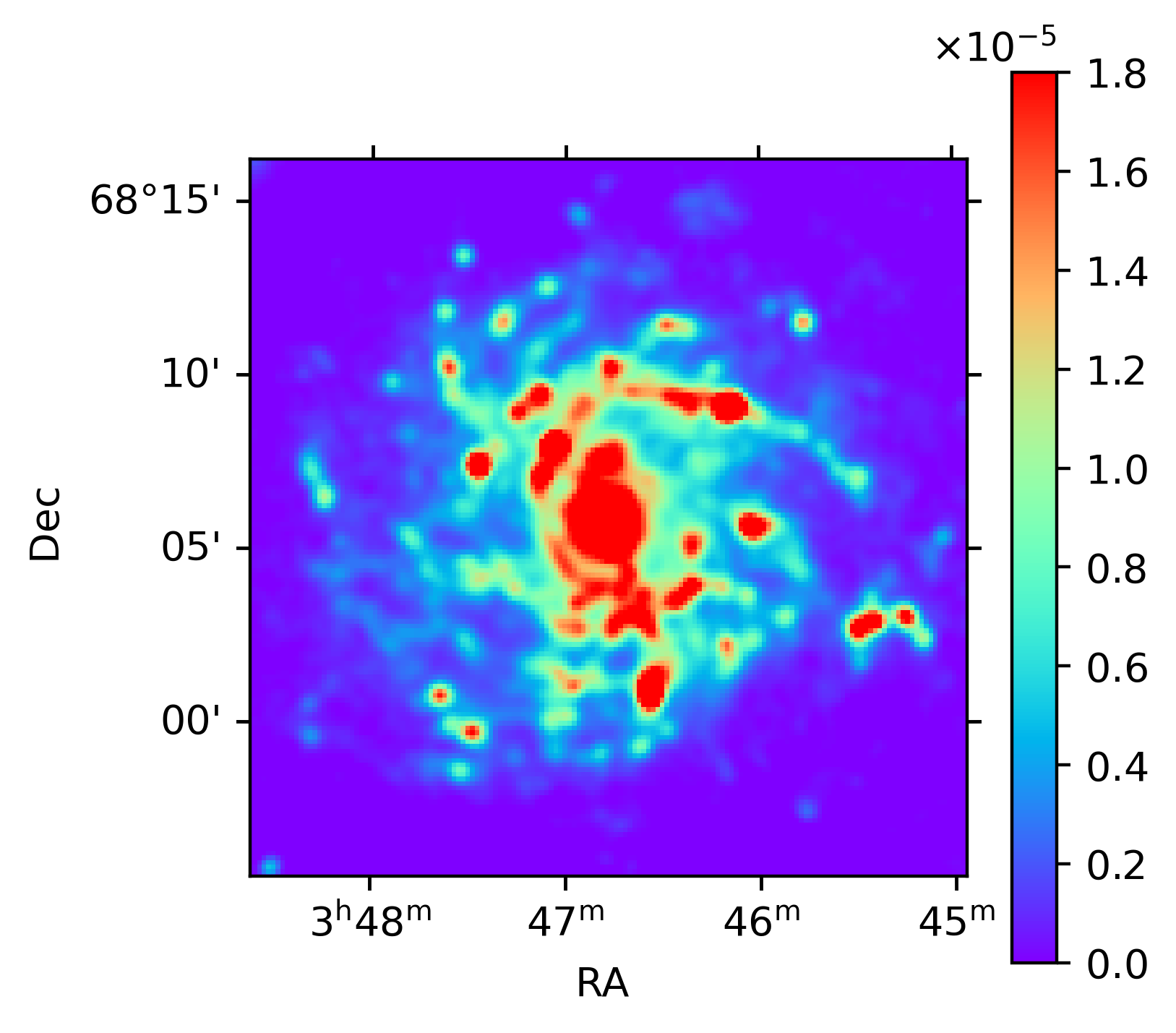}} 
 	 	  \subfigure{\includegraphics[width=8cm]{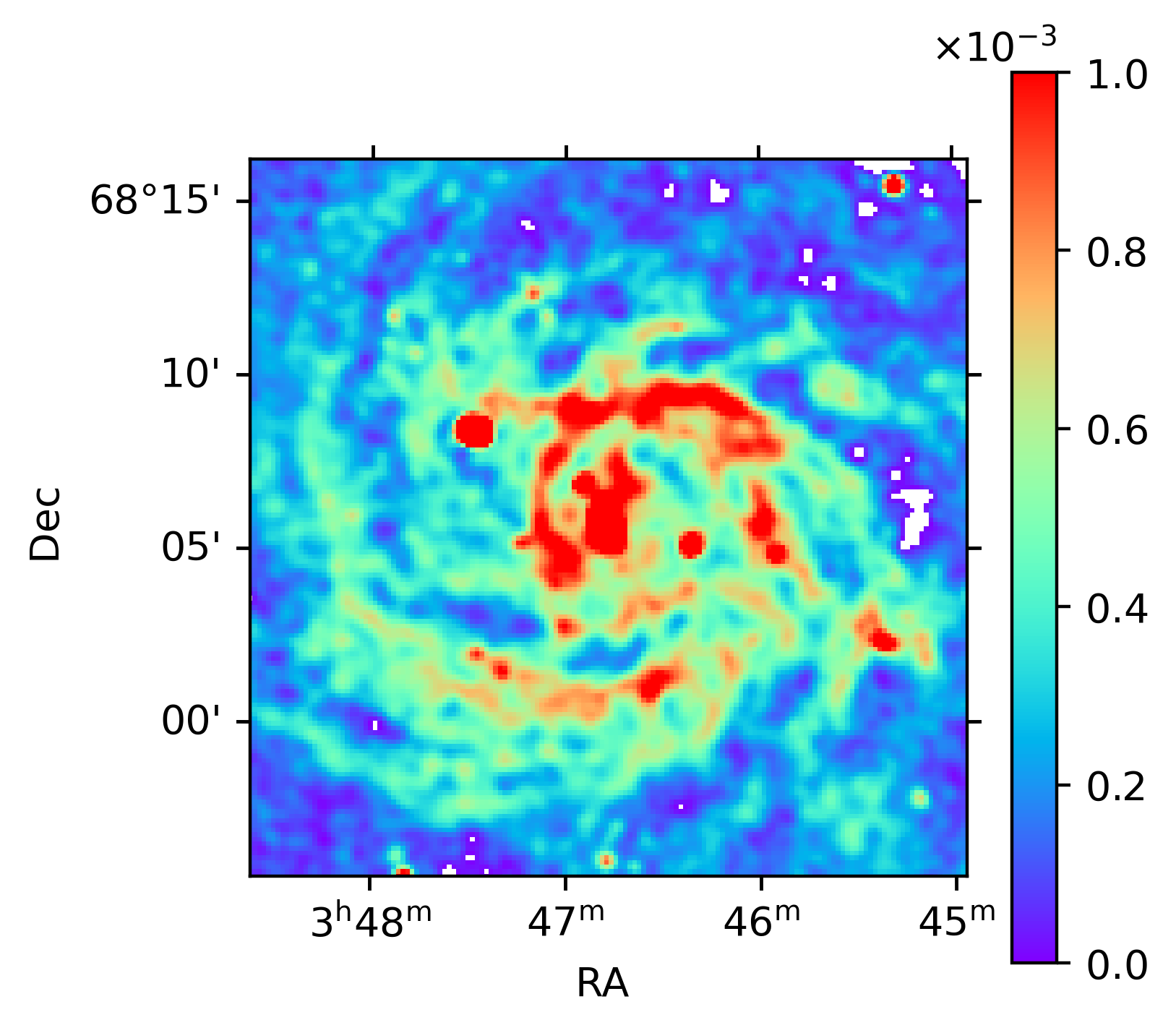}}   
     
\caption{Maps of the thermal (left) and non-thermal synchrotron (right) radio continuum emission from IC 342 at 4.8\,GHz (top row), 1.4\,GHz (middle row), and 0.144\,GHz (bottom row) frequencies, all at $25\arcsec$ angular resolution. Colour bars show the intensities in units of Jy/pix.}

    \label{fig:4.8th}
\end{figure*}

We need a separation method with no prior assumption about the non-thermal spectral index. In an ionized gas, ideal tracers of the free-free emission are recombination lines with H$\alpha$ emission as the brightest. Using a  de-reddened H$\alpha$ map has been the most feasible technique to study the thermal/nonthermal emission from different galactic regions (e.g., arms vs inter-arm regions, inner vs outer disks) down to very low surface brightness levels \citep{Spec1&taba,intro12&taba,intro15&taba,intro24&taba,intro14&taba,Sep2&Hamid,Spec2&Heesen}. {To our knowledge, IC~342 has a lack of complete and reliably calibrated H$\alpha$ map available which can be because of its low-declination (viewed through the Milky-Way disk) and low-surface brightness. Therefore, we have to use another thermal tracer for this galaxy. 
Several studies have explored the use of the mid-IR emission at 22-24\,$\mu$m wavelength to estimate the thermal free-free flux density \citep[e.g.][]{Sep3&Murphy,thermal3&Basu}. These studies show that, as an SFR tracer, the mid-IR emission can also provide an accurate free-free template in star forming regions. 
For IC~342, we opt to use the WISE 22\,$\mu$m data to map the thermal radio emission. Following \citep{Sep3&Murphy}, }
the radio thermal emission flux density, $S_{\nu}^{\rm th}$, in Jy (per pixel) is related to the $24\mu m$ flux, $f_{\rm 24}$, as
\setlength{\abovedisplayskip}{6pt}
\setlength{\belowdisplayskip}{6pt}
\begin{equation}\label{Eq1}
\left(\frac{S_{\nu}^{\rm th}}{\rm Jy}\right) \sim 7.93\times 10^{-3} \left(\frac{T}{\rm 10 ^{4} K}\right)^{0.45} \left(\frac{\nu}{\rm GHz}\right)^{-0.1} \left(\frac{f_{\rm 24}}{\rm Jy}\right),
\end{equation}
where $\nu$ is the radio frequency and $T$ the thermal electron temperature which is typically $10^{4}$\,K.  After convolving the maps to the same resolution and normalizing them to the same pixel size and geometry, we first obtain the thermal radio flux densities per pixel using Eq~\ref{Eq1}.  The nonthermal emission is then computed for each pixel by subtracting the thermal flux from the observed radio continuum flux at each frequency ($S_{\nu}^{\rm nt}= S_{\nu}^{\rm obs} - S_{\nu}^{\rm th}$).
Figure \ref{fig:4.8th} shows the resulting maps of the thermal and nonthermal emission.%
The nonthermal maps exhibit a pronounced large-scale spiral pattern unlike the thermal maps, although both are brighter in the center and star-forming regions.  
{The two background radio sources,} one near the center of the galaxy at distance of 2.4 kpc (RA = $3^{h}$ $47^{m}$ $27^{s}$ and DEC = $68^{\circ}$ $8{\arcmin}$  $17{\arcsec}$), and the other at a distance of 6.3 kpc from the center in the northeast of the galaxy (RA = $3^{h}$ $45^{m}$ $19^{s}$ and DEC = $68^{\circ}$ $16{\arcmin}$  $23{\arcsec}$), {are the brightest sources in the nonthermal map}.

{Obtaining} the thermal fraction defined as $S_{\nu}^{\rm th}/S_{\nu}^{\rm obs}\times100$, provides more detailed insights into the relative contributions of the thermal and nonthermal processes across the galaxy. {We calculate the thermal fraction for the entire galaxy and for two different regimes of star forming ISM, separately. We utilize the 22\,$\mu m$ map to delineate a densely star-forming regime (R1) and a diffuse and weakly star-forming regime (R2) as explained in  Appendix~\ref{app:star_formation_regimes}.}
 {At 144 MHz, the thermal fraction in R1 and R2 are $5.7\% \pm 0.5$ and $1.1\% \pm 0.1$, respectively, increasing to  $13.5\% \pm 1.0$ and $3.5\% \pm 0.3$ at 1.4\,GHz.}
%
At 4.8 GHz, the thermal fraction reaches $29.7\% \pm 2.1 $ and $8.7\% \pm 0.6$ in R1 and R2, respectively. These systematic variations demonstrate that the thermal processes are more tied to massive star formation dominating in the spiral arms {than in diffuse ISM. On average, the nonthermal emission dominates the thermal emission in both R1 and R2 even at 4.8\,GHz in IC~342. At all three frequencies investigated (4.8 GHz, 1.4 GHz, and 0.144 GHz), the highest thermal fraction is consistently observed in the central region of the galaxy, with a fraction of $63\% \pm 4.7$, $35\% \pm 2.5$, and $20\% \pm 1.6$,  respectively, due to its high SFR activity and dense ionized gas.}
%
Table~\ref{tab:fluxes} lists the integrated flux densities and thermal fractions for the entire galaxy and R1 and R2. The reported values for the entire galaxy are obtained by integrating the observed and thermal RC maps in the plane of the galaxy (i = $31^{\circ}$) around the center out to a radius of $R<10$ kpc. 

\begin{center}
\begin{table*}[!htb]
\caption{Radio continuum flux densities and thermal fractions of IC~342 and its regions.}
\label{tab:fluxes}
\centering
\begin{tabular}{cccccc}
\hline\hline
Region & Frequency & Observed flux & Thermal flux & Thermal Fraction \\
       & (GHz)     & (Jy)    & (Jy)  & (\%)             \\
\hline
       & 4.8       & $0.89 \pm 0.04$    & $0.14 \pm 0.01$  & $15.7 \pm 1.1$             \\
IC~342  & 1.4       & $2.99 \pm 0.16$   & $0.16 \pm 0.01$  & $5.3 \pm0.4$             \\
       & 0.1     & $8.87 \pm 0.63$    & $0.21 \pm 0.01$  & $2.3 \pm0.2$              \\
\hline
       & 4.8       & $0.37 \pm 0.01$    & $0.11 \pm 0.01$  &$ 29.7 \pm 2.1 $            \\
R1 & 1.4       & $0.96 \pm 0.05$    & $0.13 \pm 0.01$  &$ 13.5 \pm 1.0$              \\
       & 0.1     & $2.78 \pm 0.18$   &  $0.16 \pm 0.01$  & $5.7 \pm 0.5 $              \\
\hline
       & 4.8       & $0.23 \pm 0.01$     & $0.02 \pm 0.01$  & $8.7 \pm 0.6 $            \\
R2 & 1.4       & $0.84 \pm 0.05$    & $0.03 \pm 0.01$  & $3.5 \pm 0.3 $              \\
       & 0.1     & $2.65 \pm 0.15$    & $0.03 \pm 0.01$  & $1.1 \pm 0.1$              \\
\hline\hline
\end{tabular}

\begin{tablenotes}
\footnotesize
\item\tablefoot{The radio continuum flux densities and thermal fractions for the entire IC~342 galaxy and its regions of star-forming (R1) and diffuse ISM (R2). The frequencies are given in GHz and the flux densities in Jy.}
\end{tablenotes}
\end{table*}
\end{center}

\section{Synchrotron spectral index}
\label{sec:alpha}

\begin{figure}[h!]
    \centering
{\includegraphics[width=8.5cm]{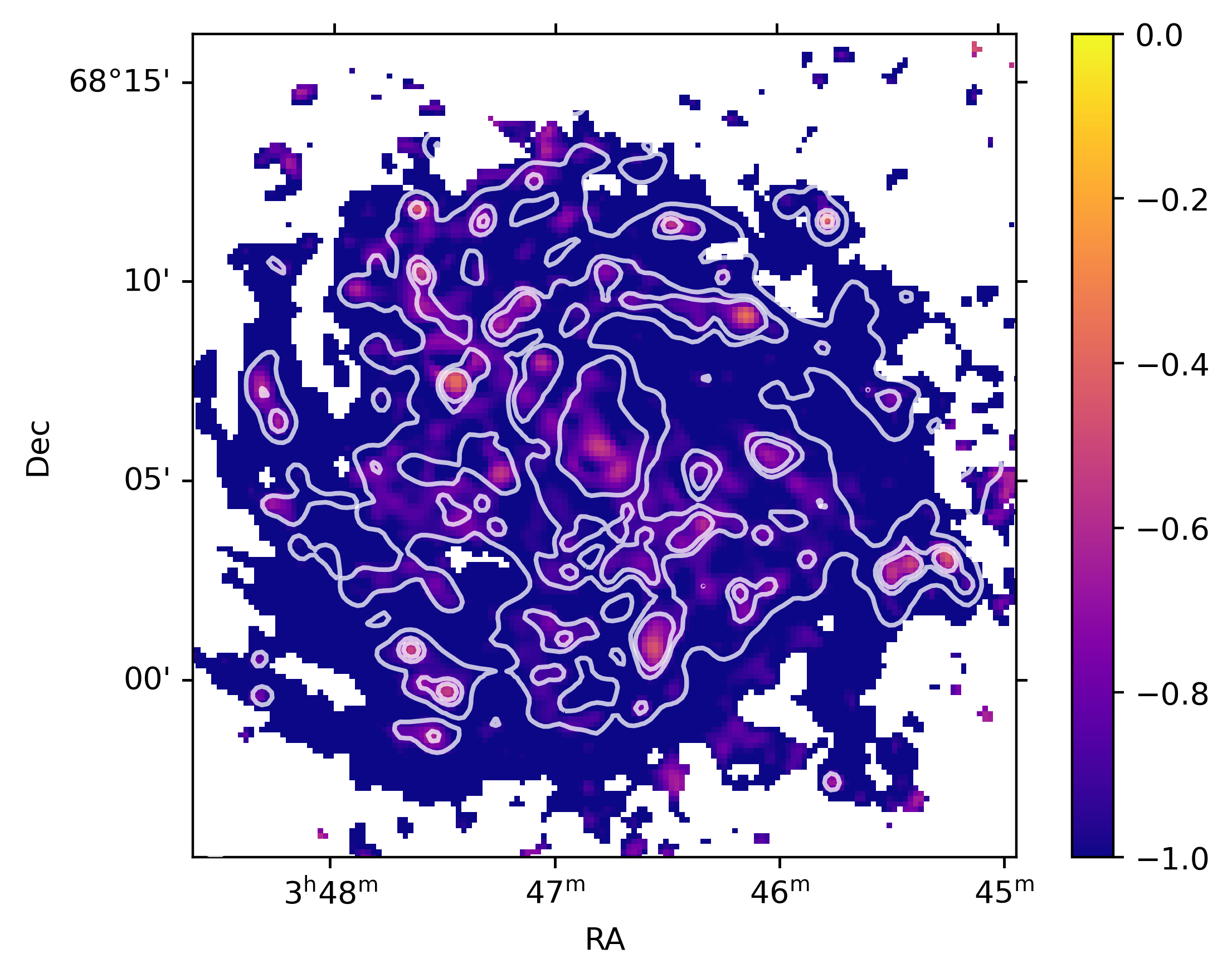}}

{\includegraphics[width=8.5cm]{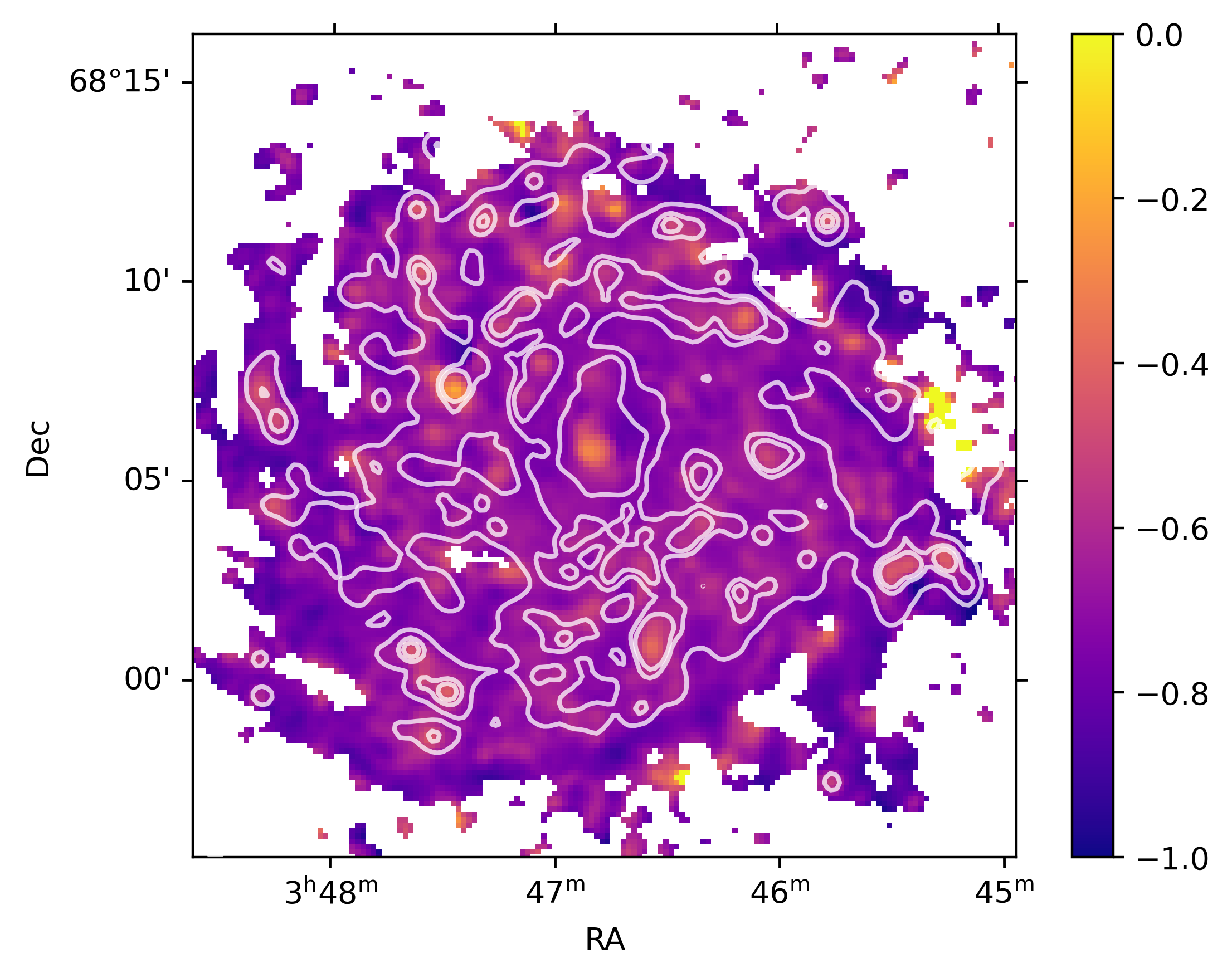}}

{\includegraphics[width=8.5cm]{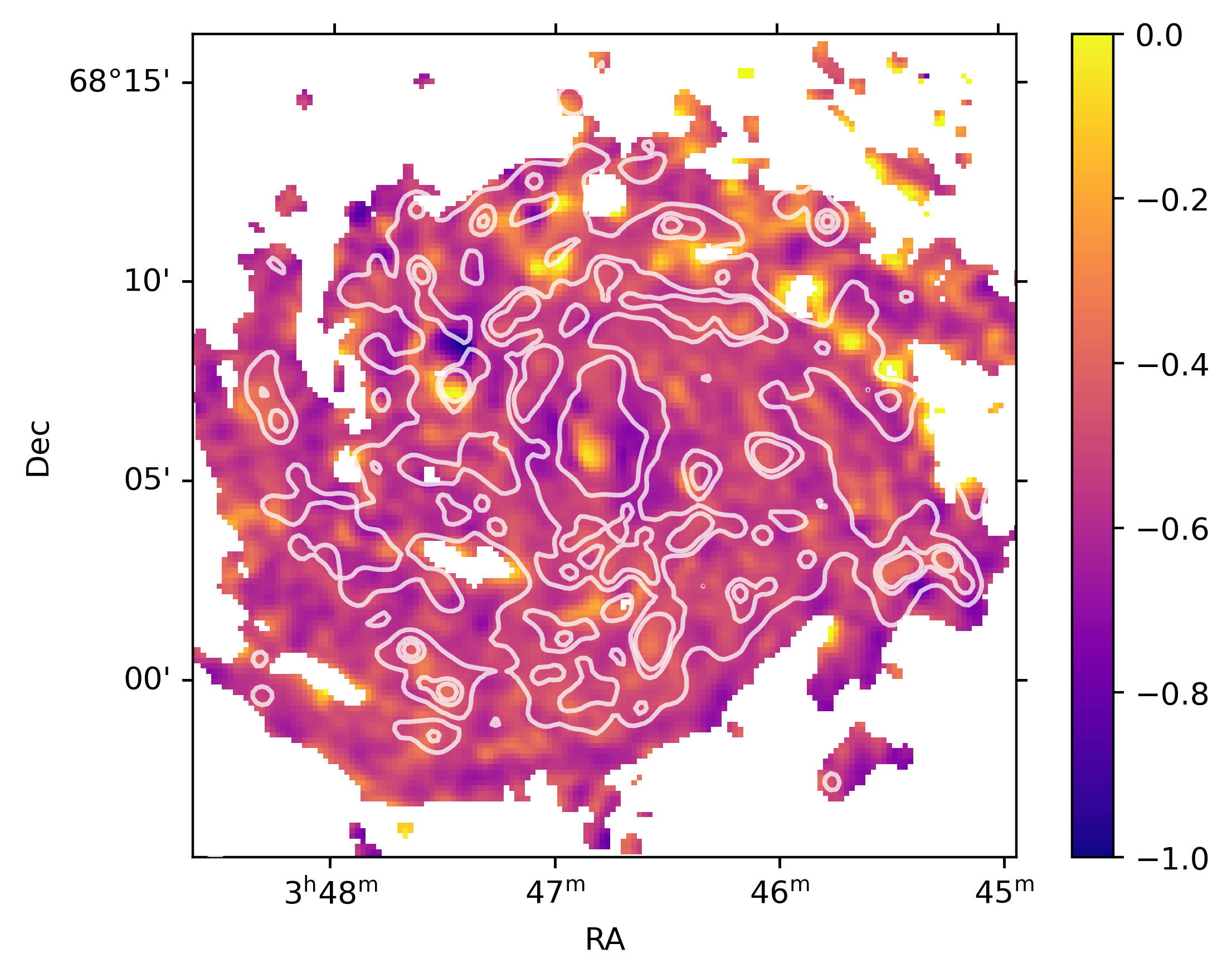}}
 \caption{Synchrotron spectral index maps of IC 342 for $\alpha_{[1.4-4.8]}$ ({\it top}), $\alpha_{[0.14-4.8]}$ ({\it middle}) and $\alpha_{[0.14-1.4]}$ ({\it bottom}) overlaid with contours of the 22\,$\mu$m emission indicating star forming regions. Contour levels are 2, 4, 6, 8, and 10 $\times$ $10^{-4}$ Jy/pix.}
	\label{fig:spect}

\end{figure}

\begin{figure}
    \centering
    {\includegraphics[width=9.5cm]{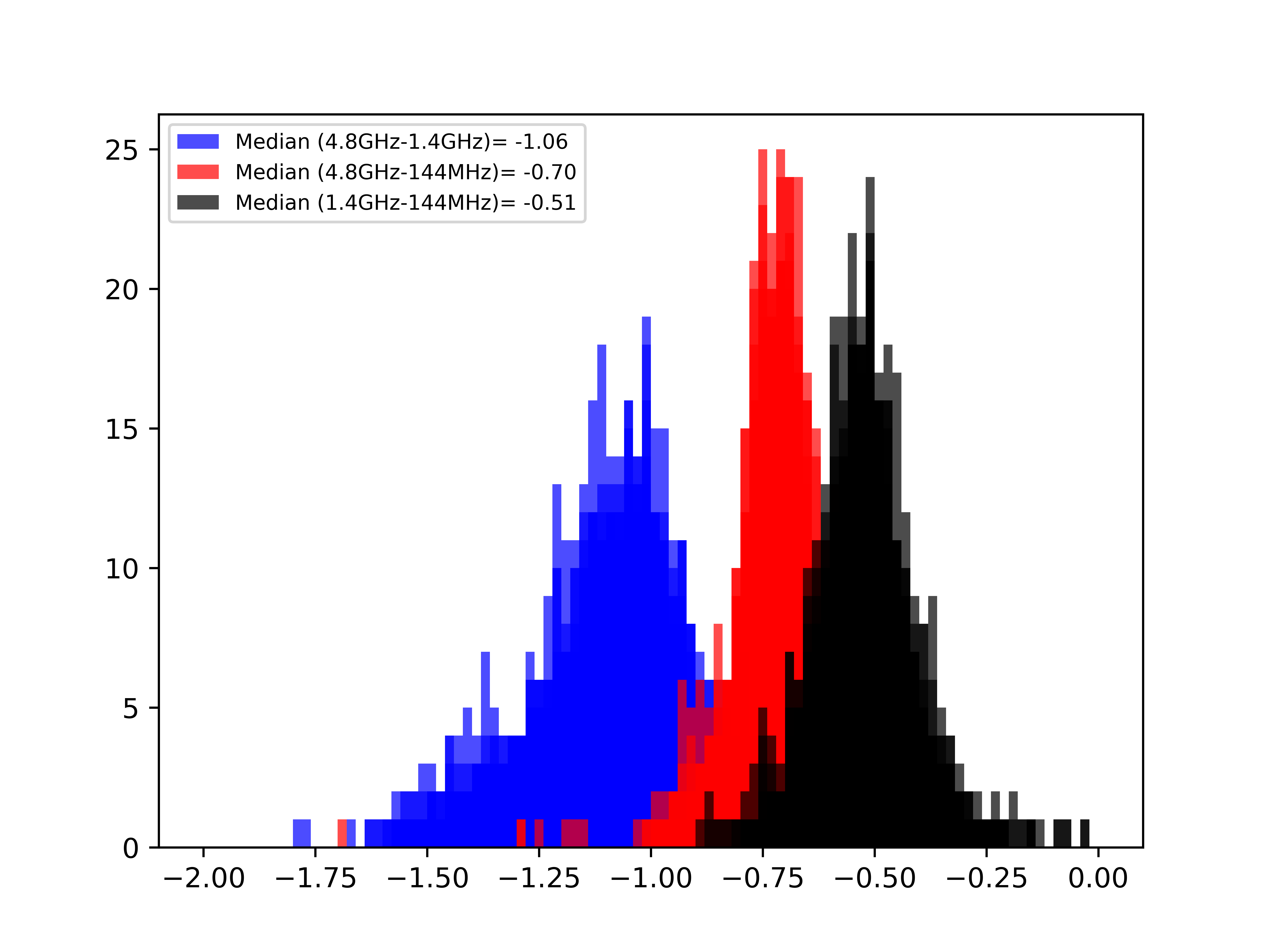}}
\caption{Histogram of the synchrotron spectral index in the high frequency interval of $1.4 \leq \nu \leq 4.8$\,GHz ({\it blue}), the entire frequency interval of $0.14 \leq \nu \leq 4.8$\,GHz ({\it red}), and low frequency interval of $0.14 \leq \nu \leq 1.4$\,GHz ({\it black}). } 
    \label{fig:histo}

\end{figure}

The nonthermal radio maps derived in Sect.~\ref{sec:thermal} were used to obtain the synchrotron spectral index ($\alpha, ~S\propto \nu^{\alpha}$) over the entire frequency range of 0.14 to 4.8\,GHz, $\alpha_{[0.1-4.8]}$, at 25\arcsec~angular resolution (Fig.~\ref{fig:spect}). This was obtained only for pixels with intensities above $3\sigma$ rms noise levels. We find that the spectral index changes across the disk. It is flatter in star forming regions in the galaxy center and in the spiral arms than in other places in the galaxy. The distribution of the synchrotron spectrum across the disk agrees with injection of  CREs  in star forming regions and their energy loss propagating away from these regions towards the inter-arm and outer disk. This agrees with previous findings in other nearby galaxies such as M33 and NGC6946 \citep{Spec1&taba,intro12&taba,Sep2&Hamid,intro14&taba}. The median value of $\alpha_{[0.1-4.8]}$ across IC342 is $-0.70$ with a dispersion of $0.06$ as shown in the histogram representation (Fig.~\ref{fig:histo}).
This is flatter than the typical synchrotron spectral index found for galaxies in the mid-radio frequencies of 1-10\,GHz  \citep[$\simeq -0.97\pm 0.16$][]{intro24&taba}. To be more certain about the flattening, we  derive the spectral index separately for two frequency intervals of lower and higher than 1.4\,GHz, i.e, $\alpha_{[0.1-1.4]}$ and $\alpha_{[1.4-4.8]}$. Figure~\ref{fig:histo} shows that the spectral index is on average flatter in the low-frequency interval ($\alpha_{[0.1-1.4]}=-0.51\pm 0.09$) than in the high-frequency interval ($\alpha_{[1.4-4.8]}=-1.06\pm 0.19$).  Thus, while the spectral index agrees with that of \cite{intro24&taba} at high-frequency interval, it becomes flatter at low frequencies $\nu<1.4$\,GHz.  A similar flattening is reported for other LoTSS galaxies  by \cite{Spec2&Heesen}. It is important to note that this low-frequency flattening occurs not only in star forming regions but also in more diffuse parts of the ISM (Fig.~\ref{fig:spect}). \\

\section{Magnetic field strength}
\label{sec:magnet}
The strength of the total magnetic field $B_{\rm tot}$ is related to the nonthermal synchrotron $I_{n}$ intensity. Assuming that the energy densities of the magnetic field and cosmic rays are equal
$(\varepsilon_{\rm CR} = \varepsilon_{B_{\rm tot}} = B_{\rm tot}^{2} / 8\pi)$:
\begin{equation}\label{Eq. 2}
B_{\rm tot} = C(\alpha_{n},K,L)[I_{n}]^{\dfrac{1}{\alpha_{n}+3}},
\end{equation}
where C is a function of the synchrotron spectral index $\alpha_{n}$, K the ratio between the number densities of cosmic ray protons and electrons, and L the path length in the synchrotron emitting medium \citep[see][]{MapsMF1&Beck}. 
Assuming that the magnetic field is parallel to the plane of the galaxy (inclination of $i = 31^{\circ}$ and position angle of the major axis of $PA = 39^{\circ})$, $B_{\rm tot}$ is mapped across the galaxy using the nonthermal emission at 1.4\,GHz  and $\alpha_{n}=\alpha_{[1.4-4.8]}$ which is close to the theoretical synchrotron loss spectral index. The usage of the low-frequency part of the spectrum ($\nu<1.4$\,GHz)  in Eq.~\ref{Eq. 2} can lead to an incorrect result because  the low-frequency flattening  (Sect.~\ref{sec:alpha}) can be linked to unrelated energy loss mechanisms or absorption effects. 
In our calculations, we used $K \simeq 100$ \citep{MapsMF1&Beck} and $L \simeq$~1~kpc/cos(i). 

\begin{figure}[htpb]
    
     {\includegraphics[width=8.5cm]{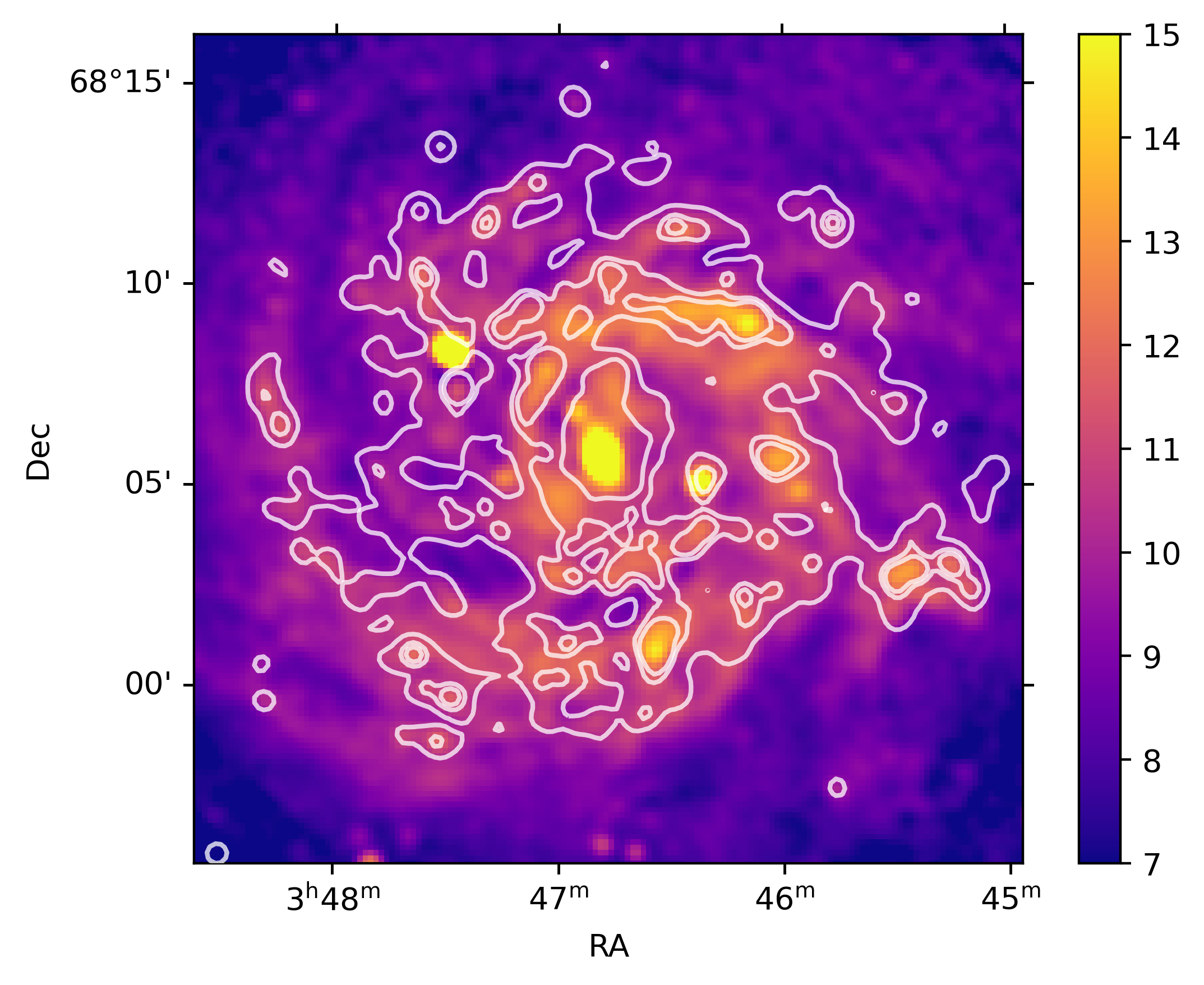}} 
    {\includegraphics[width=8.5cm]{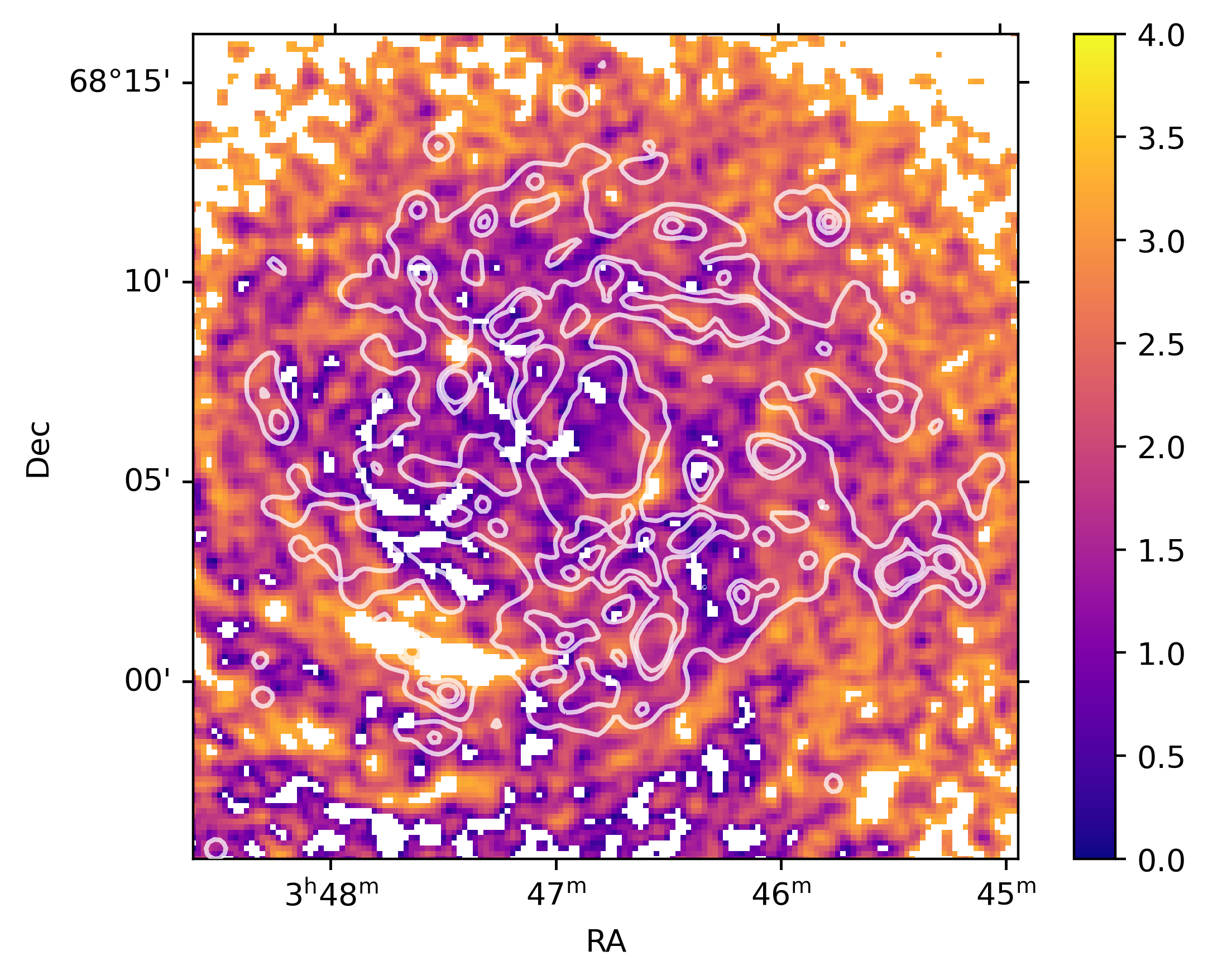}} 
    {\includegraphics[width=8.5cm]{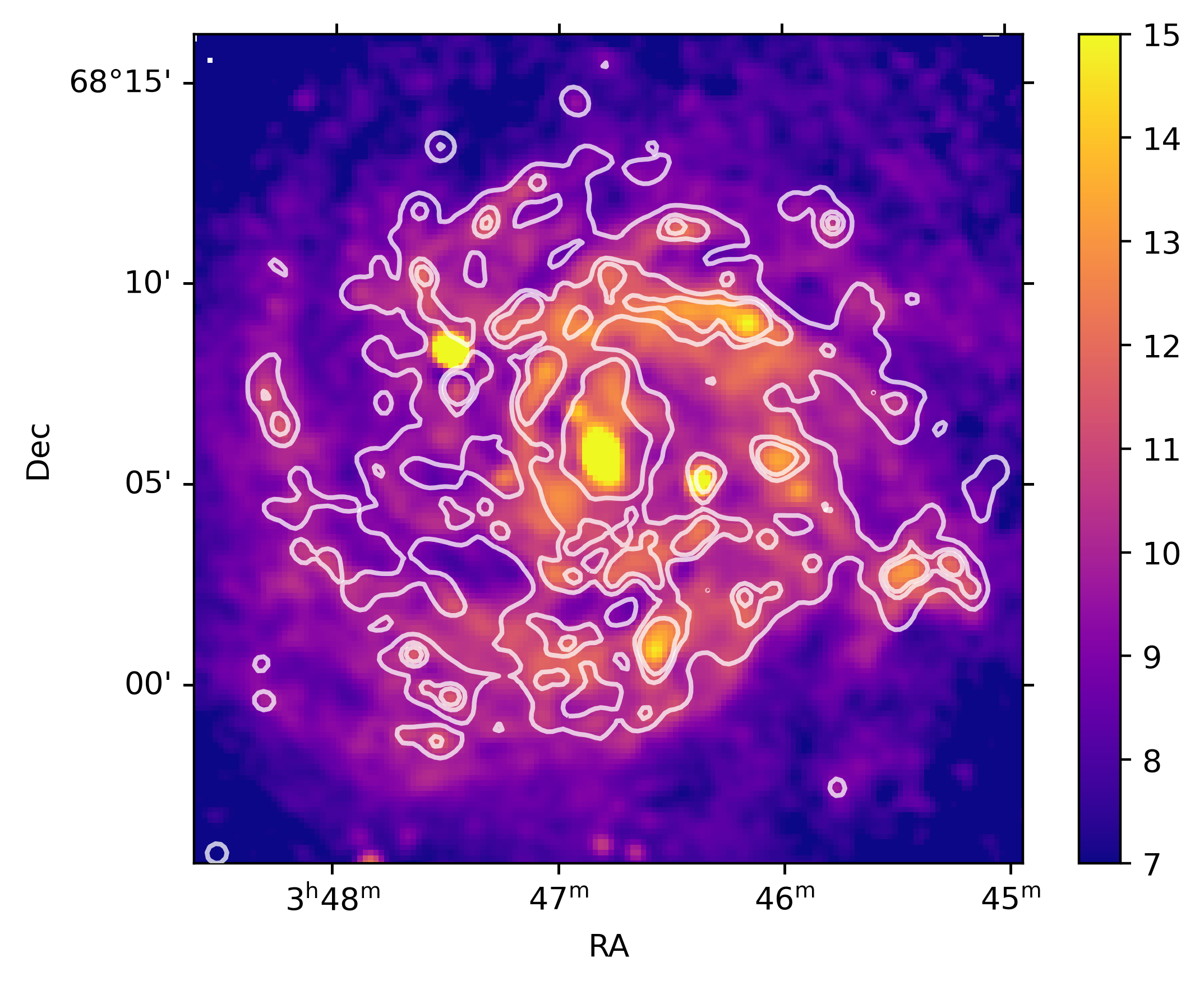}}
\caption{Strength of the total magnetic field $B_{tot}$ obtained using the synchrotron emission at  1.4\,GHz in IC~342 at $25"$ angular resolution ({\it top}).  Also shown are maps of the ordered $B_{ord}$ ({\it middle}) and turbulent $B_{tur}$ ({\it bottom}) magnetic field strengths. The bars on the right of each image show the magnetic field strength in $\mu$G. Contours show the 22\,$\mu$m emission at levels of 2, 4, 6, 8, and 10 $\times$ $10^{-4}$ Jy/pix.}
    \label{fig:magnet}
\end{figure}

Figure \ref{fig:magnet} shows that the magnetic field is the strongest in the center of the galaxy ($B_{\rm tot}\simeq 20~\mu$G), {while it is $\simeq$~10\,$\mu$G in other star forming regions and spiral arms.} On average, we find that $B_{\rm tot}=10.23~\pm~1.30~\mu$G over the disk of IC~342 which is about $20$\% weaker than that obtained by \citep[][]{beck_15}. This discrepancy can be attributed to the utilization of distinct methodologies for separating the thermal and nonthermal emission, notably our approach which refrains from relying on any a priori assumptions regarding the non-thermal spectral index.

As the fraction of the polarized intensity PI is related to the strength of the ordered magnetic field, and the nonthermal intensity $I_{n}$ to the total magnetic field in the plane of the sky, $I_{n}$ - $(PI/0.75)$ gives the nonthermal emission due to the turbulent magnetic field $B_{\rm tur}$ \citep{refId0}. Using this intensity in Eq.~\ref{Eq. 2} yields the distribution of $B_{\rm tur}$ across the galaxy. We also map the strength of the ordered magnetic field defined as $B_{\rm ord}^2= B_{\rm tot}^2-B_{\rm tur}^2$. As shown in Fig.~\ref{fig:magnet}, the ordered field is stronger in the outer disk than in the inner disk. It is inferred that the strong $B_{\rm tot}$ in star forming regions is mainly due to turbulence as $B_{\rm tur}$ is also stronger there. It is also likely that, in star forming regions, the ordered field has different directions (or tangled) and their polarization  signals cancel out each other within the beam size. As such the structure of $B_{\rm tur}$ can be a combination of a random, isotropic component and a field tangled on unresolved scales.
%
In other words, we cannot distinguish between unresolved structures of the ordered field and truly turbulent fields for structures smaller than the beam. 

\section{Radio--FIR correlation}
\label{sec:radio-fir}
Investigating the  nonthermal radio--FIR correlation,  we invoke various approaches used often in resolved {studies} such as the classical pixel wise correlation, the FIR-to-radio ratio mapping as well as the multi-scale analysis. We perform the first two approaches at all 3 radio frequencies at 25\arcsec~angular resolution (that of the 4.8\,GHz). As a higher resolution is preferred for the multi-scale  analysis, this method uses only the 0.14 and 1.4\,GHz maps at 15\arcsec~($\simeq~0.2$\,kpc) resolution.  The Herschel PACS FIR data are used at different bands of 70, 100 and 160\,$\mu$m to investigate possible influence of dust temperature (warmer dust is traced at a shorter FIR wavelength). These approaches and their results are presented as follows. 
\\

\subsection{Pixel-by-pixel correlation}
\label{sec:pix}
The Pearson's linear correlation coefficient between two images of identical pixels $i$, $S_{1}(x,y)$ and $S_{2}(x,y)$, is given by:
\begin{equation}
r_{p} = \dfrac{\sum_{i=0}^{\rm n} (S_{1i}-\langle S_{1}\rangle)(S_{2i}-\langle S_{2}\rangle)}{\sqrt{\sum_{i=0}^{\rm n}(S_{1i}-\langle S_{1}\rangle)^2\sum_{i=0}^{\rm n} (S_{2i}-\langle S_{2}\rangle) ^{2}}},
\end{equation}
{where 
 \(S_{1i}\) and \(S_{2i}\) represent the intensity of the $i$-th pixel in images 1 and 2, respectively. Mean of the pixel intensity values in each image is denoted as $\langle S_1 \rangle = \frac{1}{n} \sum_{i=0}^n S_{1i}$ and $\langle S_2 \rangle = \frac{1}{n} \sum_{i=0}^n S_{2i}$, with \(n\)  the total number of pixels.}

In the case of a perfect correlation, $r_{p} = 1$ and for a perfect anti-correlation,  $r_{p} = -1$. If the two images are completely uncorrelated, then  $r_{p} = 0$. {The statistical error in $r_{p}$ is given by $\triangle r_{p}$ = $\sqrt{1 -r_{p}^{2}}/({\sqrt{n-2}})$ which depends on the strength of the correlation and the number of pixels n. }

To extend the analysis, Pearson correlation coefficients were computed for the radio continuum and FIR bands at 70, 100, and 160 $\mu m$, focusing on regions delineated as R1 (star-forming) and R2 (diffuse) as discussed in Sect.~\ref{sec:thermal}. Sets of independent data points ($n$) were established, ensuring a beam area overlap of less than $20\%$ and selecting pixels spaced by more than one beamwidth. A Student's t-test was subsequently applied to assess the statistical significance of the correlations.  It is noteworthy that both the observed radio continuum maps at 1.4 GHz and 4.8 GHz, as well as their synchrotron components, exhibit strong correlations with the FIR maps, particularly within the star-forming regime (R1), with correlation coefficients ($r_{p}$) exceeding $\approx 0.90$  (Table~4). Conversely, in the diffuse regime (R2), correlations are weak, typically below $r_{p} \approx 0.5$. {This underscores the variability in the strength of the radio--FIR correlations across different environments within the galaxy. }
{For the entire galaxy, the correlations at 0.144\,GHz are not as tight as those at higher frequencies  ($r_{p} \lesssim 0.70$). However, the correlations are much improved  in the R1 regime, i.e., in denser parts of the ISM. This shows the significance of the diffuse component at 0.144\,GHz causing the relatively weak radio--FIR correlation for the entire galaxy.

It is interesting to note that the correlations at different FIR bands all agree withing the errors irrespective of the region. This indicates that dust temperature does not play a role in this correlation in IC~342 unlike in M33 \citep{Spec1&taba,2007A&A...466..509T}. It is likely that the warm and cold dust components are well mixed or that the radiation field does not change much across the galaxy. The fact that this galaxy does not host giant HII complexes such as those in M33 producing energetic ionizing UV photons means that dust is heated mainly by almost a uniform interstellar radiation field (ISRF) at least down to the resolution of this study ($\simeq 200$\,pc). }

We note that almost all correlations follow a sub-linear radio vs FIR relations (or a super-linear FIR vs radio), for example, the  synchrotron emission at 1.4\,GHz is correlated with the FIR emission at 70\,$\mu$m with a slope of $0.79\pm0.03$ (in log-log plane) with a t-test value large enough to be taken as statistically significant ($t>2$). A similar sub-linearity between the radio continuum emission and different SFR tracers was reported in other resolved studies \citep[e.g.][]{SFR3&Heesen} which is linked to diffusion of  CREs  \citep[e.g.][]{Berkhuijsen:2013uaa}.

\subsection{FIR-to-radio ratio map}
\label{sec:qmap}

\begin{figure}[htbp]
    \centering
    {\includegraphics[width=8.7cm]{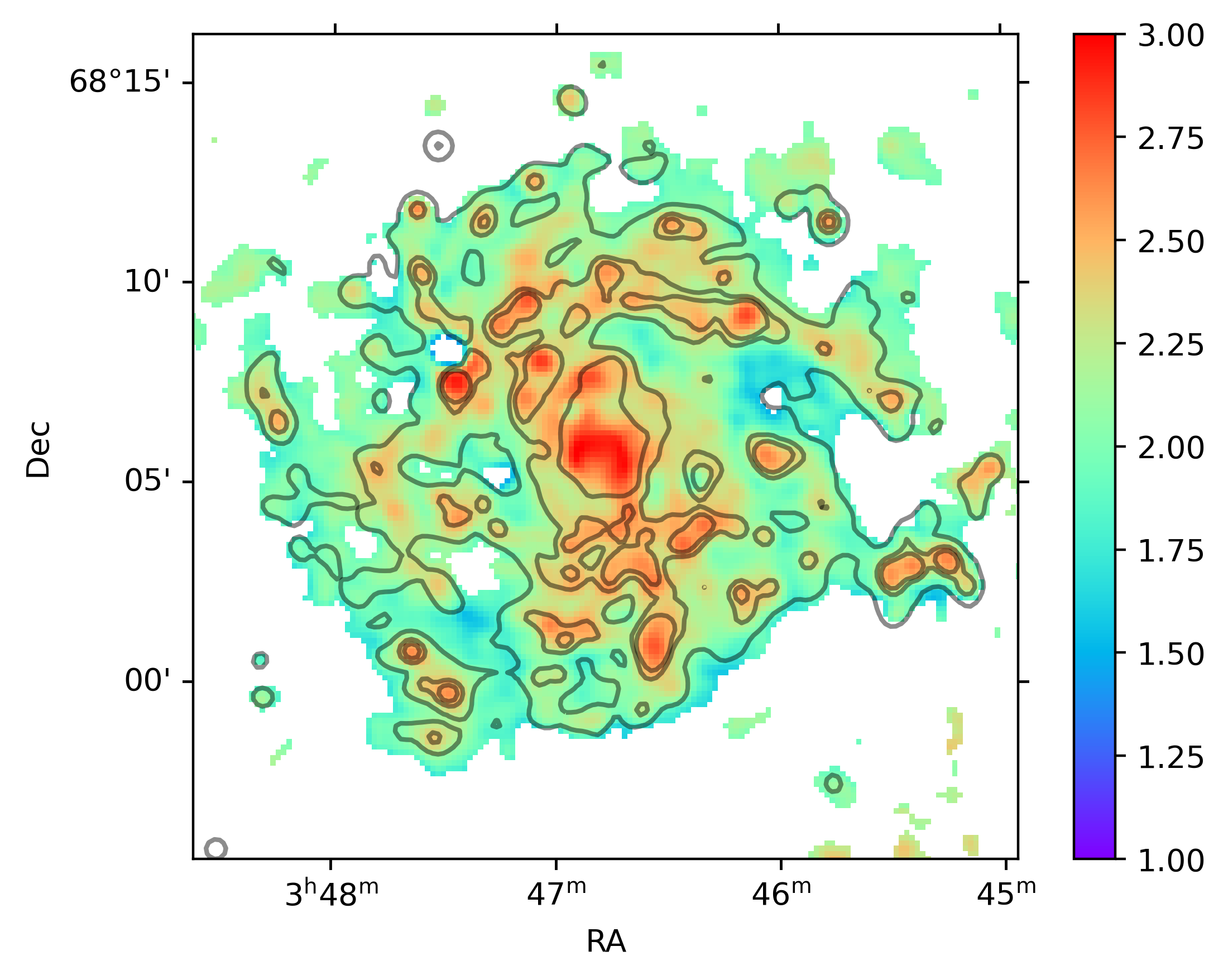}}     
    {\includegraphics[width=8.7cm]{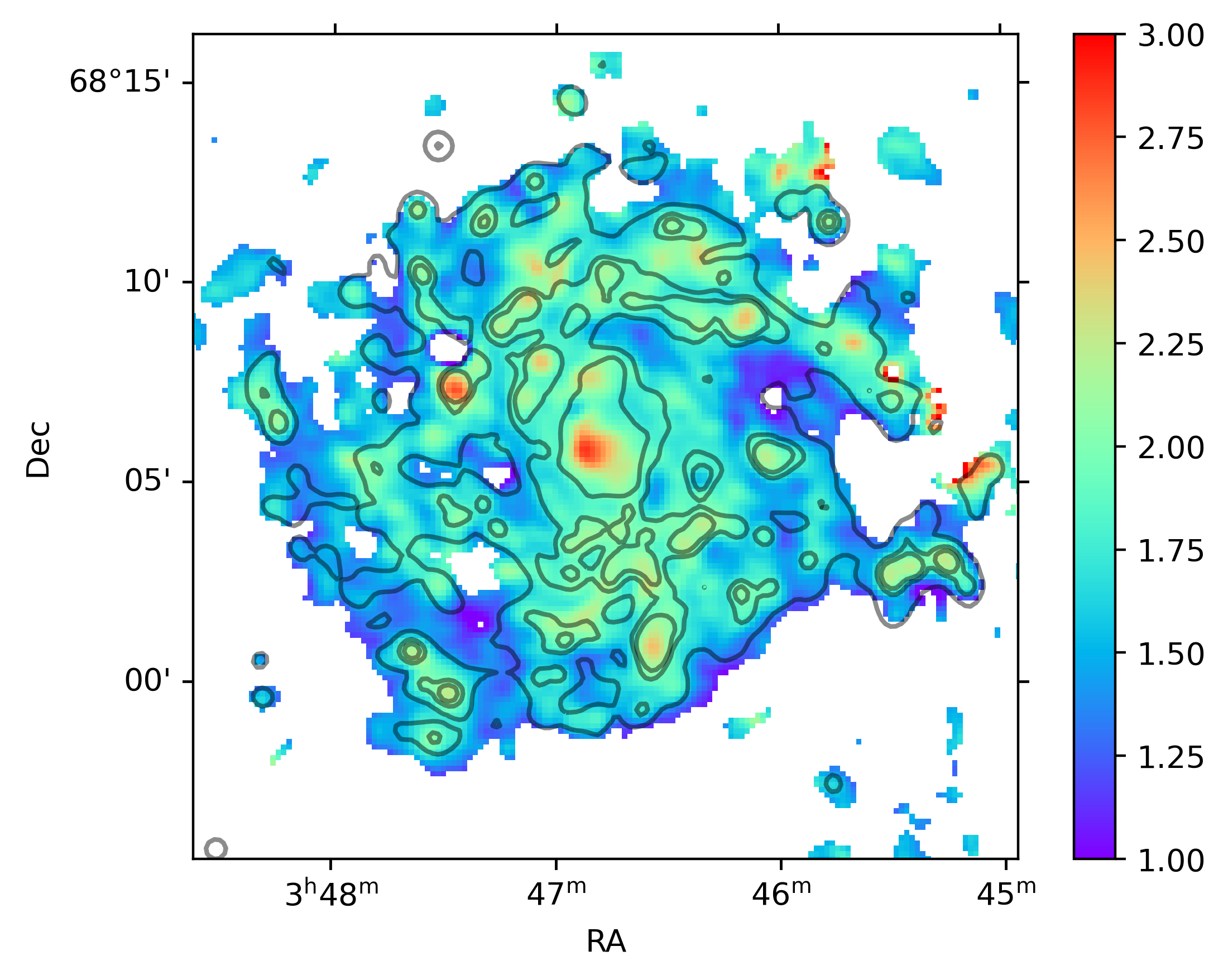}}
\caption{Logarithmic ratio of the 70$\mu$m emission to the radio synchrotron intensity at 1.4\,GHz ($q_{1.4}$, {\it top}) and 0.14 GHz ($q_{0.14}$, {\it bottom}) at $25^{\arcsec}$ angular resolution.  Contours show the 22\,$\mu$m emission at levels of 2, 4, 6, 8, and 10 $\times$ $10^{-4}$ Jy/pix. }
    \label{fig:q}
\end{figure}

\begin{figure}[htbp]
    \centering    
  {\includegraphics[width=10.0cm]{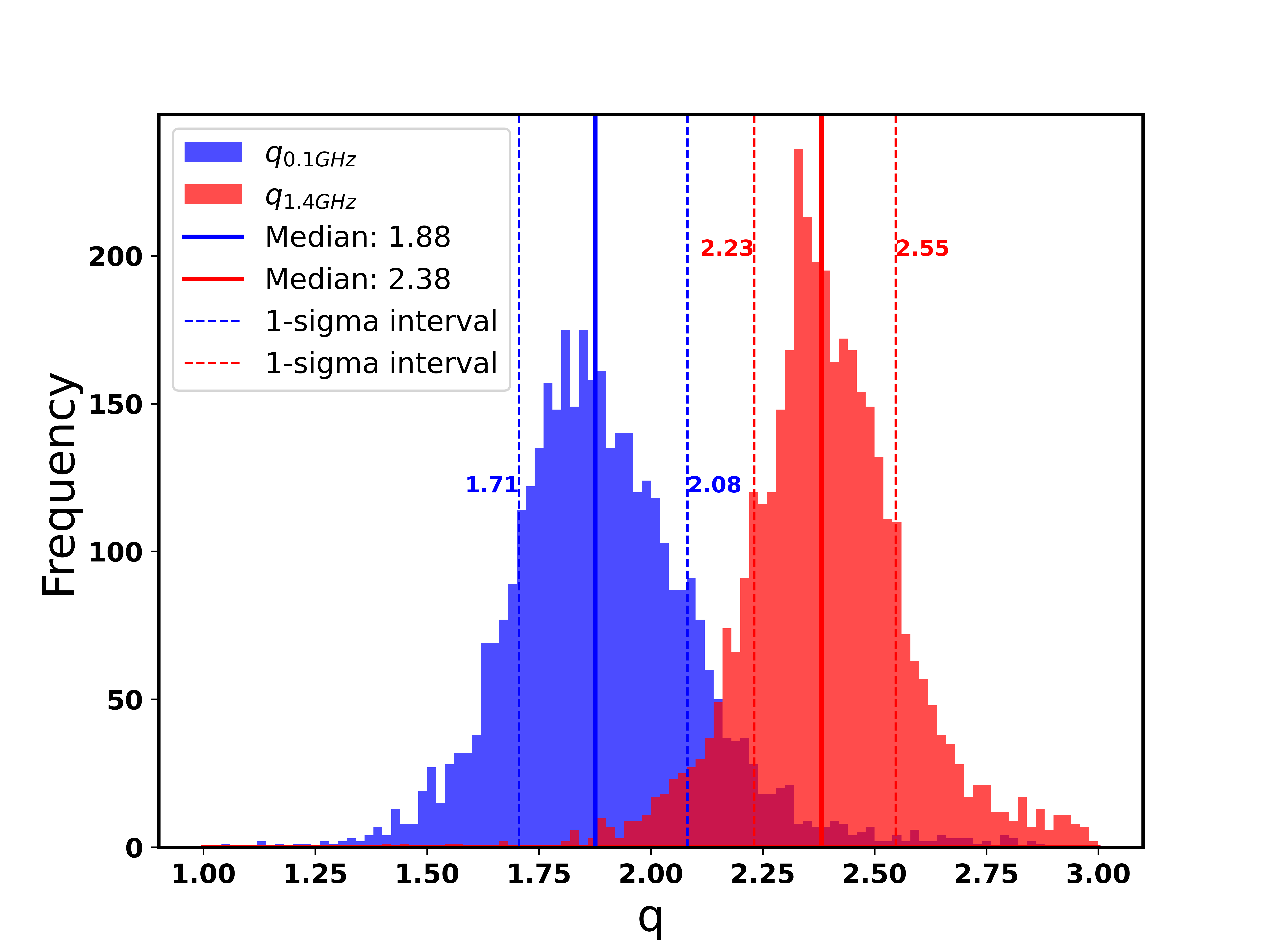}} \caption{Histogram of $q_{1.4}$ ({\it red}) and $q_{0.14}$ ({\it blue}) in star-forming regime of the ISM (R1) in IC~342. Median value of each distribution ({\it solid lines}) and their 1-sigma intervals ({\it solid lines}) are also indicated.}
    \label{fig:hist-q}
\end{figure}

{As already shown in the pixel-by-pixel correlation analysis, the radio-FIR correlation changes depending on the level of star formation activity in IC~342. Another method to assess the relative variations of the radio and FIR emission locally 
is to map the ratio} of the FIR to radio fluxes in logarithmic scale \citep[called as q-parameter, see e.g., ][]{Murphy_2006,intro11&Hughes,intro15&taba}. This will further help to investigate if variations in the radio–FIR correlation correspond to particular physical structures in IC~342. {Calculating the q-parameter, it is common to use a  combination of the IRAS bands at 60\,$\mu$m and 100\,$\mu$m as a proxy of the integrated FIR emission in the frequency range of 42-122\,$\mu$m \citep{q1&Helou}, a single IR band at 24\,$\mu$m or 70\,$\mu$m \citep[][]{Appleton_2004,Murphy_2006}, or direct integration of the IR SEDs \citep{intro15&taba,intro24&taba}. In this study, considering the available data\footnote{We note that the 22\,$\mu$m IR data is used as a tracer of the free-free emission (and SFR) and, hence, it has not been utilized in calculating the q-parameter to avoid any artificial correlation when comparing it with $\Sigma_{\rm SFR}$ in Sect.~\ref{sec:corVSmf}.}, we opt to use the single-band 70\,$\mu$m data to map the q-parameter following the convention by \cite{Appleton_2004} and \cite{Murphy_2006}:}  
\begin{equation}
\label{Eq4}
q \equiv \log \,\left(\dfrac{S_{70 \mu m}}{S_{\rm radio}}\right).
\end{equation}
In this definition, $S_{\rm radio}$ is the radio flux density and $S_{70 \mu m}$ is the monochromatic 70 $\mu m$ flux density. 
%
We obtained the $q$-parameter for the nonthermal synchrotron emission (as $S_{\rm radio}$ in Eq.~\ref{Eq4}) at the conventional radio frequency of 1.4\,GHz ($q_{1.4}$). For comparison, we also mapped this quantity at the low radio frequency of  0.14\,GHz ($q_{0.1}$). The maps are constructed excluding pixels below the $3\sigma$ rms noise level in the synchrotron and FIR maps. As shown in Fig.~\ref{fig:q}, the $q$-parameter is generally higher in the spiral arms and star forming regions irrespective of the radio frequency in agreement with \cite{intro15&taba}. Fig.~\ref{fig:hist-q} also presents a histogram of $q_{1.4}$ and $q_{0.1}$ in the star-forming regime (R1) in IC 342. 
{On average,} $q=2.16 \pm 0.23$ using the 1.4 GHz synchrotron emission which agrees with previous studies. A lower value is obtained using the radio synchrotron emission at lower frequency of 0.144\,GHz ($q=1.68 \pm 0.28$) consistent with \citep[]{q2Yun,intro16&Read}, while it is higher using the  4.8 GHz emission ($q=2.73 \pm 0.22$).

\subsection{Multi-scale correlation}
\label{sec:wavelet}
The pixel-by-pixel correlation can be biased  toward a specific scale if its corresponding structure contains a relatively more weight (brightness) in a map.  Hence, we also use a scale-by-scale analysis by means of a 2D Wavelet decomposition which consists of convolution of an image with a set of self-similar basis functions that vary with scale and location. In general, the wavelet transform can be understood as a generalization of the Fourier transform. The oscillatory function is both localized in time and in frequency in the wavelet transform. Based on a mother function, called the analyzing wavelet, the family of basis functions is generated. Throughout this paper, we employ the continuous wavelet transform in two dimensions:
\begin{equation}
W(a,x) = \dfrac{1}{a^{\kappa}} \int\limits_{-\infty}^\infty f(x') \psi^{*} \left(\dfrac{x'- x}{a}\right) dx' 
\end{equation}

A wavelet is a multidimensional function representing the position of a wavelet, where  $f(x)$ is a two-dimensional function (an image), $\psi(x)$ is the analyzing wavelet, the symbol $\ast$  is the complex conjugate and x = (x, y)  represents the position and a indicates its scale. An energy normalization parameter, $\kappa$ is applied (we use $\kappa = 2$). { Following \cite{wavlet1&Frick}, to achieve both good spatial and scale resolution, we use the "PetHat" function as the processing wavelet $\psi(x)$.} For two identical images with the same size and linear resolution, the wavelet cross-correlation coefficient at scale $a$ is defined as follows:
\begin{equation}
\label{Eq6}
r_{w}(a) =\dfrac{\int\int W_{1}(a,x)\,W^{*}_{2}(a,x)\,dx}{[M_{1}(a)\,M_{2}(a)]^{1/2}},
\end{equation}
with M(a)=$\int\limits_{-\infty}^\infty \, \int\limits_{-\infty}^\infty \, \vert W(a,x)\vert^{2}$ the wavelet equivalent of the power spectrum in Fourier space. The value of $r_{w}$ varies between $-1$ (total anti correlation) and +1 (total correlation). In the case of a given scale, a correlation coefficient of $|r_{w}| = 0.5$ represents a marginal value for accepting the correlation between the structures \citep{intro12&taba}. Error in  $r_{w}$ is given by $\triangle r_{w} =\sqrt{1 -r_{w}^2}/{\sqrt{n-2}}$ { with $n$ the number of points at each specific scale}.


We decomposed all the radio and FIR maps into 50 angular scales between $7^{\arcsec} \sim 111pc$  \citep[half of the beam width, following][]{wavlet1&Frick} and $1007^{\arcsec} \sim 16$~kpc to a map of third of the image size to avoid boundary effects { (a third of the image size to avoid boundary effects)}. Then the wavelet coefficients W(a, x) of the radio and FIR maps are driven and used in Eq.~\ref{Eq6} to calculate the wavelet correlation coefficient $r_{w}$ for every scale of decomposition.

\begin{figure}[!h]
    \centering

    \subfigure{\includegraphics[width=9.4cm]{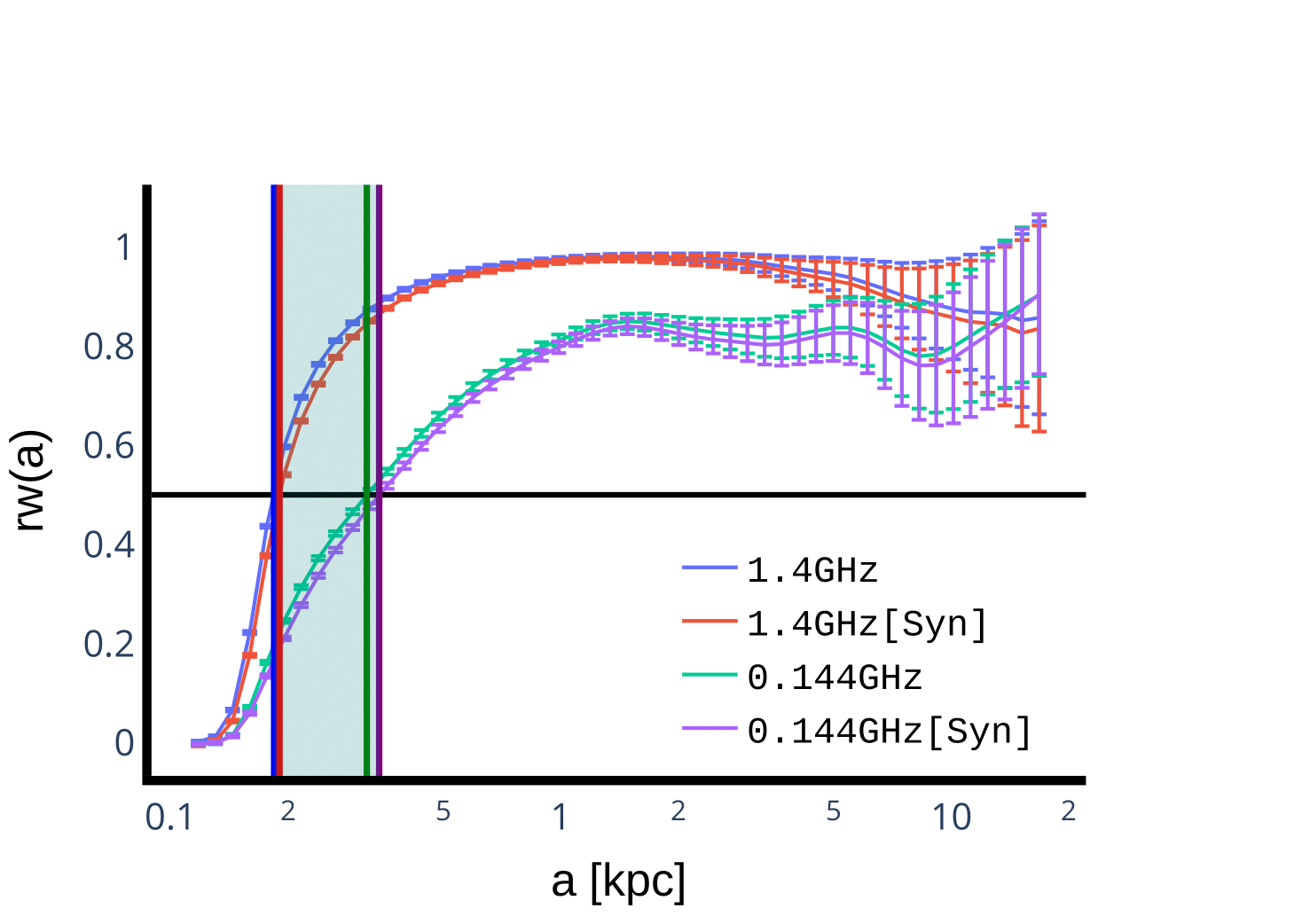}}  
    \subfigure{\includegraphics[width=9.4cm]{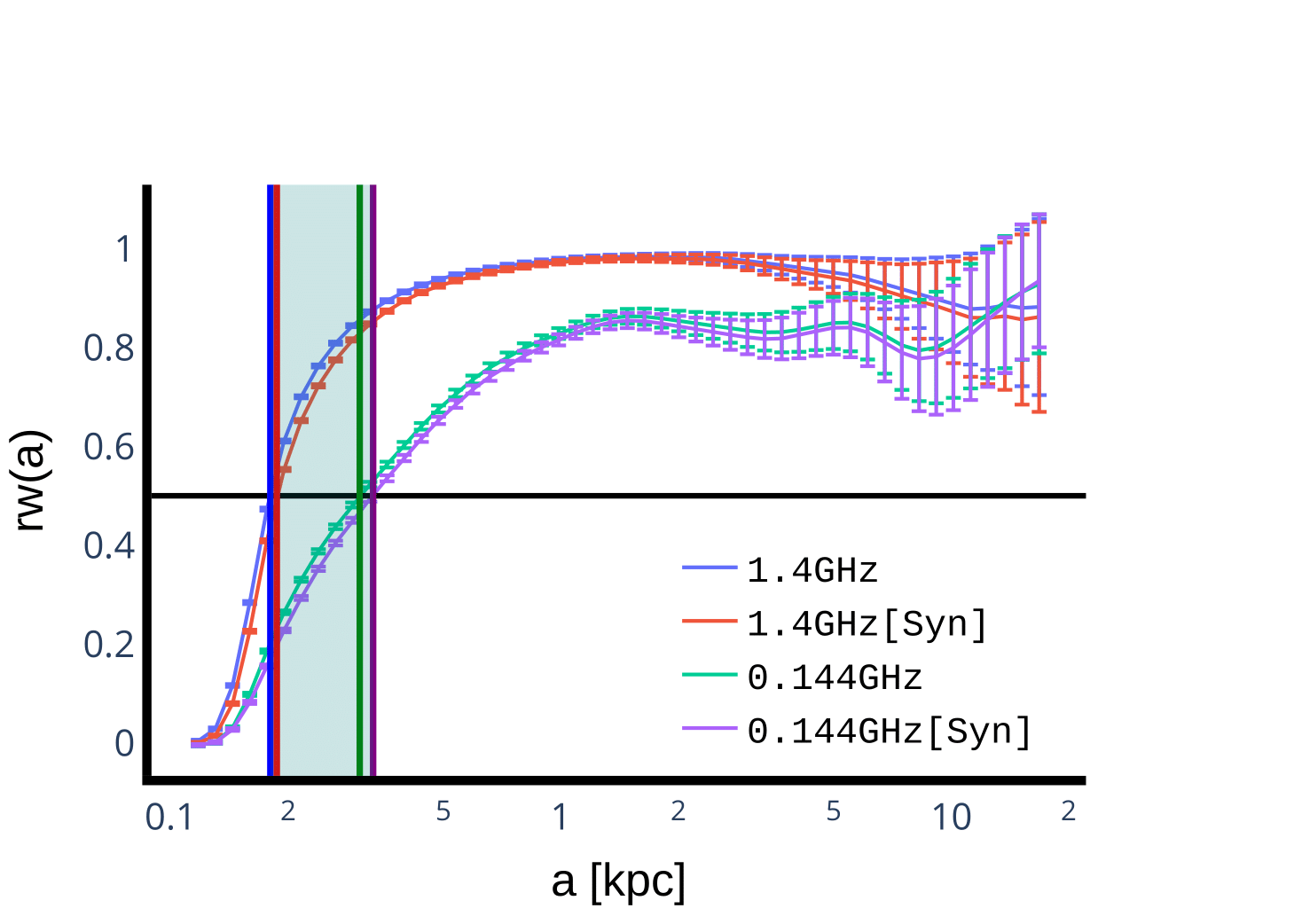}} 
    \subfigure{\includegraphics[width=9.4cm]{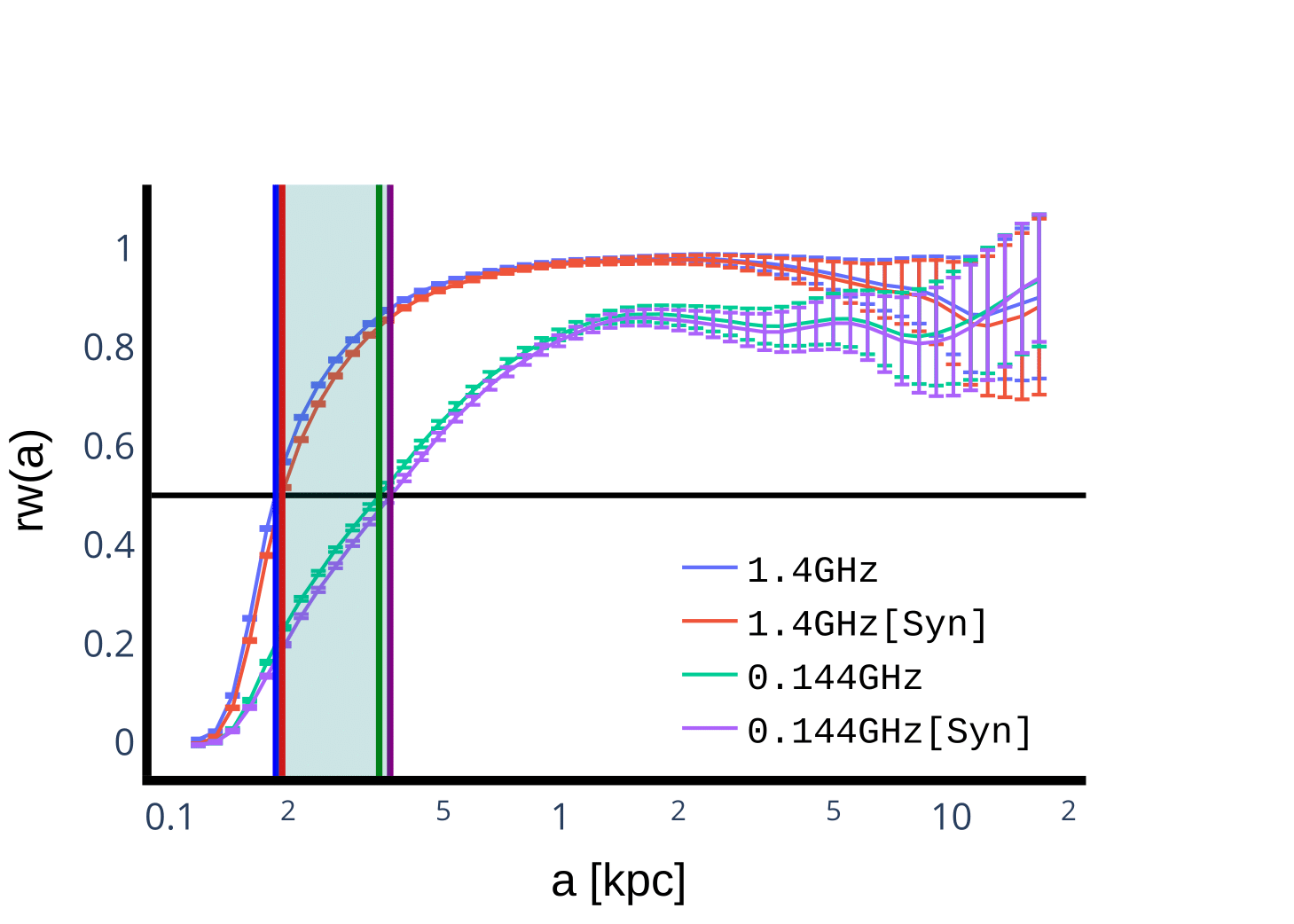}} 
\caption{Multi-scale correlation between the radio continuum emission and the FIR bands at 70 $\mu$m (top), 100 $\mu$m (middle), and 160 $\mu$m (bottom). {At each panel,  horizontal solid line shows $r_w=0.5$ and vertical lines indicate its corresponding scales of occurrence ($l_{\rm break}$) for different radio and FIR pairs.}}
\captionsetup{font=small,labelfont={bf,sf}}
    \label{fig:wavelet} 
\end{figure}

Figure~\ref{fig:wavelet} shows the wavelet correlation coefficients $r_{w}$ vs. scale for each pair of the radio (at 0.14 and 1.4~GHz) and FIR bands at 70, 100, 160\,$\mu$m. As $r_{w}=0.5$ indicates the same number of correlated and uncorrelated structures in two decomposed maps, we consider this coefficient as a marginal condition for the acceptance of a correlation between the structures following \cite{intro12&taba}. At each FIR band shown in Fig.~\ref{fig:wavelet}, a general dropping trend is found for the correlations from large to small scales. Defining the smallest scale of the correlation, i.e., where $r_{w}=0.5$ as the break scale, $l_{\rm break}$, we find that both the observed and the nonthermal radio continuum emission at 1.4\,GHz are correlated with the FIR emission  on scales $\ge~180-200$~pc\,($=l_{\rm break,1.4}$). At the lower radio frequency of 0.14\,GHz, a similar large-to-small-scale dropping trend exists, however, the break scale occurs on scales of $\ge0.28-0.36$~pc ($=l_{\rm break,0.14}$), i.e., larger than those at 1.4~GHz. Therefore, a scale-difference is found for the correlation depending on the radio frequency. Supposedly, lower radio frequencies trace older and more diffused  CREs. Hence, a larger break scale at lower frequencies shows that a fine balance between gas and  CREs /magnetic field is indeed the most fundamental requirement for the radio-FIR correlation to be held \citep{intro12&taba}. It is also instructive to note that  CREs  traced at 0.14~GHz are propagated away on a path length about twice longer than  those traced at 1.4~GHz ($l_{\rm break,0.14}\simeq~1.8~l_{\rm break,1.4}$). These are discussed further in Sect.~\ref{sec:discu}. 
Comparing different FIR bands shown in different panels in Fig.~\ref{fig:wavelet}, we find that the correlations are almost similar at 1.4\,GHz while a larger scatter in $l_{\rm break}$ is found at 0.14~GHz. Additionally, it should be mentioned that the break scales determined with the smoothing method \citep{2021Ap&SS.366..117H,Vollmer} are larger than the values of The radio--FIR break scales given in Table~\ref{tab:Table7}.\footnote{The magnitude of $l_{dif}$ relies on how the break scale is defined. For example due to the strong correlation in \citep{Mulcahy} study, they employed rw = 0.75 to establish the break scale. in this study If we use rw = 0.75, we get average break scales of $\sim 260$ pc for 1.4\,GHz and $\sim 620$ pc for 0.14~GHz.}

\begin{table}[H]
\caption{The radio--FIR break scales ($l_{\rm break}$) in kpc.}
\label{tab:Table7} 
\centering
\begin{tabular}{c c c c} 
\hline\hline
1.4\,GHz & Observed & Synchrotron \\[0.5ex] 
70 $\mu$m & $0.18\,\pm 0.01$ &  $0.20\,\pm 0.01$ \\ 
100 $\mu$m & $0.18\,\pm 0.01$ & $0.19\,\pm 0.01$ \\ 
160 $\mu$m & $0.18\,\pm 0.02$ & $0.19\,\pm 0.02$ \\ 
\hline
0.14\,GHz & Observed & Synchrotron \\
70 $\mu$m & $0.31 \pm 0.01$ &  $0.32 \pm 0.01$ \\ 
100 $\mu$m & $0.28 \pm 0.02$ & $0.29 \pm 0.02$  \\ 
160 $\mu$m & $0.34 \pm 0.02$ & $0.36 \pm 0.02$ \\ 
\hline
\hline
\end{tabular}
\end{table}

\section{Discussion}
\label{sec:discu}
{As follows, we discuss the variations in the radio--FIR correlation in the disk of IC~342 and investigate the role of the propagation of  CREs  in this correlation. Then possible  CREs  cooling mechanisms, diffusion, and their related time scales are investigated and compared with other galaxies.} 

\subsection{Sources of variation in the radio-FIR correlation}
\label{sec:corVSmf}

The radio--FIR correlation assessed via the FIR-to-radio ratio ($q$)  is often used to calibrate the SFR globally in galaxy samples because it does not change systematically with the SFR itself  in star forming galaxies \citep{q2Yun}. However, $q$ can change locally inside galaxies  as shown in a sample of the SINGS galaxies \citep[][]{Murphy_8}, NGC~6946 \citep{intro15&taba}, the LMC \citep{intro11&Hughes}, and also in IC~342 (Sect.~\ref{sec:qmap}).  A first possible reason for a change in $q$ can be different origins of the nonthermal radio and thermal FIR emission in some locations in a galaxy.  For instance, we can refer to diffuse parts of the ISM from where the FIR emission can be due to dust heated by non-ionizing photons from old solar-mass stars while the synchrotron radiation is generated by secondary  CREs  propagated away in the ISM.  In this case, no radio--FIR correlation is in fact expected. Accordingly, we can explain the maximum dispersion from the mean value of $q$ ($<2$) in IC~342 as it occurs in R2 where indeed no correlation holds (Sec.~\ref{sec:pix}). 
On the other hand, smaller variations are even found in the star forming regime  R1: Although the correlation holds tightly, $q$ still changes locally there.  In this case, a common source, e.g., massive stars, can be responsible for both the nonthermal radio and the thermal FIR  emission but following different dependencies.  This is further investigated by focusing on the R1 regime. 
\begin{figure}[htbp]
    \centering    
{\includegraphics[width=9cm]{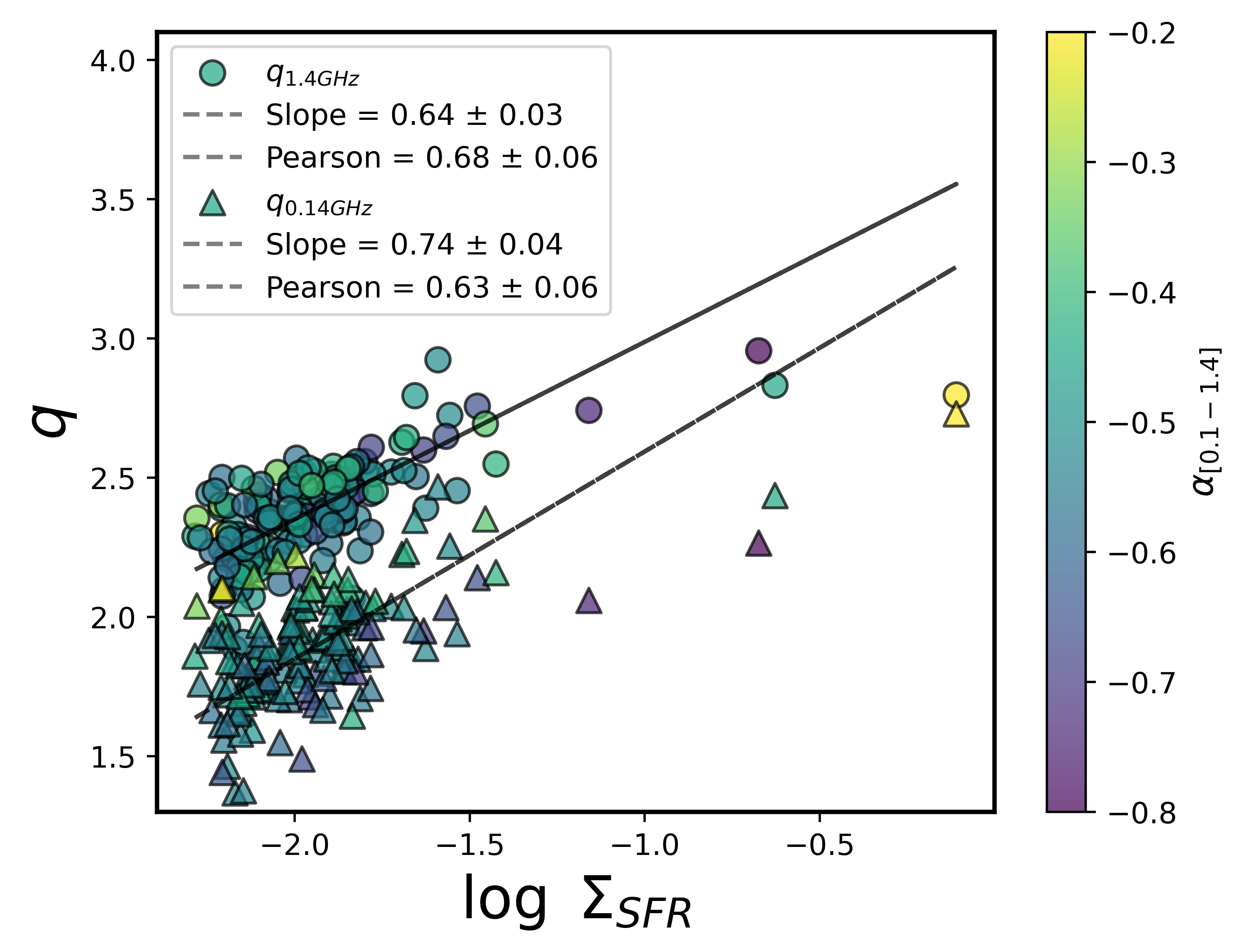}} 
\caption{The logarithmic ratio of the 70$\mu$m to the radio synchrotron emission  at 1.4\,GHz ($q_{1.4}$, circles) and 0.14\,GHz ($q_{0.14}$, triangles) versus  star formation surface density $\Sigma _{\rm SFR}$ ($\rm M_{\odot}\,\text{yr}^{-1}\,\text{kpc}^{-2}$) in regime R1. Solid lines show bisector OLS fits. The color bar shows the spectral index between 0.14 and 1.4\,GHz $\alpha_{[0.1-1.4]}$.}
    \label{fig:radio-q}
\end{figure}
%

\begin{figure}[htbp]
    \centering    
        {\includegraphics[width=8.0cm]{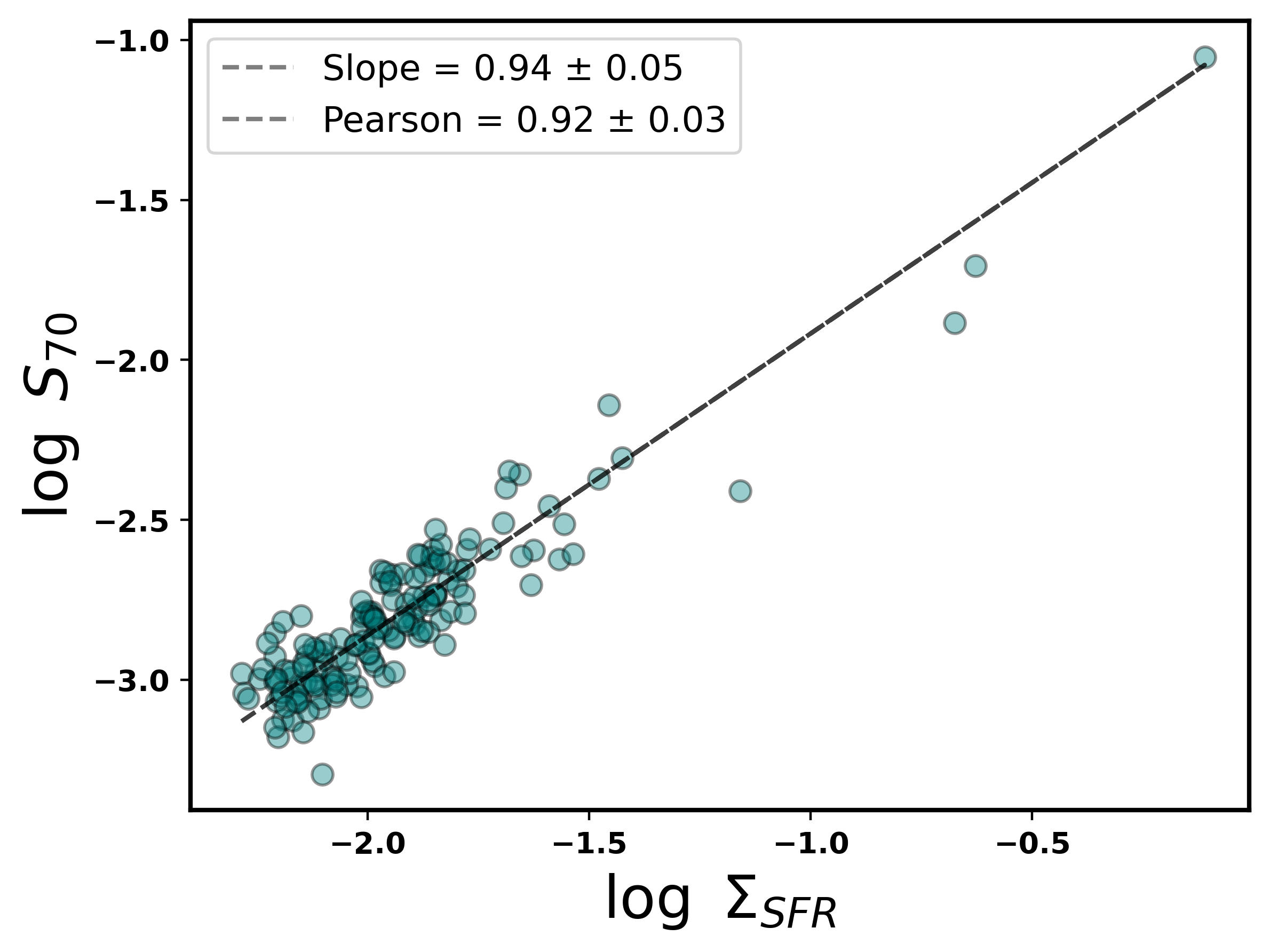}} 
     {\includegraphics[width=8.0cm]{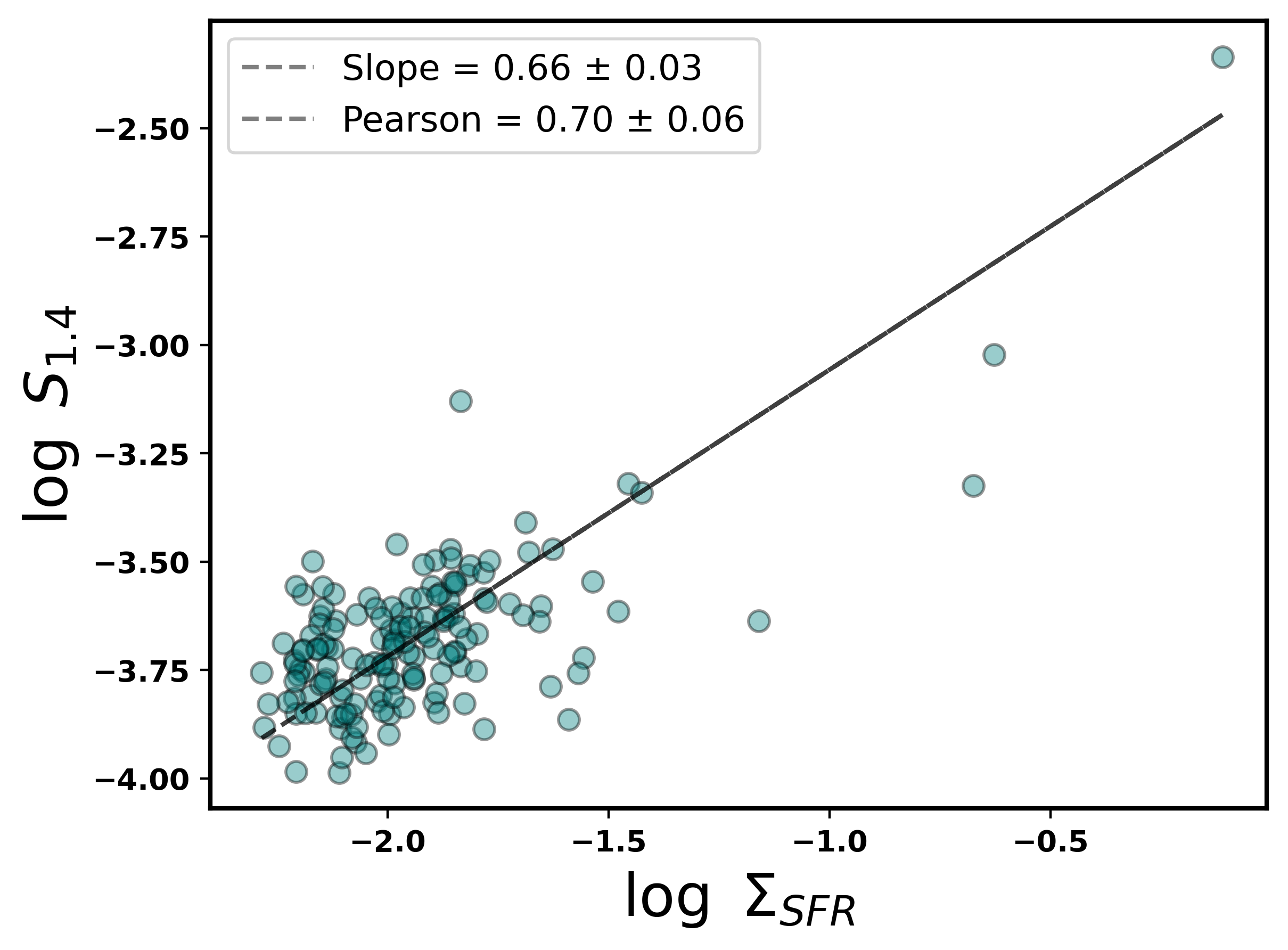}} 
 \caption{The FIR emission at 70$\mu$m ({\it top}) and synchrotron emission at 1.4\,GHz ({\it bottom})  vs star formation rate surface density $\Sigma _{\rm SFR}$ ($\rm M_{\odot}\,\text{yr}^{-1}\,\text{kpc}^{-2}$) in the star forming regime (R1). Solid lines show bisector OLS fits.}
  \label{fig:vssfr}
\end{figure}

We first derive the star formation surface density $(\Sigma_{\rm SFR})$  using a combination of the FUV and mid-infrared (MIR) emission presented by \citep{Bigiel_2008&SFR} as a hybrid tracer,
\begin{multline}
\label{eq:SFR}
\left(\frac{\Sigma_{\rm SFR}}{\rm M_{\odot}\,\text{yr}^{-1}\,\text{kpc}^{-2}}\right) =8.1 \times 10^{-2} \left(\frac{I_{\rm FUV}}{\rm MJy\,\text{sr}^{-1}}\right) + 3.2 \times 10^{-3} \left(\frac{I_{\rm MIR}}{\rm MJy\,\text{sr}^{-1}}\right) 
\end{multline}
In the above calibration relation, the intensity of the MIR (22$\mu$m) emission is used as a proxy for the extinction of the FUV emission.
%
%
%
In Fig. \ref{fig:radio-q}, the FIR-to-synchrotron ratio at 1.4\,GHz ($q_{1.4}$) and 0.14\,GHz ($q_{0.14}$) is plotted against $\Sigma_{\rm SFR}$ taking into account only pixels above $3\sigma$ rms noise level. With a Pearson correlation coefficient of $\simeq$ 0.7 (0.6), $q$ increases with $\Sigma_{\rm SFR}$ as
\begin{equation}
q_{1.4} \propto (0.64 \pm 0.03) \log \Sigma _{\rm SFR},
\end{equation}
 and
\begin{equation}
q_{0.14} \propto (0.74 \pm 0.08)  \log \Sigma _{\rm SFR},
\end{equation}
at 1.4\,GHz and 0.14\,GHz,  respectively. A tight positive correlation was also found in NGC~6946 although with a much flatter slope of $\simeq$ 0.1  \citep{intro15&taba}. Assuming that the FIR emission (here the 70$\mu$m emission) is linearly correlated with the SFR, this increase must be due to a sub-linear correlation of the synchrotron emission and SFR. As shown in Fig.~\ref{fig:vssfr}, this is exactly what occurs in the star-forming ISM (R1) of IC~342. This means that the synchrotron emission is not produced as efficiently as the FIR emission with SFR. However, as shown in Fig.~\ref{fig:magnet}, it is expected that the magnetic field becomes stronger in star-forming regions which should result in the synchrotron cooling of  CREs  via the synchrotron radiation. The non-efficient synchrotron emission observed in the star forming regions can have two main reasons: 1) a non-efficient amplification of the magnetic field and 2) decoupling and escape of the  CREs  via winds in these regions \citep{intro14&taba}. These are addressed in Sects.~\ref{sec:d-mag}~and~\ref{sec:enrgySPEC}.


\subsection{Magnetic field and star formation}
\label{sec:d-mag}
\begin{figure}[htbp]
    \centering    
        {\includegraphics[width=8.5cm]{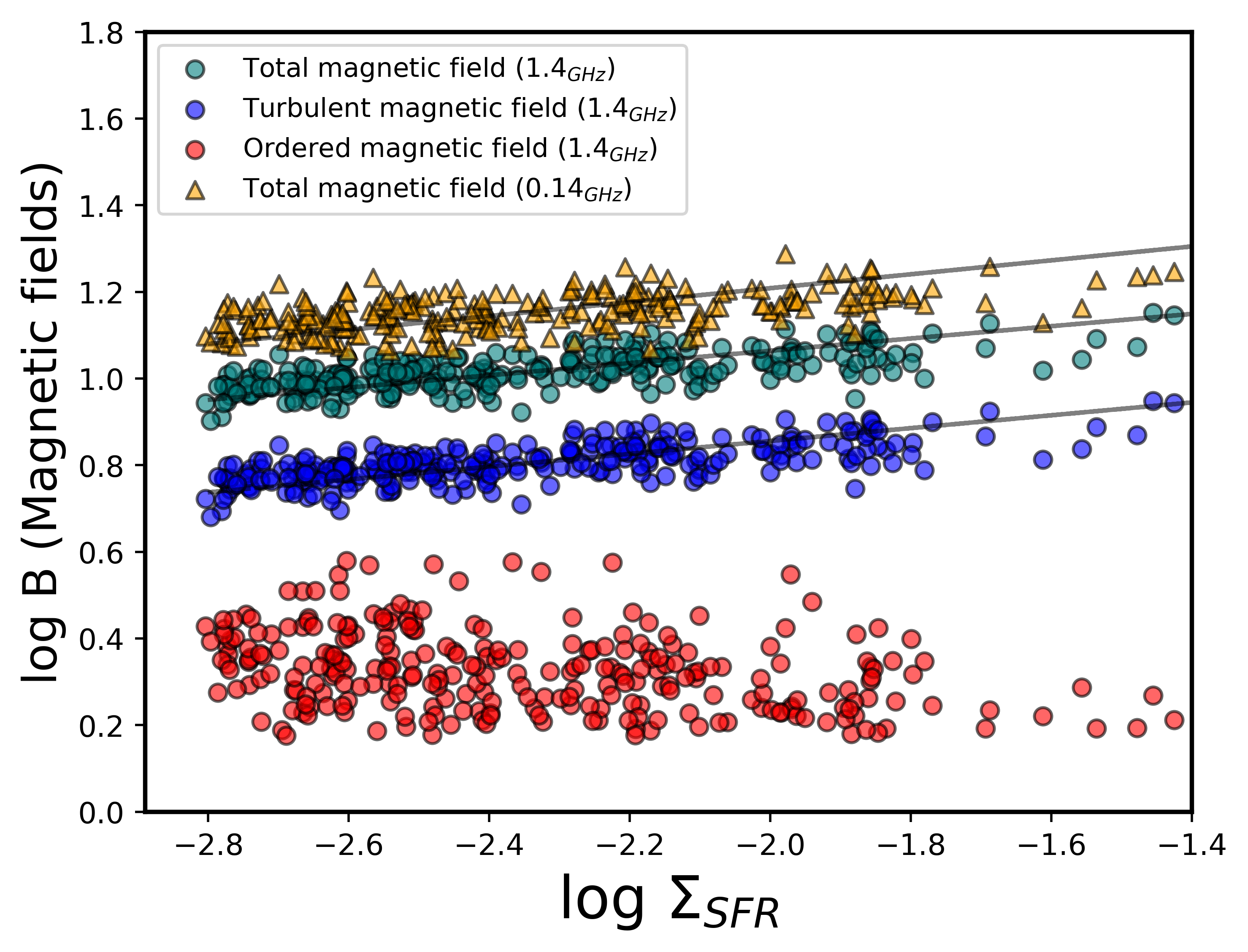}} 
    
     \caption{Total, turbulent and ordered magnetic field strength obtained using the synchrotron emission at 1.4\,GHz (circles) and 0.14\,GHz (triangles) versus star formation rate surface density $\Sigma _{\rm SFR}$ ($\rm M_{\odot}\,\text{yr}^{-1}\,\text{kpc}^{-2}$). Solid lines show bisector OLS fits. { For a better illustration, the turbulent magnetic field strength is shifted by 0.2 along the y-axis.}}
  \label{fig:sfrB}
\end{figure}

{As mentioned in Sect.~\ref{sec:corVSmf}, the magnetic field is expected to be amplified due to star formation activities. One of the most important mechanisms often considered is supernova-driven turbulence \citep{Gressel}. The energy density of the turbulent gas motions is then converted to the magnetic energy density by a small-scale dynamo mechanism. This will lead to a power-law relation between the magnetic field strength and star formation rate surface density with an index of 0.3 ($B \propto \Sigma_{\rm SFR}^{0.3}$) assuming equipartition between the magnetic field and cosmic rays \citep{Schleicher}. 
We investigate whether such a correlation holds in  IC~342.  The equipartition magnetic field strength obtained using the 1.4\,GHz synchrotron emission (Sect.~\ref{sec:magnet}), shows a modest correlation with $\Sigma _{\rm SFR}$ with a Pearson correlation coefficient of $\simeq 0.55$ for both total and turbulent fields (Fig.~\ref{fig:sfrB}). For the entire galaxy, the correlation between the total magnetic field strength and $\Sigma _{\rm SFR}$ can best be explained by the following power-law relationship obtained using OLS bisector regression: 
\begin{equation}\label{b-sfr}
B_{\rm tot} \propto \Sigma _{\rm SFR}^{(0.14 \pm 0.01)}.
 \end{equation}  
The same relation is obtained between the turbulent field and star formation rate surface density. The ordered field shows a decreasing trend with $\Sigma _{\rm SFR}$ but with a large scatter ($r_p=-0.3$). This is expected because the field becomes more turbulent or tangled in the star-forming regions. Table~\ref{tab:magnetic_field_data} summarizes $r_p$ and the fit power-law exponents obtained. 

\begin{table}[!htb]
    \centering
    \caption{Correlation coefficients and power-law exponents for magnetic field strengths.}
    \label{tab:magnetic_field_data}
    \begin{tabular}{ccccc}
        \toprule
        Y & X & $b$ & $r_p$ & {n} \\
        \midrule
        \textbf{$B_{\rm tot}$} & $\Sigma_{\rm SFR}$ & $0.14 \pm 0.01$ & $0.62 \pm 0.04$ & 274 \\
        \textbf{$B_{\rm tur}$} & $\Sigma_{\rm SFR}$ & $0.14 \pm 0.01$ & $0.64 \pm 0.04$ & 274 \\
        \textbf{$B_{\rm ord}$} & $\Sigma_{\rm SFR}$ & $-0.42 \pm 0.02$ & $-0.30 \pm 0.02 $ & 274 \\
        \bottomrule
    \end{tabular}
    \begin{tablenotes}
    \footnotesize
    \item\tablefoot{Pearson correlation coefficients ($r_p$) and the fitted power-law exponent $b$ are listed for the total ($B_{\rm tot}$), turbulent ($B_{\rm tur}$), and ordered ($B_{\rm ord}$) magnetic field strength against the star formation rate surface density ($\Sigma_{\rm SFR}$). n indicates the number of independent data points.}
    \end{tablenotes}
\end{table}

 
As noted by \cite{Sep2&Hamid} and \cite{intro14&taba}, the slope of the $B_{\rm tot}-\Sigma_{\rm SFR}$ relation can be flat due to a contamination by diffuse ISM. Focusing on the star forming regime R1, we obtain
\begin{equation}
B_{\rm tot} \propto \Sigma _{\rm SFR}^{(0.18 \pm 0.01)},
 \end{equation}  
which is slightly steeper than that given in Eq. (\ref{b-sfr}). 
We note that the same correlation and power-low exponent is obtained if the low-frequency synchrotron emission at 0.14\,GHz to estimate the magnetic field strength. 

%
%
Compared with similar studies in other nearby galaxies, the B--SFR correlations are generally weaker in IC~342. The exponents obtained for the entire galaxy agree with those derived in NGC\,6946 \citep[$0.14 \pm 0.01$,][]{intro15&taba} and NGC~4254 \citep[$0.18 \pm 0.01$,][]{chyzy08} within uncertainties but flatter than those in the LMC and SMC \citep[$\simeq 0.25-0.30$, ][ which is closer to the theoretical small-scale dynamo value]{Sep2&Hamid} perhaps due to intervening amplification mechanisms such as gas compression in the ISM of spiral galaxies which is denser than that in dwarf irregulars.  }


\subsection{Effect of star formation and magnetic field on energy spectrum of  CREs }
\label{sec:enrgySPEC}

	Energy of  CREs  follows a power-law distribution of the form $E_{\nu}\propto \nu^{p}$, with $p=(2\alpha -1)$ the energy index and $\alpha$ the radiation spectral index ($S_{\nu}\propto \nu^{\alpha}$). Theoretically,  CREs  are injected with a flat power-law index of $p=-2$ or a spectral index of $\alpha=-0.5$ from their sources in star-forming regions (e.g., SNRs). As they propagate away, they can loose their energy as a result of interaction with matter (via ionization/adiabatic losses), magnetic fields (via synchrotron loss), and radiation (via inverse Compton loss).  The spectral index maps presented in Figs.~\ref{fig:spect} can hence be used to infer the locations in the galaxy where  CREs  gain or lose energy and further study the role of the SFR and magnetic fields.

\begin{table}[!htb]
    \centering
    \caption{Correlation coefficients and power-law exponents for spectral indices.}
    \label{tab:SI}
    \begin{tabular}{ccccc}
        \toprule
        Y & X & $b$ & $r_p$ & {n} \\
        \midrule
        \textbf{$\alpha_{[0.1-1.4]}$} & $\Sigma_{\rm SFR}$ & ... & $0.21 \pm 0.02$ & 388 \\
        \textbf{$\alpha_{[0.1-4.8]}$} & $\Sigma_{\rm SFR}$ & $0.31 \pm 0.01$ & $0.45 \pm 0.02$ & 412 \\
        \textbf{$\alpha_{[1.4-4.8]}$} & $\Sigma_{\rm SFR}$ & $0.58 \pm 0.01$ & $0.47 \pm 0.02 $ & 412 \\
        \bottomrule        
        \textbf{$\alpha_{[0.1-1.4]}$} & ${B_{\rm tot}}$ & ... & $0.11 \pm 0.02$ & 391 \\
        \textbf{$\alpha_{[0.1-4.8]}$} & ${B_{\rm tot} }$ & ... & $0.30 \pm 0.02$ & 386 \\
        \textbf{$\alpha_{[1.4-4.8]}$} & ${B_{\rm tot} }$ & ... & $0.29 \pm 0.02 $ & 417 \\
        \bottomrule
    \end{tabular}
    \begin{tablenotes}
    \footnotesize
    \item \tablefoot{Pearson correlation coefficients ($r_p$) and the fitted power-law exponent $b$ are listed for the spectral indices $\alpha_{[0.1-1.4]}$, $\alpha_{[0.1-4.8]}$, and $\alpha_{[1.4-4.8]}$ against the star formation rate surface density ($\Sigma_{\rm SFR}$) and the total magnetic field strength ($B_{\rm tot}$). n represents the number of independent data points. Curves are fitted only in cases of reasonable correlation ($r_p > 0.30$).}
    \end{tablenotes}
\end{table}

\begin{figure}[htbp]
    \centering
    {\includegraphics[width=8cm]{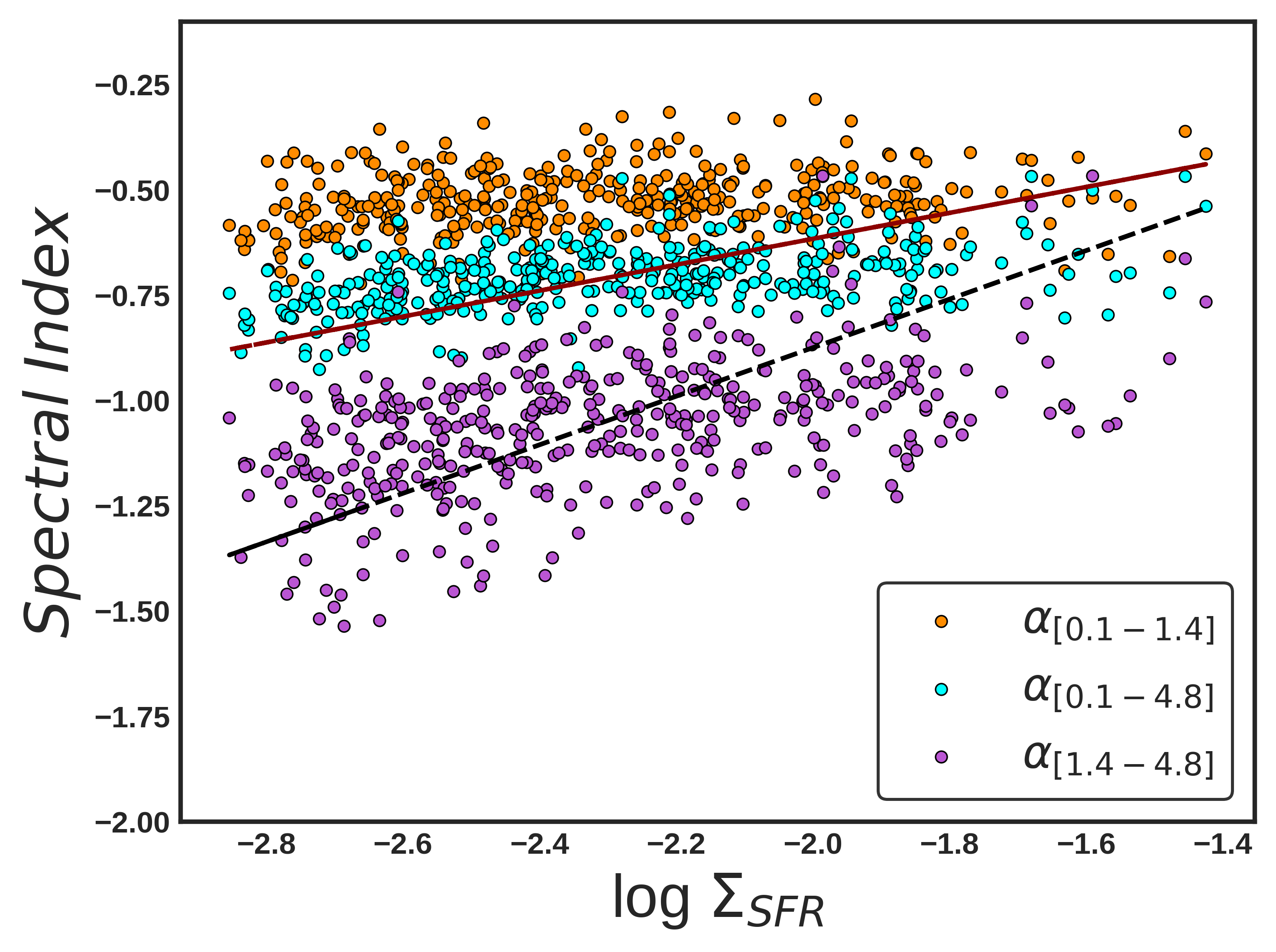}} 
     {\includegraphics[width=8cm]{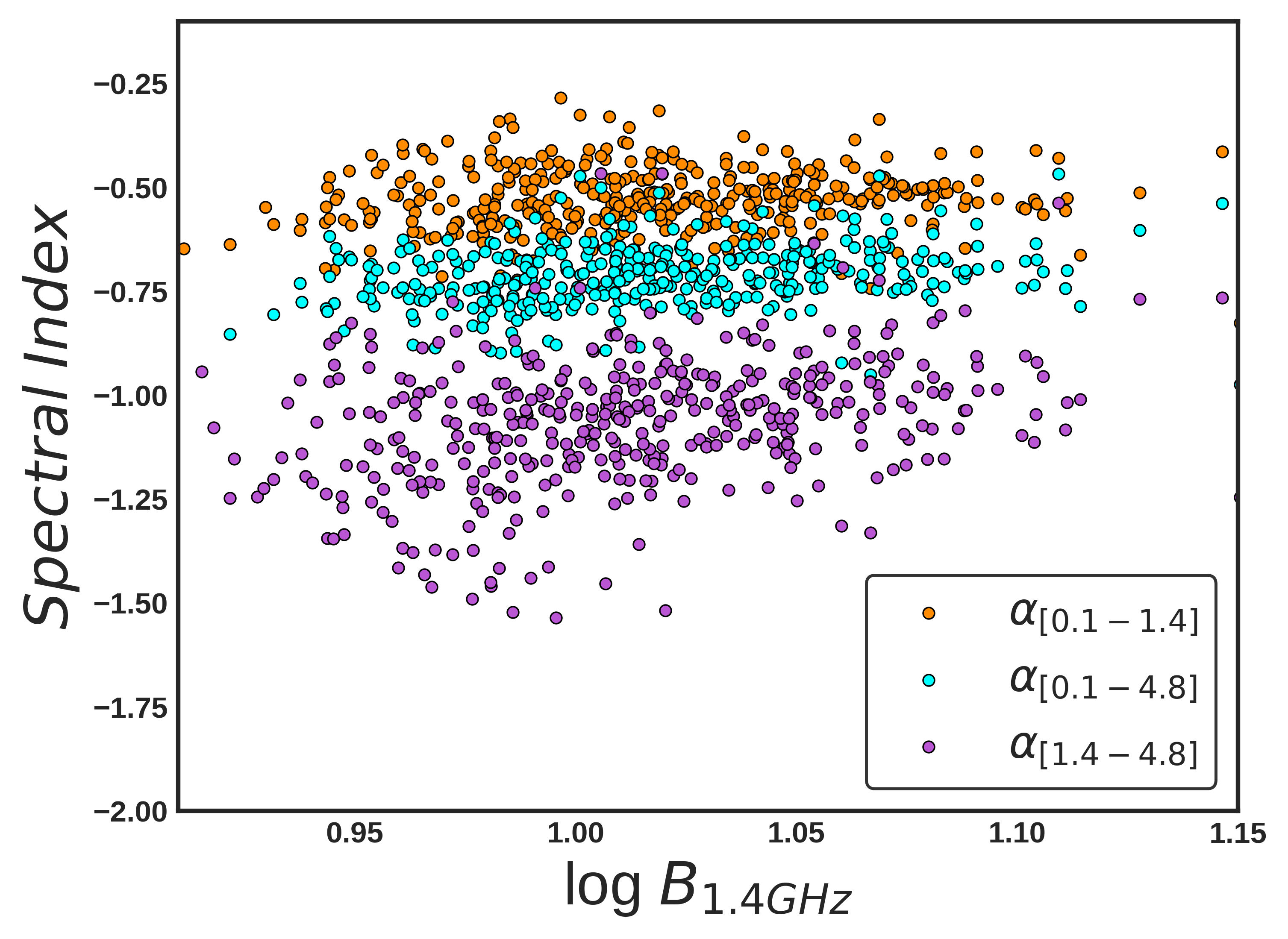}}
     {\includegraphics[width=8cm]{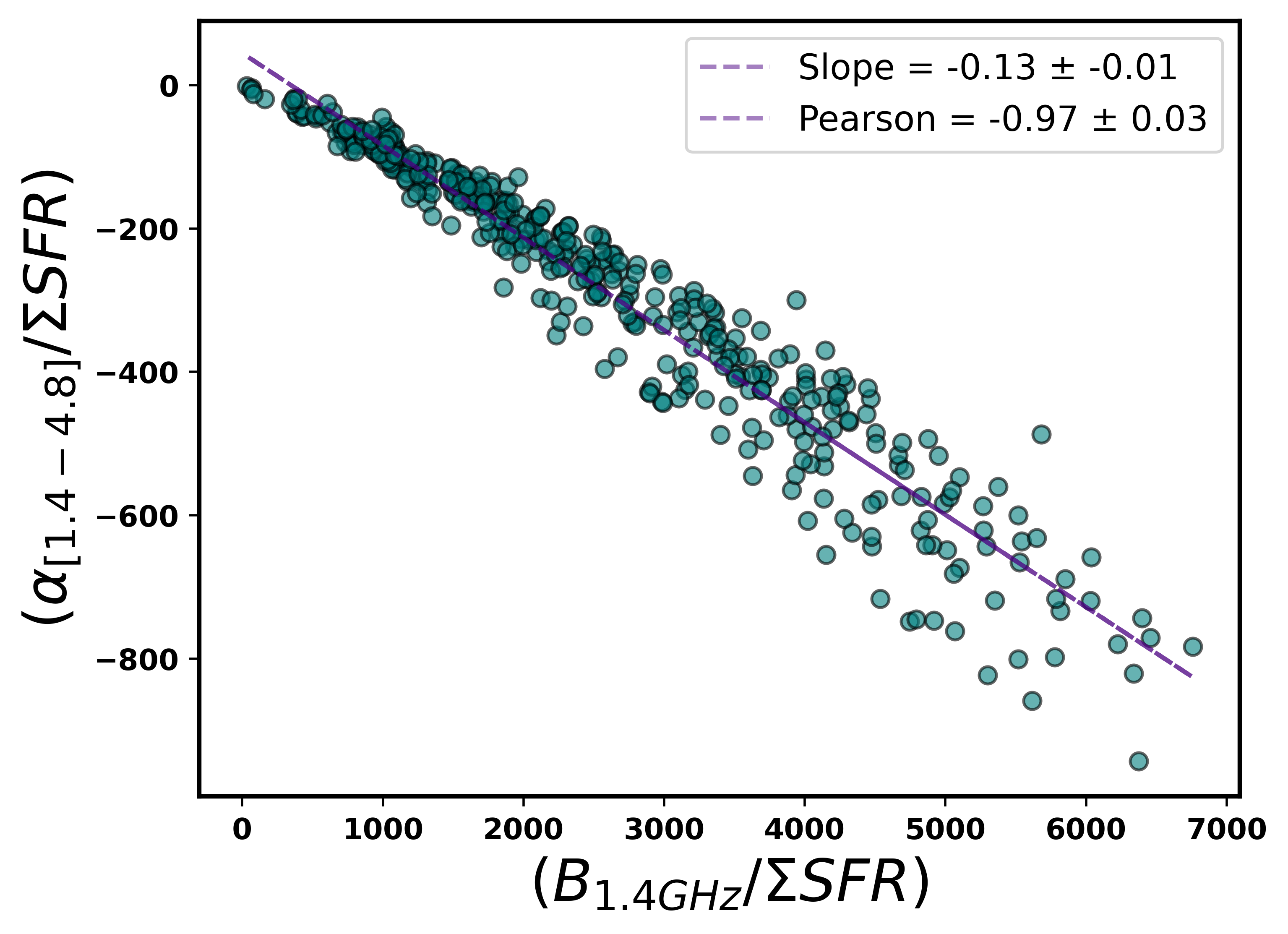}}
\caption{{\it Top:} Synchrotron spectral index obtained at different frequency intervals ($\alpha_{[0.1-1.4]}$,~$\alpha_{[1.4-4.8]}$ and $\alpha_{[0.1-4.8]}$) versus star formation surface density $\Sigma _{\rm SFR}$ ($\rm M_{\odot}\,\text{yr}^{-1}\,\text{kpc}^{-2}$). Lines show bisector OLS fits.
{\it Middle:} Same spectral indices vs the total magnetic field strength. {\it Bottom:} Synchrotron spectral index $\alpha_{[1.4-4.8]}$ versus  magnetic field after compensating both for the effect of the SFR.}
  \label{fig:sfr-allspec}
\end{figure}

Theoretically, the energy spectrum of cosmic rays can be affected by both massive star formation activity (a flattening or energy gain) and magnetic fields (a steepening or energy loss) in an opposite way. Observations in other galaxies already confirm the expected trend with $\Sigma_{\rm SFR}$, i.e., flattening of the spectral index. However, in spite of a steeper $\alpha$ along the ordered magnetic field observed in NGC~6946 \citep{intro15&taba}, no clear trend  with the total magnetic field strength is reported in the literature.  In fact, separating the two effects together can be difficult if the magnetic field is also amplified/generated in star-forming regions which is often the case \citep{energyspec1&taba}. These are re-visited in IC~342. Figure~ \ref{fig:sfr-allspec} shows the spectral index vs $\Sigma_{\rm SFR}$ and $B_{\rm tot}$. {Table~\ref{tab:SI} lists the corresponding correlation coefficients $r_p$ and the fitted power-law exponents for the case of a reasonable correlation ($r_p>0.30$).}
Remarkably, the analysis reveals a distinct correlation between $\Sigma_{\rm SFR}$ and the spectral index {measured between 4.8\,GHz and the other two frequencies, $\alpha_{[1.4-4.8]}$ and $\alpha_{[0.1-4.8]}$}. This correlation suggests a potential influence of massive star formation activity on the energy spectrum of cosmic rays in IC~342, akin to observations in other galaxies   \citep[e.g.,][]{intro15&taba,intro14&taba}. The flattening of the spectral index with increasing $\Sigma_{\rm SFR}$ underscores the significance of star-forming regions in shaping the {spectrum of CREs traced at higher frequencies. In the low-frequency domain (between 0.1 and 1.4\,GHz), the synchrotron spectrum is flat independently of the level of star formation and therefore the correlation of $\alpha_{[0.1-1.4]}$ with $\Sigma_{\rm SFR}$ is weak ($r_p=0.21\pm0.02$). The flattening of the spectrum at low frequencies can be caused by absorption effects in  cool ionized gas \citep{Isr,Isreal} or  efficient ionization loss of CREs \citep{Hummel}. 
}

\begin{figure}[]
        \subfigure[]{\includegraphics[width=8cm]{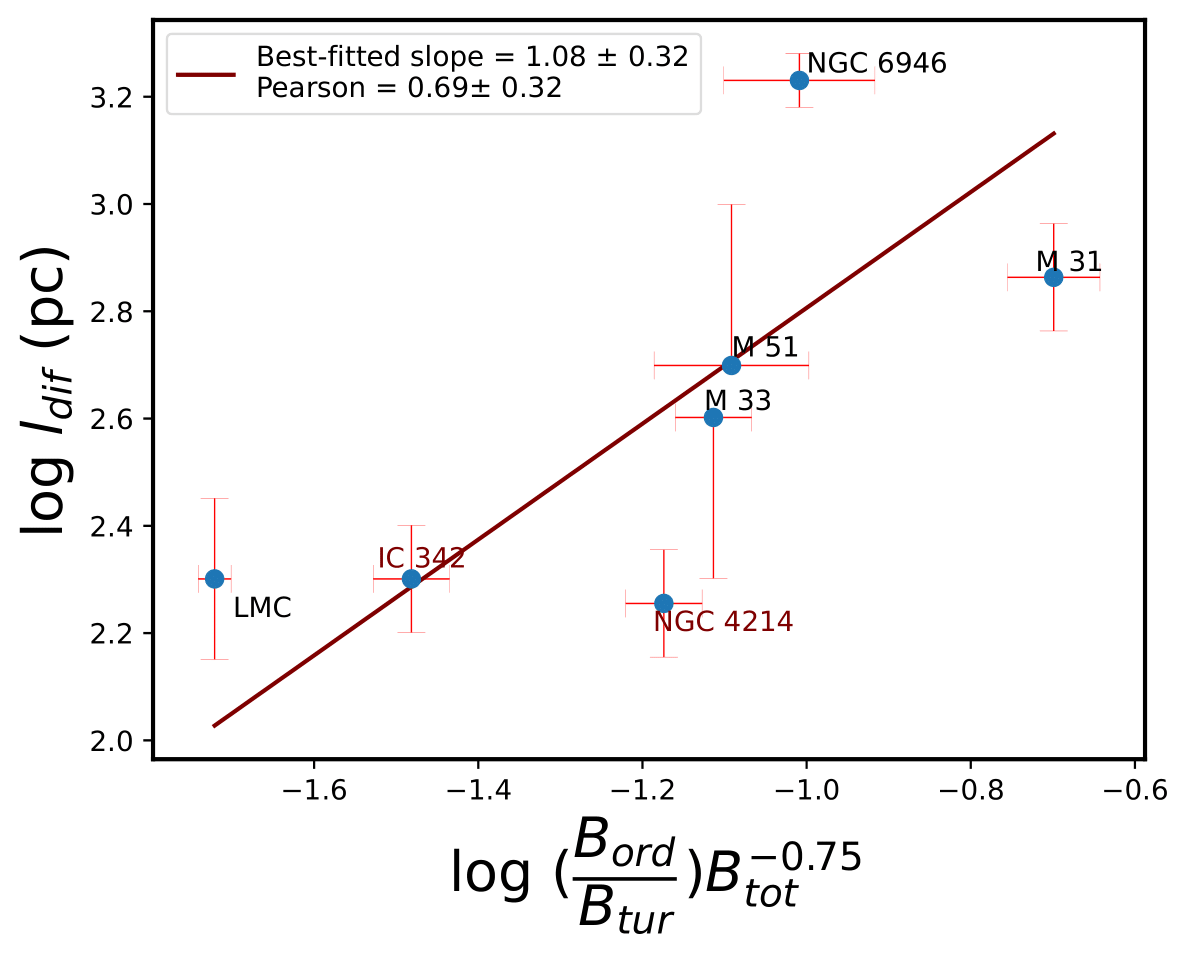}}  
    \subfigure[]{\includegraphics[width=8cm]{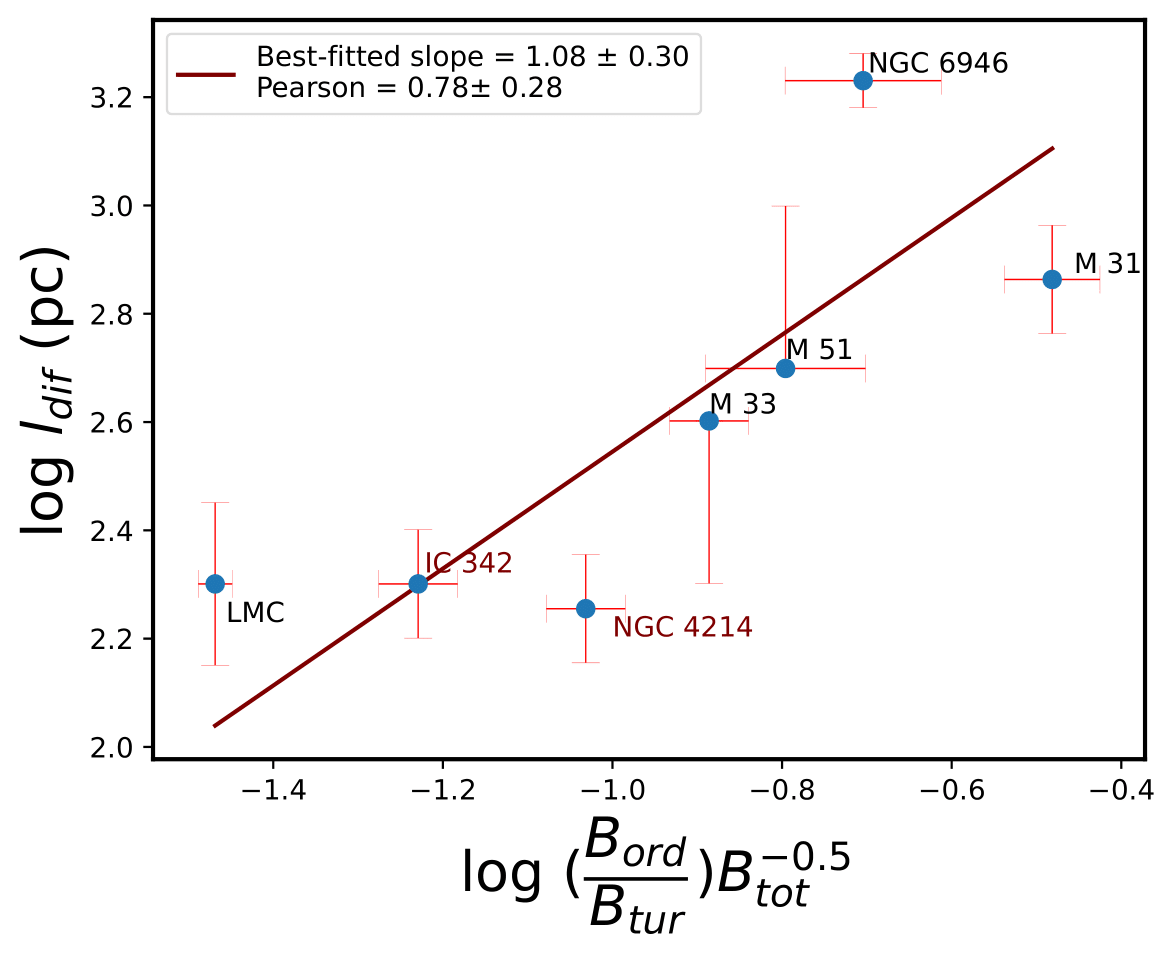}}  
\subfigure[]{\includegraphics[width=8cm]{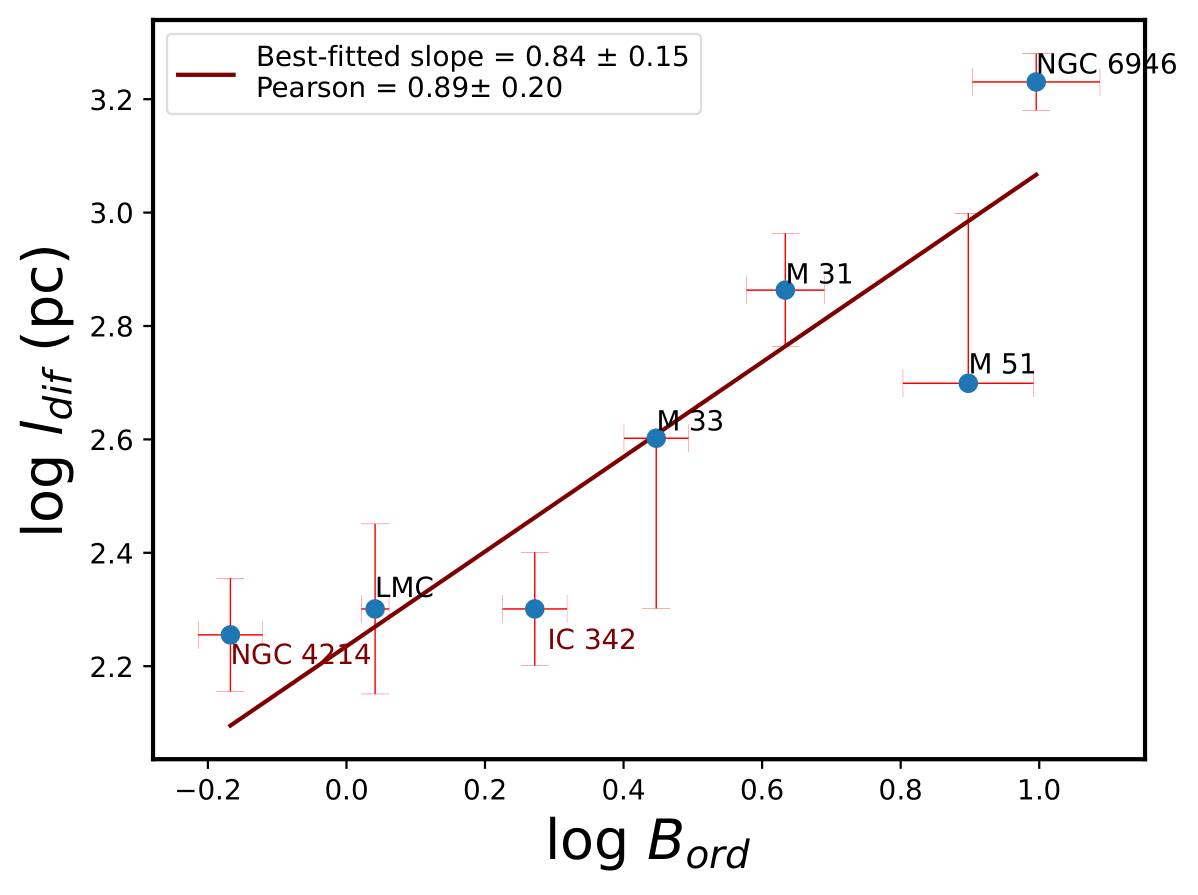}}
\caption{The diffusion length of  CREs  $l_{dif}$ obtained from observations vs that expected from the theoretical diffusion model for the case of  $t_{\rm CRE}=t_{\rm syn}$ ({\it a, Eq.~\ref{Eq. 15}}), and $t_{\rm CRE}=t_{\rm esc}$ ({\it b, Eq.~\ref{Eq. 17}}). Also shown is the correlation of $l_{dif}$ with the ordered magnetic field ({\it c}).}

  \label{fig:ldif}
\end{figure}

{Unlike with $\Sigma_{\rm SFR}$, the correlations} with $B_{\rm tot}$ is less straightforward. Contrary to expectations based on the effects of magnetic fields, there is no discernible correlation between $B_{\rm tot}$ and the spectral index across the observed data points ($r_p \leq 0.3$).
%
%
The increase of both $\alpha$ and $B_{\rm tot}$ in star-forming regions can reduce the expected {steepening of the synchrotron spectrum with increasing magnetic field strength (here, an $\alpha-B_{\rm tot}$ anti-correlation).  To compensate for the effect of star formation}, we investigate their correlation after normalizing them to $\Sigma_{\rm SFR}$. The bottom panel in Fig.~\ref{fig:sfr-allspec}, shows that per unit  $\Sigma_{\rm SFR}$, a tight linear anti-correlation holds between the spectral index and total magnetic field strength.  Hence, our observations in IC~342 confirms the theoretical expectation for synchrotron cooling of  CREs  based on the steepening of the synchrotron spectrum per unit $\Sigma_{\rm SFR}$.

\setlength{\tabcolsep}{20pt} 
\renewcommand{\arraystretch}{1.2} 
\begin{table*}[!htb] 
    \caption{Magnetic field strengths and star formation rate surface density for nearby galaxies.}
    \label{tab:Table8}
    \small\addtolength{\tabcolsep}{-15.5pt}
    \centering
    \begin{tabular}{c c c c c c c c c c} 
    \hline\hline
    Galaxy & $B_{\rm tot}$ & $B_{\rm ord}$ & $B_{\rm tur}$ & $B_{\rm ord}/B_{\rm tur}$ & $\Sigma _{\rm SFR}$ & $l_{\rm dif}$  \\
    & ($\mu$G) & ($\mu$G) & ($\mu$G) & --- & (M$_{\odot}$ Gyr$^{-1}$pc$^{-2}$) & (pc) \\
    \hline
    IC~342 & $10.2 \pm 1.3$ & $1.87 \pm 0.65$ & $10.0 \pm 1.3$ & $0.19 \pm 0.07$ & $3.5 \pm 0.3$ & $200 \pm 10$ \\
    NGC~4214 & $3.7 \pm 0.5$ & $0.68 \pm 0.27$ & $3.6 \pm 0.5$ & $0.19 \pm 0.08$ & $4.9~^{9} \pm 1.5$ & 180 \\
    M~31~$^{1}$ & $6.6 \pm 0.3$ & $4.3 \pm 0.3$ & $5.0 \pm 0.2$ & $0.86 \pm 0.07$ & $0.6 \pm 0.1$ & $730 \pm 90$ \\
    M~33~$^{1}$ & $8.1 \pm 0.5$ & $2.8 \pm 0.3$ & $7.6 \pm 0.5$ & $0.37 \pm 0.05$ & $3.0 \pm 0.6$ & $< 400$ \\
    NGC~6946~$^{2}$ & $16.0 \pm 1.5$ & $9.9 \pm 2.1$ & $12.6 \pm 2.7$ & $0.79 \pm 0.24$ & $7.2 \pm 1.8$ & $1700 \pm 200$ \\
    LMC & $10.1 \pm 1.7~^{3}$ & $1.1~^{3}$ & $10.0 \pm 1.7~^{3}$ & $0.11 \pm 0.01$ & $2.6~^{4}$ & $200 \pm 90~^{5}$ \\
    M 51 & $15 \pm 2.2~^{6}$ & $7.9 \pm 1.1~^{6}$ & $12.7 \pm 1.9~^{6}$ & $0.62 \pm 0.26~^{6}$ & $12~^{8} \pm 1.8$ & $> 500 \pm 200~^{8}$ \\
    \hline
    \end{tabular}
    \begin{tablenotes}
    \footnotesize
    \item\tablefoot{Total ($B_{\rm tot}$), ordered ($B_{\rm ord}$), and turbulent ($B_{\rm tur}$) magnetic field strengths are presented for a sample of nearby galaxies, along with the star formation rate surface density ($\Sigma_{\rm SFR}$) and the diffusion length of cosmic ray electrons ($l_{\rm dif}$.)}
    \tablebib{ $^{1}$\cite{intro12&taba}, $^{2}$\cite{intro15&taba}, $^{3}$\cite{Sep2&Hamid}, $^{4}$\cite{Whitney_LMCSFR}, $^{5}$\cite{intro11&Hughes}, $^{6}$\cite{M51mag}, $^{7}$\cite{Leroy_Ldif}, $^{8}$\cite{intro13&Dumas}, $^{9}$\cite{Calzetti_2010}.}
    \end{tablenotes}
\end{table*}



\subsection{Propagation of  CREs}


{The degree of order of the magnetic field is  believed to play a significant role in theoretical models of the propagation of  CREs, i.e., the streaming instability models \citep{CRE1&Kulsrud,CRE2&Ensslin} and diffusion models~\citep[]{CRE3&Ptuskin,CRE4&Breit,CRE5&Dogiel,CRE6&Shlachi}. Beside streaming and diffusion, escape of  CREs can affect their propagation as discussed by e.g.,  \cite{cre1&Helou} and \cite{2021Ap&SS.366..117H}. }
%
%

As a result of streaming,  CREs move along the lines of the ordered field at the Alfv\'en velocity ($\upsilon_{A}$) and propagate over a distance of $l_{prop}~=~\upsilon_{A}~t_{\rm CRE}$ during their lifetime ($t_{\rm CRE}$). Basically, streaming requires long stretches
of the ordered field to be effective, hence, it may not be
important in the turbulent thin disk of galaxies. 
%
On the other hand, {studies in the Milky Way show that the propagation of CREs} is isotropic on scales $< 1$ kpc \citep[]{CRE7&Strong,CRE8&Tsap}, which is not in accordance with streaming. Accordingly, it is assumed that diffusion dominates streaming, except in regions with highly ordered fields, where streaming may dominate.

{Diffusion occurs when irregularities in the turbulent magnetic field scatter the particles. The diffusion length is theoretically defined as $l_\mathrm{dif} = 2\,( D\,t_\mathrm{CRe})^{0.5}$, with $D$ the diffusion coefficient and $t_\mathrm{CRe}$ the life time of CREs. 
%
%
The diffusion of CREs in the ISM has been described as
\begin{equation}\label{Eq. 14}
D \propto (B_{\rm ord}/B_{\rm tur})^{2}
\end{equation}
by \cite{CRE3&Ptuskin}, \cite{CRE4&Breit}, \cite{CRE9&Yun}, \cite{CRE5&Dogiel}, and \cite{CRE6&Shlachi}. Assuming that the lifetime of CREs is limited by the synchrotron loss during the synchrotron timescale, $t_\mathrm{\rm CRE} = t_\mathrm{\rm syn} \propto B_\mathrm{\rm tot}^{-2}E^{-1} \propto B_\mathrm{\rm tot}^{-3/2} \nu^{-1/2} $, we can drive CREs diffusion length as
\begin{equation}\label{Eq. 15}
l_{\rm dif} \propto (B_{\rm ord}/B_{\rm tur}) B_{\rm tot}^{-3/4}.
\end{equation}
%


On the other hand,  taking into account the escape of CREs, the diffusion length of CREs can deviate from that given by Eq.~\ref{Eq. 15}. This is because the lifetime of CREs is not primarily determined by the synchrotron loss, but rather by their escape from the disk ($t_{\rm CRE}=t_{\rm esc}$). Following  \cite{2018MNRAS.476..158H}, the escape time inversely correlates with the escape velocity ($t_{\rm esc} \propto {\rm v}_{esc}^{-1}$), which in turn increases with the star formation rate and, consequently, the total magnetic field strength, ${\rm v}_{esc} \propto$ SFR$^{0.35}$ approximately $\propto B_{tot}$. In this case, the diffusion length is given by

\begin{equation}\label{Eq. 17}
l_{\rm dif} \propto (B_{\rm ord}/B_{\rm tur}) B_{\rm tot}^{-1/2},
\end{equation}
 which is larger than that given by Eq.(\ref{Eq. 15}).

\cite{intro12&taba} proposed an observational method to estimate the  propagation length of CREs based on the break in the synchrotron--FIR correlation. This was further strengthened by an observed increase of the break scale with both the ordered field and the degree of order of the magnetic field i.e., the ordered-to-turbulent magnetic field, $B_{\rm ord}/B_{\rm tur}$, as expected from the theoretical models. As follows, we first derive $l_{\rm dif}$ and $D$ using this observational method in IC~342. Then, we compare the theoretical models with the observations in a sample of nearby galaxies which is now larger than that presented in \cite{intro12&taba}. The results displaying the correlations based on the available data can be found in Fig. \ref{fig:ldif}-a and Fig. \ref{fig:ldif}-b}. 




%
%

\subsubsection{Diffusion length and coefficient of CREs in IC~342}

{
As discussed in Sect.~\ref{sec:wavelet}, the radio--FIR correlation to be held requires a fine pressure balance between CREs/magnetic fields and gas. Diffusion of CREs to large scales reduces their number density and violates the pressure balance on a scale that is equivalent to the diffusion length of CREs. Hence, the break scales in the synchrotron--FIR correlations obtained in Sect.\ref{sec:wavelet}, should be proportional to the diffusion length of CREs. As shown in Table~\ref{tab:Table7}, a noticeable difference in the break scale is found when using the radio emission at 1.4 GHz and 0.14\,GHz.  This can be explained by differences in the propagation of CREs traced across the frequency band. The synchrotron radiation at lower frequencies is generated by a lower energy (or older) population of CREs. This population must be the one already propagated to longer distances and have a longer diffusion length.  Hence, depending on the frequency of the synchrotron emission, the CREs diffusion length  is expected to vary.

Following the convention of \cite{intro12&taba}, we use the break scale in the correlation between synchrotron emission and FIR emission at 70\,$\mu$m (break occurs at a scale with $r_{w}\leq~0.5$) as the diffusion length of CREs. For IC~342, we obtain a diffusion length of $l_{\rm dif,1.4}= l_{\rm break,1.4}= 200\,pc $ at 1.4\,GHz. The diffusion length of CREs traced at lower radio frequency of 0.14\,GHz is larger ($l_{\rm dif,0.14} = l_{\rm break,0.14} =320\,pc$). 

Assuming that  the synchrotron and inverse Compton processes are the main sources of the energy loss of CREs \citep[e.g., ][]{intro4&Condon} and that $t_{\rm CRE}=t_{\rm syn}$, these relativistic particles diffuse over a distance equivalent to their diffusion length, which is given by $l_{dif} $ = 2 $ (D\, t_{\rm syn}) ^{0.5}$ before losing all of their energy to synchrotron and inverse Compton losses. Hence, taking into account that  
\begin{equation}
\left(\frac{t_{\rm syn}}{\rm yr}\right) = 8.352 \times 10^{9} \left(\dfrac{E}{\rm GeV}\right)^{-1}\left(\dfrac{B_{tot}}{\rm \mu G}\right)^{-2},
\end{equation}
where
\begin{equation}
\left(\frac{E}{\rm GeV}\right) = \left(\dfrac{\nu}{\rm 16.1 MHz}\right)^{0.5}\left(\dfrac{B_{tot}}{\rm \mu G}\right)^{-0.5},
\end{equation}
the diffusion coefficient $D=l_{dif}^{2}/(4\,t_{syn})$ can be estimated. 
For IC~342 by considering $B_{tot}\simeq 10~ \mu g$, we obtain $ t_{\rm syn}\simeq 2.7 \times 10^{7}$ years. Hence, $D \simeq  1.1 \times 10^{26}$\,cm$^{2}$\,s$^{-1}$ for CREs that emit at 1.4\,GHz ($l_{\rm dif}^{1.4}$= 200\,pc). This is smaller than those assumed or estimated in other galaxies ($D\simeq (1-10)\times 10^{28}$\,cm$^{2}$\,s$^{-1}$) based on theoretical  \citep[][]{Colling1&Roediger} or observational \citep[][]{CREn&Strong,CRE2n&Dahlem} studies. This implies that the assumption of  $t_{\rm CRE}=t_{\rm syn}$ may not be correct in IC~342 and other processes such as the escape of CREs which lead to a larger $l_{\rm dif}$ and $D$ are needed.

}

\subsubsection{Comparing with propagation models}
\label{sec:IC342}

{The CREs propagation models discussed can be tested in nearby galaxies if independent information about their $l_{\rm dif}$ and magnetic field components is available.  Similar multi-scale analysis of the radio--FIR correlation is previously performed for a limited number of galaxies namely M31 \& M33 \citep{intro12&taba}, NGC6946 \cite{intro15&taba}, M51 \cite{intro13&Dumas}, the LMC \cite{intro11&Hughes}, and NGC4214 (Howaida et al. in prep.).  Table~\ref{tab:Table8} lists the diffusion lengths of the galaxies along with their SFR and strengths of the magnetic field components. As the diffusion length for these galaxies is measured only at 1.4\,GHz, we consider only $l_{\rm dif,1.4}$ for IC~342 for the sake of consistency. Plotting $l_{\rm dif}$ from observations vs. that expected from the theory (Eqs.\ref{Eq. 15} and \ref{Eq. 17}) in Fig.~\ref{fig:ldif}, about a linear correlation is found for both cases of $t_{\rm CRE}=t_{\rm syn}$ and $t_{\rm CRE}=t_{\rm esc}$ although the scatter is smaller in the latter case, that is, when the escape of CREs is considered (Eq. \ref{Eq. 17}). We however find the tightest correlation between the observed $l_{\rm dif}$ and the ordered magnetic field. { A bisector OLS fit to the observed data results in
\begin{equation}\label{obs}
l_{\rm dif} \propto B_{\rm ord}^{0.84\pm0.15}.
\end{equation}
This dependency can be explained if I) the diffusion of CREs dominates along the ordered field with a diffusion coefficient of $D_\parallel \propto (B_{\rm ord}/B_{\rm tur})^2 B_{\rm ord}^{-1/3}$ \citep[][Eq. 3.41]{Shalchi} or a length of $l_{\rm dif} \propto (B_{\rm ord}/B_{\rm tur}) B_{\rm ord}^{-1/6} t_{\rm CRE}^{1/2}$ and II) the CREs lifetime is given by a confinement time (i.e., the period that the CREs
spend in the galactic disk) being similar in different galaxies\footnote{As
the scale height of the thin disk is about the same in these galaxies, the CREs
need about the same time to diffuse out of the disk.}, and III) $B_{\rm tur}$ varies less than $B_{\rm ord}$ between galaxies. In this case, we can show that $ l_{\rm dif}\propto B_{\rm ord}^{5/6}$ which is what obtained in Eq.~(\ref{obs}) observationally.

In Fig.~\ref{fig:ldif}-c, NGC~6946 shows a deviation from the best-fit relation given in Eq.~(\ref{obs}). As noted by \cite{intro12&taba}, CRE streaming can dominate diffusion in this galaxy due to the presence of a large-scale ordered field  which is currently being amplified by a
large-scale dynamo mechanism due to the disk-halo interaction \citep{beck07,heald}. In case of streaming,   $l_{\rm prop} \propto v_{A} \propto B_{\rm ord}$ which is steeper than in Eq~\ref{obs}.  
Moreover, $l_{\rm dif}$ in NGC 6946
agrees with the diffusion length expected from synchrotron \citep[$t_{\rm CRE} \sim t_{\rm syn}$, ][]{intro15&taba} unlike in IC~342 and likely more other galaxies.

As explained in Sect.~\ref{sec:magnet}, part of $B_{\rm tur}$ measured can be due to the ordered magnetic field not detected in the polarization data because of the beam depolarization effects or mixing of several uniform field components with different directions along the line of sight. This can weaken the correlations with models which include $B_{\rm ord}/B_{\rm tur}$ such as those represented by Eqs.\ref{Eq. 15} and \ref{Eq. 17}.      
}}

\section{Summary and conclusion}
\label{sec:summ}

This paper presents the LOFAR LoTSS observations of the  radio continuum emission from  the nearby galaxy IC 342 at 0.14\,GHz. Combined with the archival radio data at 1.42\,GHz and 4.85\,GHz as well as
the FIR Herschel observations at 70, 100, and 160 $ \mu m$, we study the radio-FIR correlation and its variation and dependencies on star formation rate and the ISM properties across the galactic disk.
The radio-FIR correlation is calculated using the classical pixel-by-pixel correlation, wavelet scale-by-scale correlation, and the q-method. Using WISE 22 $\mu m$ maps, the radio synchrotron emission is corrected for the thermal contamination. A de-reddened FUV map is used to trace the SFR to study its impact on the CREs  energy spectrum. The strengths of the total, ordered, and turbulent magnetic fields were also mapped to evaluate their role in the radio--FIR correlation and cooling/propagation of CREs  in this galaxy.  

We find that the radio--FIR correlation is tight in star forming regions, however, a large scatter is found in diffuse ISM of IC~342. 
This correlation is insensitive to  the FIR bands indicating that warm and cold dust components are well mixed or that the radiation field does not change much across this galaxy. 

The sub-linear radio synchrotron vs FIR correlation and the variation in the FIR/radio ratio in star forming regions shows that the synchrotron radiation is not produced as efficiently as the FIR emission in spite of the presence of a strong magnetic field in those regions. This indicates that besides the synchrotron cooling, CREs  experience other processes such as escape and diffusion. 
This is further confirmed by the multi-scale analysis of the correlation showing that the correlation breaks down on a scale which is proportional to the diffusion length of CREs  which is set by the regularity of the magnetic field. 
We also show that the scale length of CREs  decreases with radio frequency as expected theoretically. 

Cosmic ray energy spectrum can be affected by both massive star formation activity (a flattening or increased energy) and magnetic fields (a steepening or reduced energy).
Previous studies, however, had difficulty tracing and disentangle these two effects, because the magnetic field strength increases with the SFR itself. Compensating for the effect of SFR, we find, for the first time, that the energy spectrum of CREs  steepens with stronger magnetic fields and flattens with massive star formation rates.
Our study shows that the propagation of cosmic ray electrons in IC 342 is significantly influenced by the magnetic field structure. The diffusion length ($l_{\text{dif}}$), which represents the average distance CREs  travel before scattering.
The \( l_{\text{dif}} \propto B_{\rm ord}^{0.84} \) indicates that the diffusion length increases with the magnetic field strength at a sublinear rate. Physically, this means that stronger magnetic fields enhance diffusion. This finding is crucial for predicting and managing diffusion processes in magnetic environments, providing a clear, quantifiable relationship between the field strength and diffusion length.

\begin{acknowledgement}
This paper is based (in part) on data obtained with the International LOFAR
Telescope (ILT). LOFAR is the Low Frequency Array designed and constructed by
ASTRON. It has observing, data processing, and data storage facilities in several
countries, that are owned by various parties (each with their own funding sources),
and that are collectively operated by the ILT foundation under a joint scientific
policy. The ILT resources have benefitted from the following recent major funding
sources: CNRS-INSU, Observatoire de Paris and Université d'Orléans, France; BMBF,
MIWF-NRW, MPG, Germany; Science Foundation Ireland (SFI), Department of Business,
Enterprise and Innovation (DBEI), Ireland; NWO, The Netherlands; The Science and
Technology Facilities Council, UK; Ministry of Science and Higher Education, Poland. MRC gratefully acknowledges the Canadian Institute for Theoretical Astrophysics (CITA) National Fellowship for partial support; this work was supported by the Natural Sciences and Engineering Research Council of Canada (NSERC).  RJD acknowledges support by the BMBF Verbundforschung Projekt 05A23PC2.
\end{acknowledgement}
 
\bibliographystyle{aa}
\bibliography{refrence}

{
\begin{appendix}
\section{Definition of Star Formation Regimes}
\label{app:star_formation_regimes}

\begin{figure*}[!t]
    \centering
     \subfigure[]{\includegraphics[width=6.5cm]{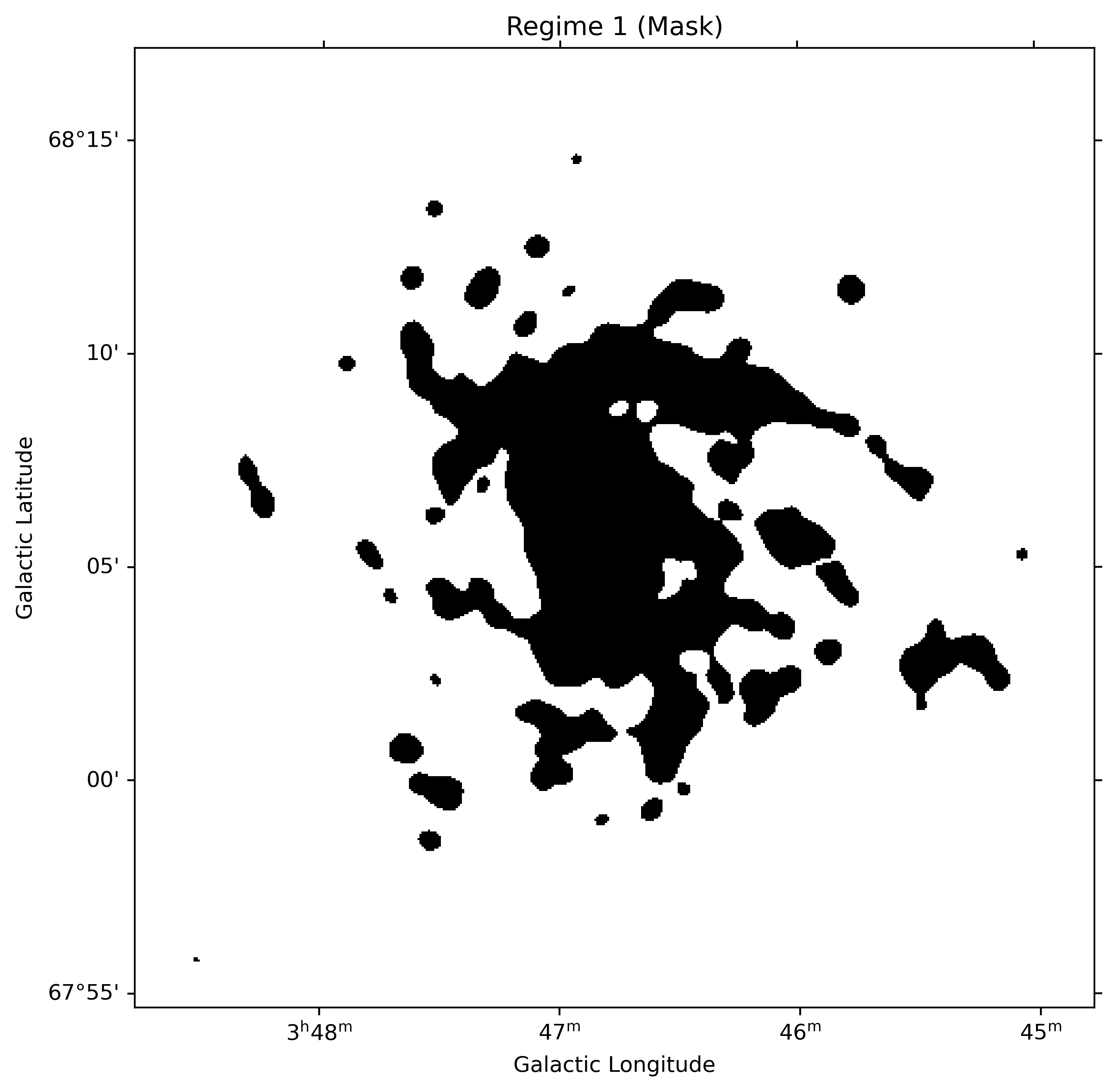}}
     \subfigure[]{\includegraphics[width=6.5cm]{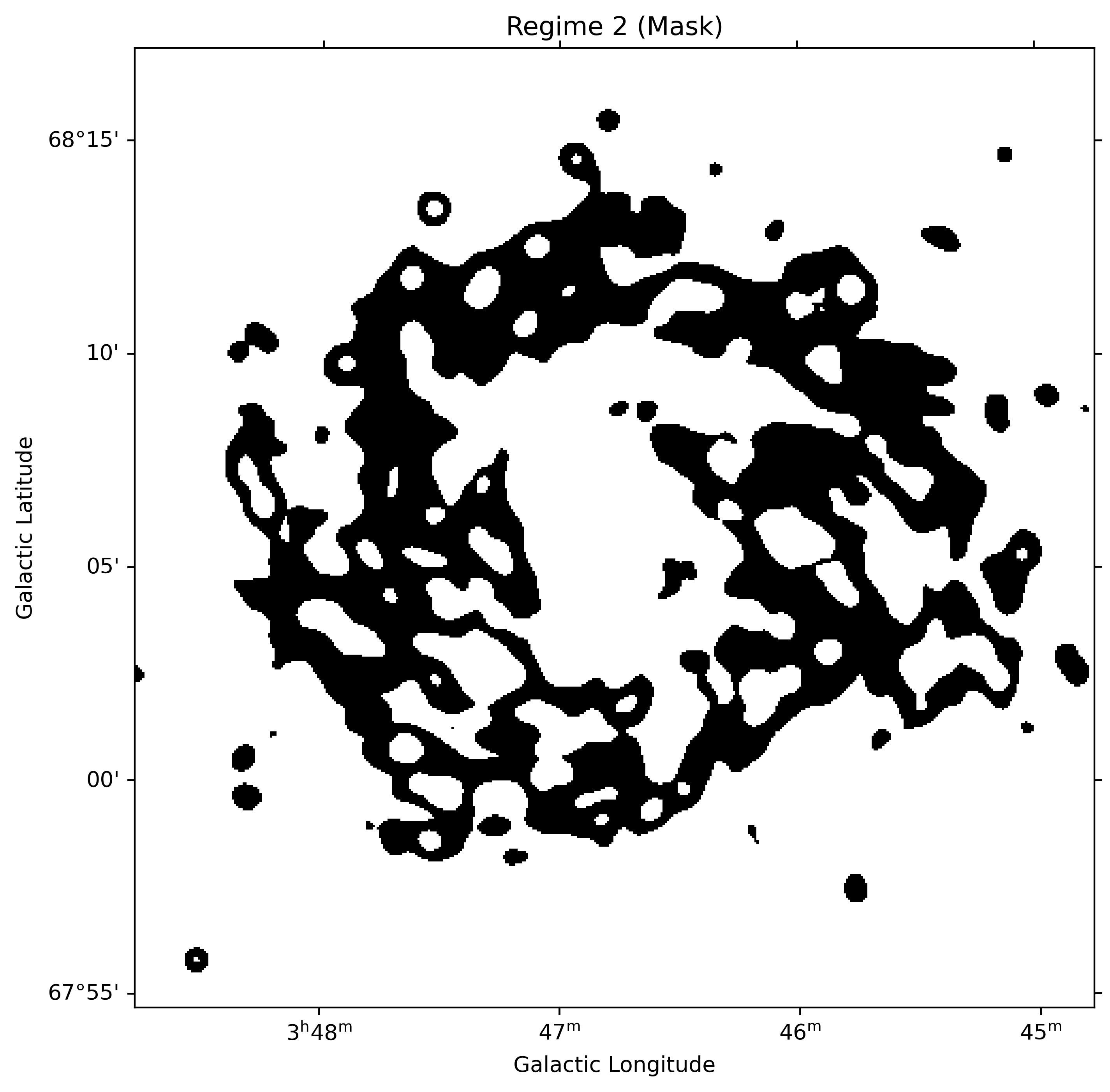}} 

\caption{{Spatial distribution of the highly star-forming R1 regime with $\Sigma_\text{SFR}>$ 3.8 M$\odot$ Gyr$^{-1}$ pc$^{-2}$ {\it (a)} and the R2 regime dominated by diffuse and weakly star forming ISM of IC~342 with $\Sigma_\text{SFR}<$ 3.8 M$\odot$ Gyr$^{-1}$ pc$^{-2}$   {\it (b)}.}}
    \label{fig:masks}
\end{figure*}

In this study, we define two distinct star formation regimes, R1 and R2, based on the star formation rate surface density ($\Sigma_\text{SFR}$) across IC~342. To measure $\Sigma_\text{SFR}$, we use the 22\,$\mu$m map with the calibration relation given by Eq.~\ref{eq:SFR}.
%
The R1 regime corresponds to regions with dense star formation activity. These areas are characterized by a $\Sigma_\text{SFR}$ greater than about 3.8 M$\odot$ Gyr$^{-1}$ pc$^{-2}$, which represents the mean $\Sigma_\text{SFR}$ value across the galaxy. This regime is typically associated with spiral arms and other dense, actively star-forming regions. The R1 regime contributes significantly to the galaxy's star formation and account for about $80\%$ of the total 22\,$\mu$m flux observed.
%
The R2 regime represents regions with more diffuse or less intense star formation activity. These areas are defined by a $\Sigma_\text{SFR}$ 
lower than the mean galaxy value of 3.8 M$_\odot$ Gyr$^{-1}$ pc$^{-2}$. The R2 regime mainly includes inter-arm regions and outer disk of the galaxy which are dominated by quiescent and diffuse ISM.
%
The spatial distributions of the R1 and R2 regimes are shown in Fig.~\ref{fig:masks}  
for a clear visual aid to understanding the spatial extent of these star-formation regimes with respect to the galactic structures. 

\begin{table*}[!htb]
\centering
\label{Table7co}
\caption{Pixel-by-pixel radio-FIR correlations for IC342}
\small
\begin{tabular}{cccccccc}
\toprule
\multirow{2}{*}{Region} & \multirow{2}{*}{Frequency} & \multirow{2}{*}{Component} & \multicolumn{4}{c}{Results} \\
\cmidrule(lr){4-7}
&&& $r_{p}$ & Slope & \(t\)-value & \(n\) \\
\midrule
 \multirow{18}{*}{IC 342}
      & \multirow{3}{*}{4.8 GHz} & OBS vs 70um &$0.87\pm0.05$ & $0.79\pm0.03$&$4.19$ & $184$ \\
      && SYN vs 70um &$0.72\pm0.06$ & $0.73\pm0.04$ & $4.24$ & $184$ \\
      && OBS vs 100um &$0.86\pm0.04$  &$0.75\pm0.02$ &$8.10$ &$427$ \\
      && SYN vs 100um &$0.75\pm0.04$ & $0.67\pm0.06$ & $7.92$ & $379$ \\
      && OBS vs 160um &$0.89\pm0.03$ & $0.69\pm0.01$& $12.27$&$678$ \\
      && SYN vs 160um &$0.74\pm0.04$ & $0.71\pm0.02$ & $11.66$ & $395$ \\
    \cmidrule(lr){2-7}
      & \multirow{3}{*}{1.4 GHz} & OBS vs 70um &$0.80\pm0.05$  & $0.70\pm0.03$&$3.84$ & $182$ \\
      && SYN vs 70um &$0.71\pm0.05$ & $0.67\pm0.03$ & $2.61$ & $182$ \\
      && OBS vs 100um &$0.79\pm0.03$ & $0.58\pm0.01$ & $7.78$ & $429$ \\
      && SYN vs 100um &$0.71\pm0.04$ & $0.57\pm0.02$ & $7.64$ & $384$ \\
      && OBS vs 160um & $0.85\pm0.04$ &$0.48\pm0.01$ & $12.13$&$699$ \\
      && SYN vs 160um &$0.72\pm0.04$ & $0.60\pm0.02$ & $11.57$ & $403$ \\
    \cmidrule(lr){2-7}
     & \multirow{3}{*} {0.1 GHz} & OBS vs 70um & $0.66\pm0.06$ & $0.74\pm0.04$ & $2.94$ & $183$ \\
      && SYN vs 70um &$0.63\pm0.06$ & $0.68\pm0.04$ & $2.99$ & $179$ \\
      && OBS vs 100um &$0.64\pm0.04$ & $0.73\pm0.03$ & $6.79$ & $390$ \\
      && SYN vs 100um &$0.62\pm0.04$ & $0.58\pm0.02$ & $6.65$ & $358$ \\
      && OBS vs 160um &$0.70\pm0.04$ & $0.70\pm0.02$ & $11.26$ & $435$ \\
      && SYN vs 160um &$0.64\pm0.04$ & $0.61\pm0.02$ & $10.94$ & $373$ \\
    \midrule

    \multirow{18}{*}{R1}
      & \multirow{3}{*}{4.8 GHz} & OBS vs 70um &$0.99\pm0.03$ & $0.86\pm0.04$&$2.57$ & $42$ \\
      && SYN vs 70um &$0.96\pm0.04$ & $0.79\pm 0.08$ & $2.61$ & $42$ \\
      && OBS vs 100um &$0.99\pm0.03$  &$0.97\pm0.06$ &$3.41$ &$42$ \\
      && SYN vs 100um &$0.95\pm0.04$ & $0.88\pm0.10$ & $4.43$ & $42$ \\
      && OBS vs 160um &$0.97\pm0.05$ & $1.06\pm0.08$& $4.45$&$42$ \\
      && SYN vs 160um &$0.93\pm0.05$ & $0.97\pm0.12$ & $4.46$ & $42$ \\
    \cmidrule(lr){2-7}
      & \multirow{3}{*}{1.4 GHz} & OBS vs 70um &$0.99\pm0.04$  & $0.81\pm0.06$&$2.51$ & $42$ \\
      && SYN vs 70um &$0.98\pm0.03$ & $0.78\pm0.07$ & $2.61$ & $42$ \\
      && OBS vs 100um &$0.98\pm0.04$ & $0.91\pm0.07$ & $3.37$ & $42$ \\
      && SYN vs 100um &$0.97\pm0.03$ & $0.87\pm0.08$ & $3.43$ & $42$ \\
      && OBS vs 160um & $0.96\pm0.04$ &$1.00\pm0.08$ & $4.43$&$42$ \\
      && SYN vs 160um &$0.95\pm0.04$ & $0.95\pm0.09$ & $4.46$ & $42$ \\
    \cmidrule(lr){2-7}
      & \multirow{3}{*} {0.1 GHz} & OBS vs 70um & $0.94\pm0.04$ & $0.68\pm0.06$ & $2.36$ & $42$ \\
      && SYN vs 70um &$0.88\pm0.07$ & $0.64\pm0.07$ & $2.60$ & $42$ \\
      && OBS vs 100um &$0.95\pm0.04$ & $0.76\pm0.06$ & $3.42$ & $42$ \\
      && SYN vs 100um &$0.90\pm0.06$ & $0.71\pm 0.07$ & $3.42$ & $42$ \\
      && OBS vs 160um &$0.96\pm0.04$ & $0.83\pm0.07$ & $4.40$ & $42$ \\
      && SYN vs 160um &$0.91\pm0.06$ & $0.78\pm0.07$ & $4.45$ & $42$ \\
 \bottomrule

\end{tabular}
\tablefoot{Pixel-by-pixel radio-FIR correlations for the observed radio continuum emission (OBS) and its synchrotron component (SYN)  for the entire IC~342 and star-forming regions (R1). $r_{p}$ is the Pearson Correlation Coefficient, \(t\)-value the Student’s t-test, and n the number of independent data points.}
\end{table*}

\end{appendix}
}

\end{document}